\documentclass[twocolumn]{aastex61}
\pdfoutput=1 %for arXiv submission
\usepackage{amsmath,amstext}
\usepackage[T1]{fontenc}
\usepackage{apjfonts} 
\usepackage[figure,figure*]{hypcap}

\newcommand{\hst}{{\sl HST}}

\shorttitle{Galaxy Redshifts near COS Sight Lines}
\shortauthors{Keeney et~al.}

\begin{document}

\title{A Galaxy Redshift Survey near \hst/COS AGN Sight Lines}

\author{Brian A. Keeney}
\affiliation{Center for Astrophysics and Space Astronomy, Department of Astrophysical and Planetary Sciences, University of Colorado, 389 UCB, Boulder, CO 80309, USA; bkeeney@gmail.com}
\author{John T. Stocke}
\affiliation{Center for Astrophysics and Space Astronomy, Department of Astrophysical and Planetary Sciences, University of Colorado, 389 UCB, Boulder, CO 80309, USA; bkeeney@gmail.com}
\author{Cameron T. Pratt}
\affiliation{Center for Astrophysics and Space Astronomy, Department of Astrophysical and Planetary Sciences, University of Colorado, 389 UCB, Boulder, CO 80309, USA; bkeeney@gmail.com}
\author{Julie D. Davis}
\affiliation{Center for Astrophysics and Space Astronomy, Department of Astrophysical and Planetary Sciences, University of Colorado, 389 UCB, Boulder, CO 80309, USA; bkeeney@gmail.com}
\affiliation{Department of Astronomy, University of Wisconsin, Madison, WI 53706, USA}
\author{David Syphers}
\affiliation{Physics Department, Eastern Washington University, Science 154, Cheney, WA 99004, USA}
\author{Charles W. Danforth}
\affiliation{Center for Astrophysics and Space Astronomy, Department of Astrophysical and Planetary Sciences, University of Colorado, 389 UCB, Boulder, CO 80309, USA; bkeeney@gmail.com}
\author{J. Michael Shull}
\affiliation{Center for Astrophysics and Space Astronomy, Department of Astrophysical and Planetary Sciences, University of Colorado, 389 UCB, Boulder, CO 80309, USA; bkeeney@gmail.com}
\author{Cynthia S. Froning}
\affiliation{Department of Astronomy, University of Texas at Austin, Austin, TX 78712, USA}
\author{James C. Green}
\affiliation{Center for Astrophysics and Space Astronomy, Department of Astrophysical and Planetary Sciences, University of Colorado, 389 UCB, Boulder, CO 80309, USA; bkeeney@gmail.com}
\author{Steven V. Penton}
\affiliation{Space Telescope Science Institute, Baltimore, MD 21218, USA}
\author{Blair D. Savage}
\affiliation{Department of Astronomy, University of Wisconsin, Madison, WI 53706, USA}

\begin{abstract}
To establish the connection between galaxies and UV-detected absorption systems in the local universe, a deep ($g\leq20$) and wide ($\sim20\arcmin$ radius) galaxy redshift survey is presented around 47 sight lines to UV-bright AGN observed by the Cosmic Origins Spectrograph (COS). Specific COS science team papers have used this survey to connect absorbers to galaxies, groups of galaxies, and large-scale structures, including voids. Here we present the technical details of the survey and the basic measurements required for its use, including redshifts for individual galaxies and uncertainties determined collectively by spectral class (emission-line, absorption-line, and composite spectra) and completeness for each sight line as a function of impact parameter and magnitude. For most of these sight lines the design criteria of $>90$\% completeness over a $>1$~Mpc region down to $\la0.1\,L^*$ luminosities at $z\leq0.1$ allows a plausible association between low-$z$ absorbers and individual galaxies. Ly$\alpha$ covering fractions are computed to approximate the star-forming and passive galaxy populations using the spectral classes above. In agreement with previous results, the covering fraction of star-forming galaxies with $L\geq0.3\,L^*$ is consistent with unity inside one virial radius and declines slowly to $>50$\% at 4 virial radii. On the other hand, passive galaxies have lower covering fractions ($\sim60$\%) and a shallower decline with impact parameter, suggesting that their gaseous halos are patchy but have a larger scale-length than star-forming galaxies. All spectra obtained by this project are made available electronically for individual measurement and use.
\end{abstract}


\section{Introduction}
\label{sec:intro}

In the past few years it has become evident that the internal structure and evolution of galaxies are affected greatly by gas surrounding these galaxies in what has come to be called the Circum-Galactic Medium \citep[CGM;][]{tumlinson11,tumlinson17}. Numerical simulations support the notion that the CGM helps to fuel and regulate star formation in the disk of spiral galaxies \citep[e.g.,][]{muratov15,muratov17,hayward17}, solving the ``G-dwarf problem'' \citep{pagel08} and maintaining high star formation rates over the luminous lifetimes of these galaxies \citep{binney87}. {At the same time, CGM gas around ``passive'' galaxies \citep{thom12} must somehow be prevented from accreting onto its associated galaxy.}

In both cases the CGM is an active element in the cosmic evolution of galaxies, although the details remain obscure. The CGM is supplied with gas by a combination of galactic outflows from nearby galaxies \citep{veilleux05}, close passages and collisions between galaxies, and direct accretion from the Inter-Galactic Medium \citep[IGM;][]{keres09}. At the outer extremes, the CGM interfaces with the IGM, which connects individual galaxies with larger structures in the distribution of galaxies like galaxy groups and filamentary large-scale structures of galaxies. We discuss these connections, whether they be the circulation of gas between the disk and CGM of an individual galaxy \citep{werk14,tumlinson17} or the larger distribution of this gas in a cosmological context. To determine how far metals spread from individual galaxies \citep{stocke06,pratt18}, two datasets are required: UV absorption-line detections of the gas, and the distribution of galaxies around the regions probed by these UV detections.

The accumulation of dozens of high-S/N, far-UV spectra of bright QSOs has been the recent work of the Cosmic Origins Spectrograph (COS) on board the {\sl Hubble Space Telescope} \citep[\hst;][]{green12}. Numerous absorption-line detections of \ion{H}{1} and metal ions are made in these spectra allowing the most sensitive probe available of tenuous, highly ionized gas in the CGM and IGM. While several major programs of COS observations have been undertaken, many of these \citep[e.g., the COS-Halos program described in][]{tumlinson11,thom12,werk14,prochaska17} use COS spectra of moderate $\mathrm{S/N}=10$-15 to probe strong absorption associated with the CGM of a single, targeted galaxy and are not of sufficient quality to detect diffuse \ion{H}{1} at $N_{\rm H\,I} \leq 10^{13.5}~\mathrm{cm}^{-2}$ (all subsequent column density values will be quoted in units of $\mathrm{cm}^{-2}$).

However, the COS Science Team (a.k.a. Guaranteed Time Observers or GTOs) obtained higher-S/N COS G130M/G160M spectra at $\mathrm{S/N}=15$-50, allowing the detection of much weaker \ion{H}{1} Lyman-series and metal lines (in some cases with limiting column densities of $\log{N_{\rm H\,I}} \leq 12.8$). These COS-GTO spectra are presented in several science papers \citep[chiefly,][]{stocke13,savage14,danforth16,keeney17} and, along with COS far-UV spectra of similar quality obtained by other observers, are archived at the Mikulski Archives for Space Telescopes\footnote{\url{https://archive.stsci.edu/prepds/igm/}} (MAST) as detailed in \citet{danforth16}. This publication and its associated archival database  \dataset[doi:10.17909/T95P4K]{http://dx.doi.org/10.17909/T95P4K} include column densities of \ion{H}{1} and metal ions detected at all redshifts along the QSO sight line.

To support a variety of scientific investigations that require the association of the UV absorption-line detection of CGM/IGM gas with nearby galaxies or the galaxy distribution around the absorber, the COS GTOs instigated a ground-based spectroscopy program to obtain redshifts and spectroscopic diagnostics of galaxies near the AGN sight lines. The purpose of this paper is to describe and present these data, which have already been used to support the scientific investigations of the low-$z$ CGM around star-forming and passive galaxies \citep{stocke13,stocke14,stocke17,savage12,savage14,keeney13,keeney17}. Since the numbers of high-S/N COS far-UV spectra has increased dramatically since the initiation of this campaign, we have restricted our survey to the original 40~COS-GTO sight lines. These were supplemented by 10~sight lines that probe SDSS groups of galaxies, which were approved for \hst\ Cycle~23 and are now being analyzed \citep[see e.g.,][]{stocke17}.

Since the COS-GTO investigations into the gas-galaxy relationship focus on the lowest redshifts ($z \leq 0.25$). We used multi-object spectroscopy (MOS) on moderate aperture (3-4~meter class) telescopes to access very wide fields ($\gtrsim20\arcmin$ radius) at moderate depth ($g \leq 20$). This MOS field-of-view allows the observation of galaxies within $\geq1$~Mpc for all absorber redshifts $z \geq 0.03$, so that virtually all absorbers have regions probed that are larger than the inferred CGM radius (assumed to be comparable to the virial radius of the associated galaxy, which is $\sim250$~kpc for an $L^*$ galaxy). The MOS depth for this survey extends a factor of $\sim5$ below the limits of the Sloan Digital Sky Survey (SDSS) spectroscopic survey and reaches $L^*$ limiting luminosities at $z \approx 0.25$ and to $0.1\,L^*$ or below at $z \la 0.1$. This survey intends to provide sufficient galaxy coverage to obtain redshifts for all plausible galaxies which can be associated with an absorber at these low redshifts. Recent work suggests that the sub-$L^*$ (i.e., $0.3 \leq L/L^* \leq 1.0$) galaxy population is the primary source for the CGM gas \citep[][{but see \citealp{johnson17} for a contrasting view}]{tumlinson05,prochaska11,burchett16,pratt18}. {Sensitivity to sub-$L^*$ galaxies is a basic design specification for this galaxy redshift survey.} 

\startlongtable
\begin{deluxetable*}{lcrccl}
\tablecaption{COS GTO Sight Lines Surveyed
\label{tab:gto_slines}}

\tablehead{ \colhead{Sight Line} & \colhead{RA}      & \colhead{Dec}     & \colhead{$z_{\rm em}$\tablenotemark{a}} & \colhead{$E(B-V)$} & \colhead{Abbrev} \\[-1ex]
                                 & \colhead{(J2000)} & \colhead{(J2000)} &                                        & \colhead{(mag)} }

\colnumbers
\startdata
1ES~1028+511      & 10:31:18.52 &   50:53:35.8 & 0.360 & 0.0117 & 1es1028  \\
1ES~1553+113      & 15:55:43.04 &   11:11:24.4 & 0.360 & 0.0448 & 1es1553  \\
1SAX~J1032.3+5051 & 10:32:16.14 &   50:51:19.7 & 0.173 & 0.0145 & 1sax1032 \\
3C~57             & 02:01:57.19 & --11:32:33.2 & 0.671 & 0.0190 & 3c57     \\
3C~263            & 11:39:57.04 &   65:47:49.4 & 0.646 & 0.0096 & 3c263    \\
FBQS~J1010+3003   & 10:10:00.69 &   30:03:21.6 & 0.256 & 0.0223 & fbqs1010 \\
H~1821+643        & 18:21:57.31 &   64:20:36.4 & 0.297 & 0.0370 & h1821    \\
HE~0153--4520     & 01:55:13.21 & --45:06:11.8 & 0.451 & 0.0128 & he0153   \\
\enddata

\tablenotetext{\mathrm{a}}{The emission-line redshift of the QSO as listed in the NASA Extragalactic Database (NED), except for HE~0435--5304, whose redshift was measured from its COS spectrum \citep{stocke13}.}
\tablenotetext{\mathrm{b}}{This sight line was never observed by \hst/COS but we include the measured galaxy redshifts for legacy value.}

\end{deluxetable*}

In this paper we present the basic survey results, including all spectra acquired in addition to the galaxy redshifts obtained, allowing subsequent analysis by future investigations. \autoref{sec:obs} includes the description of the survey design and execution for the GTO (\autoref{sec:obs:gto}) and \hst\ groups (\autoref{sec:obs:group}). \autoref{sec:results} details the procedure for determining redshifts (\autoref{sec:results:redshifts}) and includes an overall summary of the survey completeness (\autoref{sec:results:completeness}). In \autoref{sec:results:accuracy} we analyze the redshift accuracy using both internal and external comparisons. Our results are discussed in \autoref{sec:disc} and summarized in \autoref{sec:conc}. \autoref{sec:appendix} contains survey completeness tables for each sight line. Throughout this paper we adopt WMAP9 cosmological values from \citet{hinshaw13}, including $H_0 = 69.7~\mathrm{km\,s}^{-1}\,\mathrm{Mpc}^{-1}$, $\Omega_{\Lambda} = 0.718$ and $\Omega_{\rm m}=0.282$.

\section{Observations}
\label{sec:obs}

The design and execution of two separate but related galaxy redshift surveys are detailed below. The first surveys galaxies near AGN sight lines observed by the \hst/COS GTO team, and the second searches for additional group members in SDSS galaxy groups probed by \hst/COS.

\subsection{COS GTO Galaxy Redshift Survey}
\label{sec:obs:gto}

The COS GTO galaxy redshift survey was designed to obtain redshifts for all galaxies with $g<20$ that are within $20\arcmin$ of each of the 40~AGN sight lines targeted by the \hst/COS GTO team. We successfully observed galaxies near 38 of these sight lines, whose names, positions, and redshifts are listed in \autoref{tab:gto_slines} along with the color excess along the line of sight as measured by \citet{schlafly11} and the abbreviations used to identify the sight lines in the supplementary data products. The two sight lines that were observed by the \hst/COS GTO team, but not included in this survey, are (1) PKS~0003+148, which was hard to access from the southern hemisphere when we were observing sight lines with similar RAs, and (2) PKS~0405--123, which has extensive galaxy survey results to fainter magnitudes already in the literature \citep[e.g.,][]{chen09,prochaska11,johnson13}.

\begin{deluxetable*}{lccC}
\tablecaption{Journal of Imaging Observations toward COS GTO Sight Lines
\label{tab:gto_img}}

\tablehead{ \colhead{Sight Line} &  \colhead{Telescope} & \colhead{Date} & \colhead{Exposures (ksec)} }

\colnumbers
\startdata
%3C~57           & WIYN 0.9-m & 2008 Oct 2                & 3.0,3.0,3.0 \\
3C~57            & Blanco     & 2008 Sep 23               & 0.5,0.5,0.5 \\
HE~0153--4520    & Blanco     & 2008 Sep 22-23            & 0.6,0.6,0.6 \\
HE~0226--4110    & Blanco     & 2008 Sep 21               & 0.5,0.5,0.5 \\
%HE~0238--1904   & WIYN 0.9-m & 2008 Oct 3                & 3.0,3.0,3.0 \\
%HE~0238--1904   & Blanco     & 2008 Sep 22-23            & 0.6,0.6,0.6 \\
HE~0435--5304    & Blanco     & 2008 Sep 21               & 0.5,0.5,0.5 \\
HE~0439--5254    & Blanco     & 2008 Sep 23               & 0.5,0.5,0.5 \\
%HS~1939+700     & WIYN 0.9-m & 2008 Oct 2-3, 2009 Feb 21 & 3.3,3.3,3.3 \\
%PG~0832+251     & WIYN 0.9-m & 2008 Feb 8, 2009 Feb 21   & 2.9,2.9,2.9 \\
PHL~1811         & Blanco     & 2008 Sep 22               & 0.5,0.5,0.5 \\
%PKS~0003+158    & WIYN 0.9-m & 2008 Oct 3                & 1.8,1.8,1.8 \\
PKS~0003+158     & Blanco     & 2008 Sep 21               & 0.5,0.5,0.5 \\
%PKS~0405--123   & WIYN 0.9-m & 2008 Feb 8, 2008 Oct 2-3  & 5.1,5.1,5.1 \\
PKS~0405--123    & Blanco     & 2008 Sep 23               & 0.5,0.5,0.5 \\
\enddata

\tablecomments{Column~4 lists the total exposure time in the SDSS $g,r,i$ filters.}
% PMN~J1103--2329 was not observed in $i$-band.

\end{deluxetable*}

Whenever possible, we used photometry {and photometric redshifts ($z_{\rm phot}$)} from the Sloan Digital Sky Survey \citep[SDSS;][]{alam15} to choose our spectroscopic targets {(see below)}. {SDSS $z_{\rm phot}$ values have typical uncertainties of $\sigma_{\rm phot} = 0.025$ \citep{beck17} and catastrophic failure rates (differences in spectroscopic and photometric redshift determinations that exceed $3\sigma_{\rm phot}$) of $\approx1.6$\% (see \citealp{beck17} for a detailed discussion).} When SDSS imaging was unavailable (primarily in the southern hemisphere) we obtained our own images of the sight lines using the MOSAIC imagers of the Blanco 4-m telescope at Cerro Tololo Inter-American Observatory{\footnote{\url{https://www.noao.edu/ctio/mosaic/}}} ($40\arcmin\times40\arcmin$ field of view) and the WIYN 0.9-m telescope at Kitt Peak National Observatory{\footnote{\url{http://www.astro.wisc.edu/our-science/research-observatories/wiyn-09m-telescope/mosaic/}}} ($60\arcmin\times60\arcmin$ field of view).

These observations are detailed in \autoref{tab:gto_img}, which lists the sight-line name, telescope, observation date, and exposure time in the SDSS $g,r,i$ filters, respectively. {While the imaging depth achieved depends on the observing conditions, typical limiting magnitudes for point sources in Blanco data are 23.9, 24.2, and 24.2 in $g$-, $r$-, and $i$-band, respectively. In WIYN 0.9-m data we typically reach 22.1~mag in $g$-band, 22.3~mag in $r$-band, and 21.9~mag in $i$-band.}

Spectroscopic targets were identified by generating a list of all galaxies with $g<20$ located within $20\arcmin$ of the AGN sight line, and removing those galaxies with known redshifts. The remaining targets were assigned priority levels based on their brightness and photometric redshift (when available). Bright galaxies with $g<18$ were given highest priority, regardless of photometric redshift. Fainter galaxies with $18<g<20$ were given lower priority, and only {targeted} if their photometric redshifts were {no more than $\sigma_{\rm phot}$ larger than the AGN redshift}. The lowest priority targets were those that fell outside our completeness goals (i.e., galaxies with $g>20$ and/or positions $>20\arcmin$ from the AGN sight line), and were only {targeted} if their photometric redshifts were {no more than $\sigma_{\rm phot}$ larger than the AGN redshift}. We observed objects with known redshifts{, or with photometric redshifts above our thresholds,} as ``extra'' targets in a configuration only when no higher-priority galaxies could be accommodated. {In fields where photometric redshifts from SDSS are unavailable, spectroscopic target selection was based solely on apparent $g$-band magnitude and proximity to the QSO sight line.}

Multi-object spectroscopy for this survey was performed with the HYDRA spectrograph on the WIYN 3.5-m telescope at Kitt Peak National Observatory {\citep{barden93,bershady08}} or the AA$\Omega$ spectrograph on the 3.9-m Anglo-Australian Telescope at Siding Springs Observatory {\citep{sharp06}}. \autoref{tab:gto_mos} lists the sight-line name, telescope, observation date(s), and exposure time per fiber configuration {for our spectroscopic observations}.

At WIYN/HYDRA, we used the 600@10.1 grating centered at 5200~\AA\ with the blue cables and the Bench Camera, yielding a spectral resolution of $\mathcal{R}\approx1200$ over the wavelength range 3800-6600~\AA. {While the S/N achieved generally decreases as the galaxy's apparent magnitude increases, the relationship is not straightforward due to variations in observing conditions and galaxy surface brightness. The S/N of spectra obtained with this instrument is $11^{+8}_{-5}$ per pixel, and the $g$-band magnitude of targeted galaxies is $19.4^{+0.5}_{-0.7}$; quoted values are medians and ranges indicate the 16th and 84th percentile values.}

At AAT/AA$\Omega$, we used the 580V and 385R gratings centered at 4800 and 7250~\AA, respectively, yielding a spectral resolution of $\mathcal{R}\approx1300$ over the wavelength range 3700-8800~\AA. {The S/N of spectra obtained with this setup is $6^{+10}_{-3}$ per pixel, and the galaxy $g$-band magnitude range is $19.9^{+0.7}_{-1.3}$.}

With both instruments, we chose our wavelength coverage to ensure we were sensitive to \ion{Ca}{2} H \& K absorption from $z\approx0$ galaxies. Most redshifts in this survey were obtained with WIYN/HYDRA, so the wavelength of H$\alpha$ is not usually covered, and direct measurement of galaxy star formation rate is not possible.

\clearpage
\startlongtable
\begin{deluxetable*}{lllL}
\tablecaption{Journal of COS GTO MOS Observations
\label{tab:gto_mos}}

\tablehead{ \colhead{Sight Line} & \colhead{Telescope} & \colhead{Date} & \colhead{Exposures (ksec)} }

\colnumbers
\startdata
1ES~1028+511      & WIYN   & 2010 Feb 12-13                                            & 13.5,10.8               \\
1ES~1553+113      & WIYN   & 2010 Feb 12-15                                            & 10.8,8.1,8.1            \\
1SAX~J1032.3+5051 & WIYN   & 2010 Feb 15                                               & 8.1,8.1                 \\
3C~57             & AAT    & 2013 Sep 7                                                & 10.2                    \\
3C~263            & WIYN   & 2008 Feb 7-8                                              & 8.1,8.1                 \\
FBQS~J1010+3003   & WIYN   & 2008 Feb 9-11                                             & 8.1,8.1,8.1             \\
H~1821+643        & WIYN   & 2014 Jun 25-26                                            & 8.1,8.1,8.1             \\
HE~0153--4520     & AAT    & 2012 Aug 12,14                                            & 8.1,8.1                 \\
\enddata

\tablecomments{Column~4 lists the total exposure time per configuration.}

\end{deluxetable*}

\begin{figure}
  \epsscale{1.0}
  \centering\plotone{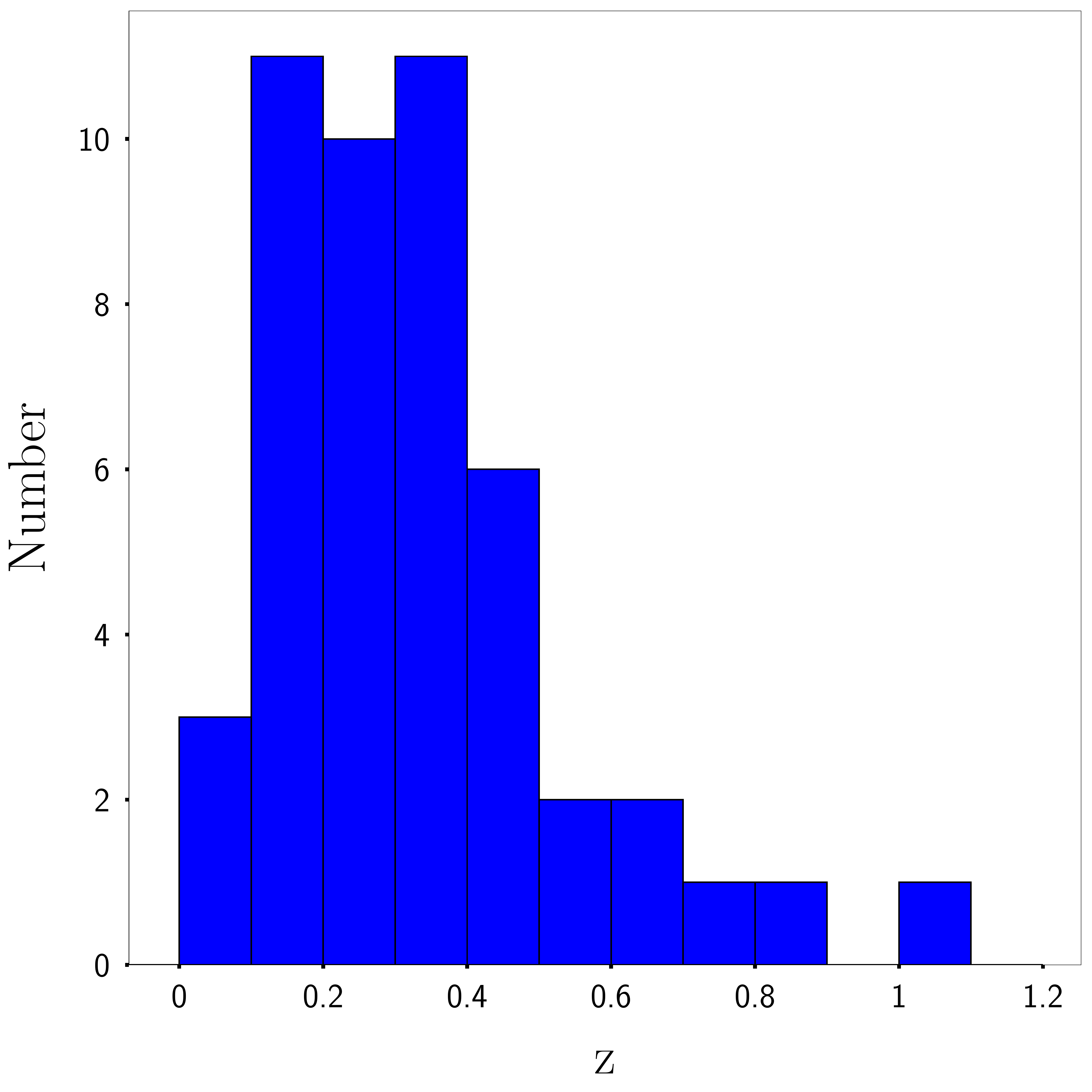}
  \vspace{-1ex}
  \caption{{Histogram showing the distribution of AGN redshifts from \autoref{tab:gto_slines} and \autoref{tab:group_slines}.}
  \label{fig:zqso}}
\end{figure}

\begin{figure}
  \epsscale{1.0}
  \centering\plotone{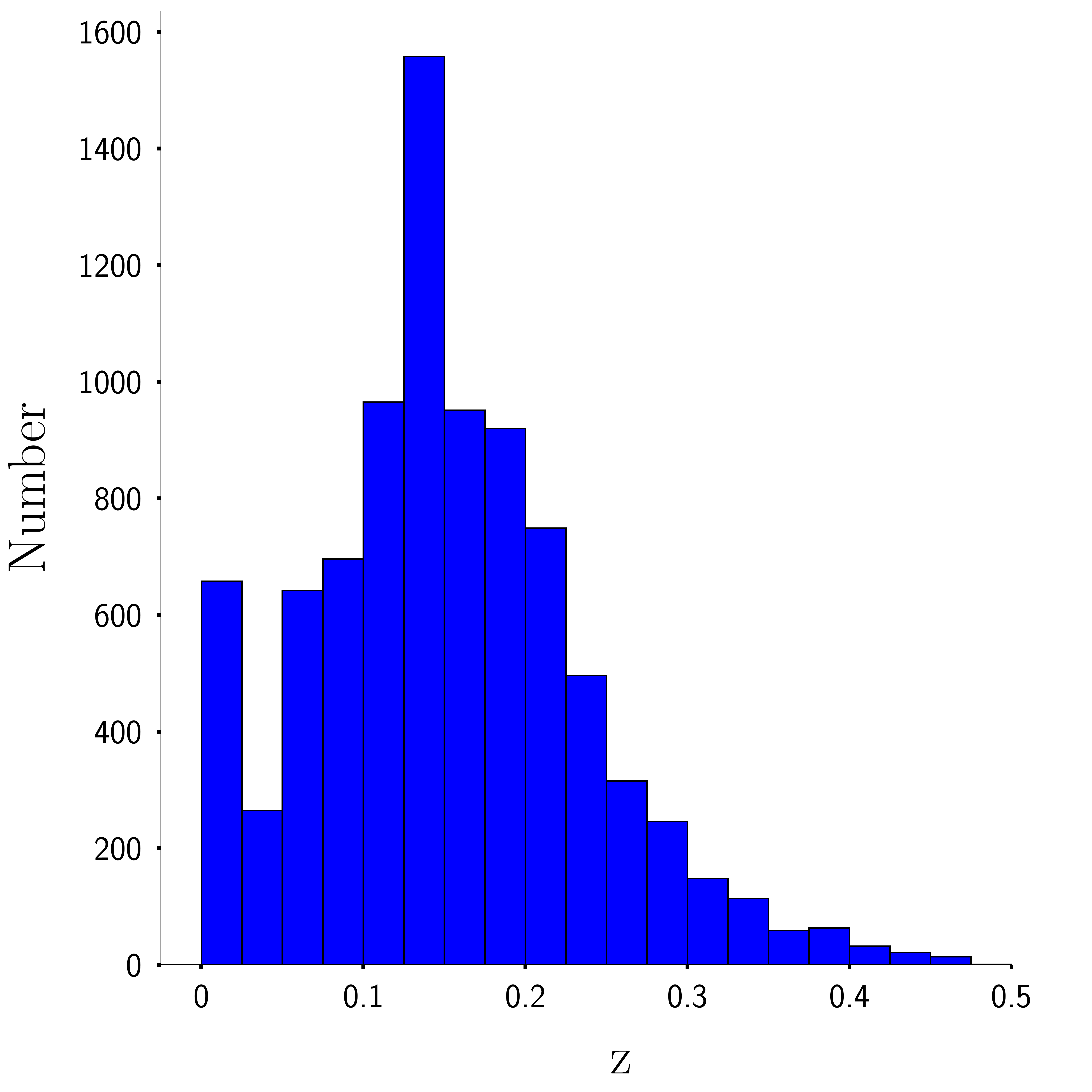}
  \vspace{-1ex}
  \caption{{Histogram showing the distribution of galaxy redshifts from \autoref{tab:redshifts}. Sixteen galaxies with $z>0.5$ are not shown, and the lowest redshift bin has a significant contribution from misclassified stars.}
  \label{fig:zgal}}
\end{figure}

\subsection{Galaxy Group Survey}
\label{sec:obs:group}

The galaxy group survey was designed to support \hst/COS observations of AGN that probe SDSS-selected galaxy groups \citep{stocke17}. These groups have modest numbers of SDSS galaxies ($N=3$-10), so we have designed a survey aimed at increasing the number of group members to $N\gtrsim20$ per group. This allows  various observed (e.g., group velocity dispersion) and inferred (e.g., group halo mass) group properties to be well-constrained \citep{stocke18}. The 10~sight lines observed as part of this study are listed in \autoref{tab:group_slines}, along with their positions, redshifts, color excesses, and the abbreviations used for the sight lines in the supplementary data products.

Unlike the COS GTO galaxy redshift survey, this survey was not designed to be complete around the AGN sight line. Instead, we placed fibers on galaxies with unknown redshift near the SDSS group centers. Bright galaxies ($g<18$) within $20\arcmin$ of the group center were given highest priority regardless of their photometric redshift, followed by fainter galaxies ($18<g<20$) with photometric redshifts within $2\sigma$ of the SDSS group redshift, then galaxies with $g<20$ that were located $20\arcmin$-$30\arcmin$ from the group center and had photometric redshifts within $2\sigma$ of the group redshift. We observed fewer configurations per sight line for this survey because our goal was not high completeness, but rather a fair sampling of potential group members. We were not concerned with identifying all possible group members so long as we found enough so that the group is well-characterized.

Multi-object spectroscopy for this survey was performed with WIYN/HYDRA and MMT/Hectospec {\citep{fabricant05}}. A journal of our observations is shown in \autoref{tab:groups_mos}, which lists the sight line name, telescope, observation date(s), and exposure time per fiber configuration. Four of the 10 sight lines in this survey were observed with both WIYN/HYDRA and MMT/Hectospec.

\begin{deluxetable*}{lcrccl}
\tablecaption{SDSS Group Sight Lines Surveyed
\label{tab:group_slines}}

\tablehead{ \colhead{Sight Line} & \colhead{RA}      & \colhead{Dec}     & \colhead{$z_{\rm em}$\tablenotemark{a}} & \colhead{$E(B-V)$} & \colhead{Abbrev} \\[-1ex]
                                 & \colhead{(J2000)} & \colhead{(J2000)} &                                        & \colhead{(mag)} }

\colnumbers
\startdata
B~1612+266        & 16:14:10.62 &   26:32:50.5 & 0.395 & 0.0361 & b1612    \\
CSO~1022          & 13:53:26.12 &   36:20:49.5 & 0.285 & 0.0130 & cso1022  \\
CSO~1080          & 15:05:27.60 &   29:47:18.4 & 0.526 & 0.0189 & cso1080  \\
FBQS~J1030+3102   & 10:30:59.09 &   31:02:55.7 & 0.178 & 0.0168 & fbqs1030 \\
FBQS~J1519+2838   & 15:19:36.15 &   28:38:27.6 & 0.270 & 0.0227 & fbqs1519 \\
RBS~711           & 08:36:58.91 &   44:26:02.3 & 0.255 & 0.0244 & rbs711   \\
SBS~0956+509      & 09:59:31.67 &   50:44:49.1 & 0.143 & 0.0154 & sbs0956  \\
SDSS~J1028+2119   & 10:28:14.56 &   21:19:55.1 & 0.374 & 0.0201 & sdss1028 \\
SDSS~J1333+4518   & 13:33:00.83 &   45:18:09.0 & 0.320 & 0.0136 & sdss1333 \\
SDSS~J1540--0205  & 15:40:19.46 & --02:05:05.4 & 0.320 & 0.1309 & sdss1540 \\
\enddata

\tablenotetext{\mathrm{a}}{The emission-line redshift of the QSO as listed in the NASA Extragalactic Database (NED).}

\end{deluxetable*}
\begin{deluxetable*}{lllL}
\tablecaption{Journal of SDSS Group MOS Observations
\label{tab:groups_mos}}

\tablehead{ \colhead{Sight Line} & \colhead{Telescope} & \colhead{Date} & \colhead{Exposures (ksec)} }

\colnumbers
\startdata
B~1612+266        & MMT    & 2016 May 10                                               & 3.6                     \\
                  & WIYN   & 2017 Mar 4                                                & 7.2                     \\
CSO~1022          & MMT    & 2016 Jun 7,9                                              & 3.6,3.6                 \\
CSO~1080          & MMT    & 2016 Jun 7,9                                              & 3.6,3.6                 \\
FBQS~J1030+3102   & MMT    & 2017 Feb 24-25                                            & 3.6,3.6                 \\
FBQS~J1519+2838   & MMT    & 2016 Jun 7,8,28                                           & 3.6,3.6,3.6             \\
RBS~711           & MMT    & 2017 Feb 2                                                & 3.6,3.6                 \\
                  & WIYN   & 2016 Mar 12 ,2017 Mar 3-4                                & 8.1,8.1,2.7             \\
SBS~0956+509      & MMT    & 2016 Jun 9, 2017 Feb 21                                   & 3.6,3.6                 \\
SDSS~J1028+2119   & MMT    & 2017 Feb 23                                               & 3.6                     \\
                  & WIYN   & 2016 Mar 12, 2017 Mar 4                                   & 1.5,8.1                 \\
SDSS~J1333+4518   & MMT    & 2016 May 7, 2016 Jun 28                                   & 3.6,3.6                 \\
                  & WIYN   & 2016 Mar 12                                               & 4.5                     \\
SDSS~J1540--0205  & MMT    & 2016 May 8                                                & 3.6                     \\
\enddata

\tablecomments{Column~4 lists the total exposure time per configuration.}

\end{deluxetable*}
\clearpage
\movetabledown=1in

\begin{rotatetable}
\begin{deluxetable*}{lDDCcccDcDDcC}
\tablecaption{Galaxy Redshift Measurements
\label{tab:redshifts}}

\tablehead{ \colhead{Galaxy} & \twocolhead{RA}      & \twocolhead{Dec}     & \colhead{$z_{\rm gal}$} & \colhead{$g$}   & \colhead{$r$}   & \colhead{$i$}   & \twocolhead{$L$}     & \colhead{$R_{\rm vir}$} & \twocolhead{$\log{M_{\rm h}}$} & \twocolhead{$\log{M_*}$} & \colhead{$\rho$} & \colhead{Abs. Flag} \\
            \colhead{}       & \twocolhead{(J2000)} & \twocolhead{(J2000)} & \colhead{}             & \colhead{(mag)} & \colhead{(mag)} & \colhead{(mag)} & \twocolhead{($L^*$)} & \colhead{(kpc)}        & \twocolhead{}                 & \twocolhead{}            & \colhead{(kpc)}  & \colhead{} }

\decimalcolnumbers
\startdata
1es1028\_114\_31     & 157.838763 &  50.888281 & 0.35949\pm0.00007 & 19.8 & 18.6 & 18.0 &     7.8 &     360 &   12.42 &   11.26 &     157 &  0 \\
1es1028\_230\_207    & 157.772796 &  50.847103 & 0.17212\pm0.00009 & 19.8 & 18.6 & 18.1 &     1.0 &     185 &   11.55 &   10.64 &     610 &  3 \\
1es1028\_282\_227    & 157.731542 &  50.912542 & 0.17881\pm0.00004 & 20.1 & 19.5 & 19.3 &    0.57 &     150 &   11.28 &    9.60 &     690 &  3 \\
1es1028\_79\_284     & 157.946487 &  50.917817 & 0.17215\pm0.00008 & 19.7 & 18.6 & 18.1 &     1.1 &     189 &   11.58 &   10.65 &     838 &  1 \\
1es1028\_129\_287    & 157.905429 &  50.830550 & 0.10874\pm0.00005 & 19.7 & 18.8 & 18.4 &    0.32 &     125 &   11.04 &   10.00 &     573 &  3 \\
1es1028\_35\_374     & 157.893050 &  50.988531 & 0.13707\pm0.00004 & 18.8 & 18.1 & 17.8 &     1.1 &     189 &   11.58 &   10.25 &     912 &  2 \\
1es1028\_42\_418     & 157.916971 &  50.994692 & 0.13692\pm0.00005 & 19.9 & 19.5 & 19.3 &    0.35 &     129 &   11.08 &    9.32 &    1019 &  1 \\
1es1028\_137\_447    & 157.927413 &  50.786206 & 0.00000\pm0.00021 & 18.3 & 16.9 & 16.3 &     .   & \nodata &     .   &     .   & \nodata & -1 \\
1es1028\_269\_550    & 157.585000 &  50.887650 & 0.20929\pm0.00006 & 19.7 & 19.1 & 18.8 &     1.2 &     190 &   11.59 &   10.02 &    1888 &  3 \\
1es1028\_323\_555    & 157.722146 &  51.032556 & 0.04439\pm0.00004 & 19.7 & 19.0 & 18.7 &   0.044 &      73 &   10.34 &    8.93 &     488 &  1 \\
1es1028\_113\_557    & 158.031367 &  50.807325 & 0.21438\pm0.00006 & 19.4 & 18.0 & 17.6 &     3.1 &     267 &   12.03 &   11.05 &    1954 &  0 \\
1es1028\_193\_609    & 157.790308 &  50.725447 & 0.25776\pm0.00009 & 19.8 & 18.6 & 18.1 &     3.3 &     269 &   12.04 &   10.95 &    2454 &  3 \\
\enddata

\tablecomments{All masses are in units of $M_{\Sun}$, and $\rho$ is the impact parameter between the galaxy and AGN sight-line positions. Absorption flag values have the following meanings: $-2$ = no COS spectrum available; $-1$ = object has $z<0.001$ and is likely a star; 0 = galaxy is not within $1000~\mathrm{km\,s^{-1}}$ of an absorber; 1 = galaxy is within $1000~\mathrm{km\,s^{-1}}$ of an absorber but is not the closest galaxy; 2 = galaxy is closest in this Table to an absorber, but a closer galaxy is known from SDSS or other sources; 3 = closest known galaxy to an absorber. This table is available in its entirety in machine-readable form.}

\end{deluxetable*}
\end{rotatetable}
\clearpage

At WIYN/HYDRA we again used the 600@10.1 grating centered at 5200~\AA\ ($\mathcal{R}\approx1200$ from 3800-6600~\AA); {the S/N of these spectra is $9^{+5}_{-3}$ per pixel, and the galaxy $g$-band magnitude range is $19.4^{+0.5}_{-0.7}$.} At MMT/Hectospec, we used the 270~gpm grating to achieve $\mathcal{R}\approx1000$ from 3700-9100~\AA; {the S/N of these spectra is $7^{+6}_{-3}$ per pixel, and the galaxy $g$-band magnitude range is $19.8^{+0.7}_{-0.8}$.} As before, we chose our wavelength coverage to ensure sensitivity to \ion{Ca}{2} H \& K absorption at $z\approx0$.

\section{Analysis and Results}
\label{sec:results}

{\autoref{fig:zqso}-\ref{fig:zgal} show the distribution of AGN redshifts from \autoref{tab:gto_slines} and \autoref{tab:group_slines}, and galaxy redshifts from \autoref{tab:redshifts}, respectively. All but seven of the AGN sight lines have $z_{\rm em}<0.5$, and most of the galaxies whose redshifts we retrieve (see \autoref{sec:results:redshifts}) have $z_{\rm gal}<0.2$.}

Most of our sight lines are in the SDSS footprint, which allowed us to use photometry and photometric redshifts from the SDSS SkyServer\footnote{\url{http://skyserver.sdss.org}} for spectroscopic target selection. For other sight lines, reduced and co-added versions of MOSAIC images taken at CTIO were provided by the National Optical Astronomy Observatory (NOAO) Science Archive\footnote{\url{http://archive.noao.edu}}. MOSAIC images obtained at the WIYN 0.9-m telescope were reduced and co-added using the mosaic reduction package ({\tt mscred}) of IRAF {\citep{valdes98}}. Galaxy positions and photometry were extracted from co-added MOSAIC images using the Picture Processing Package \citep[PPP;][]{yee91} for use in spectroscopic target selection for these sight lines.

All HYDRA data were reduced and extracted using the NOAO's {\tt hydra} IRAF package{\footnote{\url{http://iraf.noao.edu/tutorials/dohydra/dohydra.html}}}. Data from AAT/AA$\Omega$ were reduced and extracted using the Australian Astronomical Observatory's stand-alone reduction software {\tt 2dfdr}{\footnote{\url{https://www.aao.gov.au/science/instruments/current/AAOmega/reduction}}}. Finally, MMT/Hectospec data were reduced in IDL using the {\tt HSRed} reduction pipeline{\footnote{\url{http://www.mmto.org/node/536}}}. The end product of all of these reduction procedures are one-dimensional, wavelength-calibrated text files for each galaxy observed as part of a configuration; if an individual galaxy was observed more than once then it will have an extracted spectrum for each observation. A rudimentary spectrophotometric flux calibration is applied for HYDRA data, but AA$\Omega$ and Hectospec data have extracted fluxes and uncertainties in units of $\mathrm{counts\,s^{-1}}$.

\subsection{Redshift Measurement}
\label{sec:results:redshifts}

Galaxy redshifts are assigned from a by-eye verification and correction of initial automated redshifts. The automated routine takes three approaches. The first is a cross-correlation with SDSS spectral templates\footnote{\url{http://www.sdss.org/dr7/algorithms/spectemplates/}}, but this was generally found to be an ineffective method due to the poor spectrophotometry of the spectra, particularly near the bandpass edges. More effective are the two line-search methods, the first of which looks for emission lines (\ion{Mg}{2} 2799~\AA, [\ion{O}{2}] 3728~\AA, H$\beta$, [\ion{O}{3}] 4959/5007~\AA, H$\alpha$) and the second of which looks for absorption lines (\ion{Ca}{2} H \& K, G-band, \ion{Mg}{1} b, \ion{Na}{1} D). An \`a trous wavelet \citep*[wavelet ``with holes''; i.e., edge avoiding;][]{starck97} is applied to the spectra, with parameters optimized on a training set for our resolution and lines of interest. Cuts are made to find lines above the S/N. The routine searches for doublet lines (\ion{Ca}{2} H \& K, [\ion{O}{3}] 4959/5007~\AA), and searches for line identifications that maximize other detected lines matching to known lines. Consistent redshifts from several lines together are given higher probability, as are stronger lines.

This approach is generally effective for good S/N galaxies, particularly those with emission lines. However, its accuracy is not enough to enable raw use of the automated results. Sometimes lines are present, beneath our S/N threshold, or the automated guess for the strongest line was mistaken. Noise vectors are also sometimes incorrect. The results of the automated routine are thus verified and corrected by eye, with the program's guesses for each method (cross-correlation, emission lines, and absorption lines) presented to the user to accept or correct. If the redshift needed correction, the user would correct a known line to its proper location, and the program would center on this line.

The final object redshift estimates are made by using whichever method was verified as ``good'', combining emission and absorption redshifts if they were both good and agreed with each other to within $\delta z = 0.002$. If they disagreed, emission redshifts were preferred. Uncertainty was estimated from agreement of all lines, with an additional 15~$\mathrm{km\,s^{-1}}$ added in quadrature to account for wavelength calibration uncertainty.

\autoref{tab:redshifts} lists basic and derived information for the nearly 9,000~galaxies observed as part of the COS GTO or galaxy group redshift surveys for which we have redshift determinations. The basic information includes the galaxy name, sky position, redshift, and apparent $g$-, $r$- and $i$-band magnitudes. In addition, the galaxy's luminosity, virial radius, halo mass, stellar mass, and impact parameter with respect to the QSO sight line are listed. {We are disseminating all of the individual galaxy spectra we have acquired \dataset[doi:10.17909/T9XH52]{http://dx.doi.org/10.17909/T9XH52} as a MAST High-Level Science Product\footnote{\url{https://archive.stsci.edu/prepds/igm-gal/}}.}

The galaxy name is a combination of the sight-line abbreviation (\autoref{tab:gto_slines} and \ref{tab:group_slines}), the galaxy's position angle with respect to the sight line (in degrees), and the galaxy's angular distance from the sight line (in arcsec)\footnote{For example, the first galaxy in \autoref{tab:redshifts} is located $31\arcsec$ from the 1ES~1028+511 sight line at a position angle of $114\degr$.}. The tabulated luminosities are calculated in the rest-frame $g$-band \citep[$M_g^*=-20.3$;][]{montero-dorta09} using $K$-corrections from \citet*{chilingarian10} and \citet{chilingarian12}. {A galaxy's virial radius and halo mass are estimated from its rest-frame $g$-band luminosity using the prescription of \citet{stocke13}, and the stellar mass is calculated from the galaxy's rest-frame $i$-band luminosity using Equation~8 of \citet{taylor11}}.

For the luminosity calculations, the apparent magnitudes in \autoref{tab:redshifts} are corrected for the effects of Galactic foreground extinction using the sight-line color excesses (\autoref{tab:gto_slines} and \autoref{tab:group_slines}) and the reddening law of \citet{fitzpatrick99} with $R_V=3.1$. The analytic $K$-corrections of \citet{chilingarian10} are only defined for $z<0.5$, so the rare objects in \autoref{tab:redshifts} at larger redshift have their $K$-corrections evaluated at $z=0.5$ and should be treated with caution. We also constrain the galaxy colors to the range $-0.1 \leq g-r \leq 1.9$ and $0 \leq g-i \leq 3$ (in the observed frame) when evaluating the $K$-corrections and $-0.2 \leq g-i \leq 1.6$ (rest-frame) when determining the stellar mass to ensure that we are not extrapolating beyond the color range used to define the relationships.

The final column of \autoref{tab:redshifts} can be used to determine whether a galaxy is near a COS absorption-line system: a value of 3 indicates that a galaxy is the closest known galaxy within $20\arcmin$ of an IGM \ion{H}{1} absorber; a value of 2 indicates that it is the closest galaxy in \autoref{tab:redshifts} to an absorber, but a closer galaxy is known from SDSS or other sources; a value of 1 indicates that a galaxy is within $1000~\mathrm{km\,s^{-1}}$ of an absorber but is not the closest galaxy to that absorber; and a value of 0 indicates that a galaxy is not within $1000~\mathrm{km\,s^{-1}}$ of any absorber. Absorption-line redshifts are taken from the \hst/COS IGM survey of \citet{danforth16} and the galaxy groups analysis of \citet{stocke18}. Three-dimensional galaxy-absorber distances, $D$, are calculated using a reduced-Hubble-flow model with $v_{\rm red}=400~\mathrm{km\,s^{-1}}$, and the ``closest'' galaxy is defined to be the galaxy within $20\arcmin$ and $1000~\mathrm{km\,s^{-1}}$ of the absorber that has the smallest $D/R_{\rm vir}$ {(see \autoref{sec:disc:galabs} for details)}.

\begin{deluxetable*}{lCCCCCCCCC}
\tablecaption{Redshift Comparisons for Duplicate Observations
\label{tab:accuracy}}

\tablehead{ \colhead{Sample} & \multicolumn{3}{c}{Absorption-Line Galaxies} & \multicolumn{3}{c}{Composite Galaxies} & \multicolumn{3}{c}{Emission-Line Galaxies} \\[-1ex] & \colhead{$N$} & \colhead{$\overline{\Delta v}$} & \colhead{$\sigma_v$} & \colhead{$N$} & \colhead{$\overline{\Delta v}$} & \colhead{$\sigma_v$} & \colhead{$N$} & \colhead{$\overline{\Delta v}$} & \colhead{$\sigma_v$} }

\colnumbers
\startdata
AAT             &  24 &  13\pm47 & 197\pm45 &  35 &  20\pm13 & 78\pm11 &   7 &   4\pm13 & 32\pm14 \\
MMT             &  71 &  12\pm12 & 101\pm10 & 202 &   4\pm 3 & 54\pm 3 &  46 &  11\pm 8 & 52\pm 6 \\
WIYN            & 253 &  -9\pm 5 &  92\pm 4 & 544 &  -2\pm 4 & 87\pm 3 &  71 &   6\pm 7 & 60\pm 5 \\
Combined        & 348 &  -4\pm 5 & 100\pm 4 & 781 &   0\pm 3 & 81\pm 2 & 124 &  10\pm 5 & 57\pm 4 \\
\tableline
WIYN + MMT      &  37 & -49\pm20 & 118\pm18 &  60 & -39\pm10 & 76\pm 7 &  20 & -12\pm10 & 43\pm10 \\
External (SDSS) & 154 &  -9\pm 4 &  60\pm 3 &  94 & -10\pm 5 & 50\pm 4 &  12 &  -8\pm 3 & 13\pm 3 \\
\enddata

\tablecomments{The mean velocity offset, $\overline{\Delta v}$, and standard deviation, $\sigma_v$, are given in units of $\mathrm{km\,s^{-1}}$.}

\end{deluxetable*}

Despite our efforts to avoid stars when conducting our galaxy redshift surveys, they were occasionally observed. In \autoref{tab:redshifts} we assume that any object with $z<0.001$ is a star, in which case we set the final column to $-1$ and do not calculate any of the derived galaxy quantities (luminosity, virial radius, halo mass, stellar mass, or impact parameter).

Finally, one of the GTO sight lines included in our galaxy redshift survey (SDSS~J1439+3932) was never observed with COS. We include the redshift measurements for these galaxies in \autoref{tab:redshifts}, but set the final column to $-2$ since the absorber locations are unknown.

\subsection{Redshift Accuracy}
\label{sec:results:accuracy}

While the formal redshift uncertainties listed in \autoref{tab:redshifts} and described above are the only method we have to assign uncertainty to individual redshift determinations, additional estimates are possible using aggregates of galaxies with similar properties. We perform both internal and external validations of the redshifts provided in \autoref{tab:redshifts} using galaxies that we observed multiple times and galaxies that have SDSS redshift measurements, respectively.

\subsubsection{Internal Consistency}
\label{sec:results:accuracy:internal}

There are two reasons that individual galaxies were observed multiple times in our program. The first is that faint galaxies ($g>19$) were sometimes assigned to multiple configuration files to increase their S/N. The second reason is that a single configuration file was sometimes observed multiple times due to weather or instrument problems with the first observation, or because a particular sight line was preferentially positioned during an available observation window.

All told, our surveys include 2,153~galaxies that were observed twice or more (the maximum number of observations for a single galaxy is six). After removing any observations that did not result in a redshift determination, we calculated the velocity difference ($\Delta v = c\,\Delta z$) between individual redshift measurements and sorted them by telescope/instrument combination. Next, the duplicate spectra were visually classified as emission-line galaxies with strong emission and little to no absorption, absorption-line galaxies with strong absorption and little to no emission, or composite galaxies that show evidence of significant emission and absorption. Finally, we calculated the average and standard deviation of the velocity offsets for each instrument and type of galaxy. \autoref{tab:accuracy} displays the results of our analysis.

Sight lines in the COS GTO survey were never observed with more than one spectrograph (\autoref{tab:gto_mos}). This means that redshift comparisons for these sight lines primarily characterize the consistency of our redshift measurement procedure (\autoref{sec:results:redshifts}). The first three rows of \autoref{tab:accuracy} are for data taken with AAT/AA$\Omega$, MMT/Hectospec, and WIYN/HYDRA, respectively. They show that none of the instruments has a large characteristic offset between repeated redshift measurements of any galaxy type, with typical values of $|\overline{\Delta v}|<10~\mathrm{km\,s^{-1}}$. Further, the velocity distributions for emission-line galaxies ($\sigma_v \sim 50~\mathrm{km\,s^{-1}}$) are consistently narrower than the distributions for absorption-line galaxies ($\sigma_v \sim 100~\mathrm{km\,s^{-1}}$), with the composite galaxies having intermediate values. There are hints of differences in characteristic widths between the instruments, but the small sample sizes of the AAT and MMT datasets preclude us from drawing firm conclusions.

\begin{figure}
  \epsscale{1.0}
  \centering\plotone{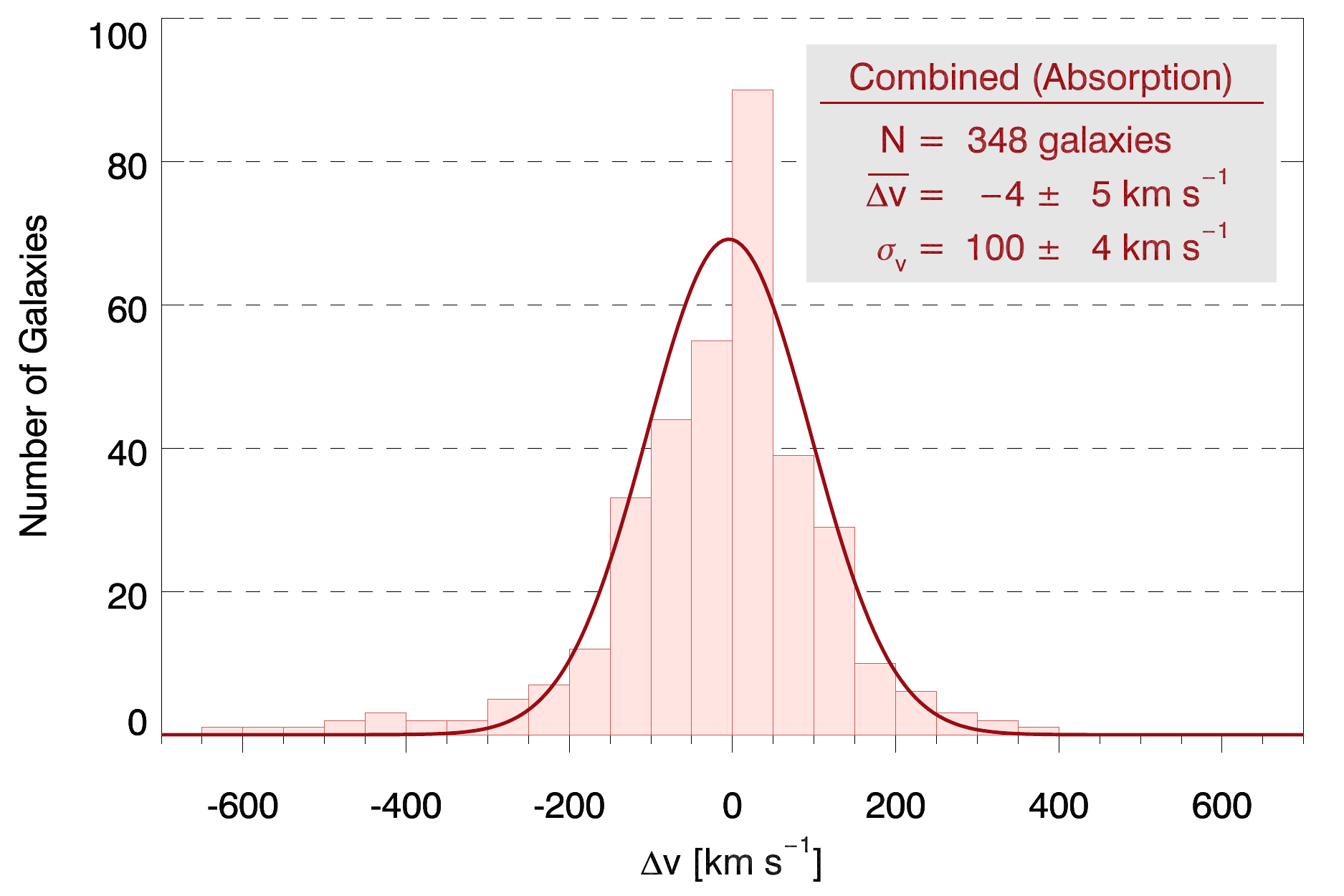}
  \centering\plotone{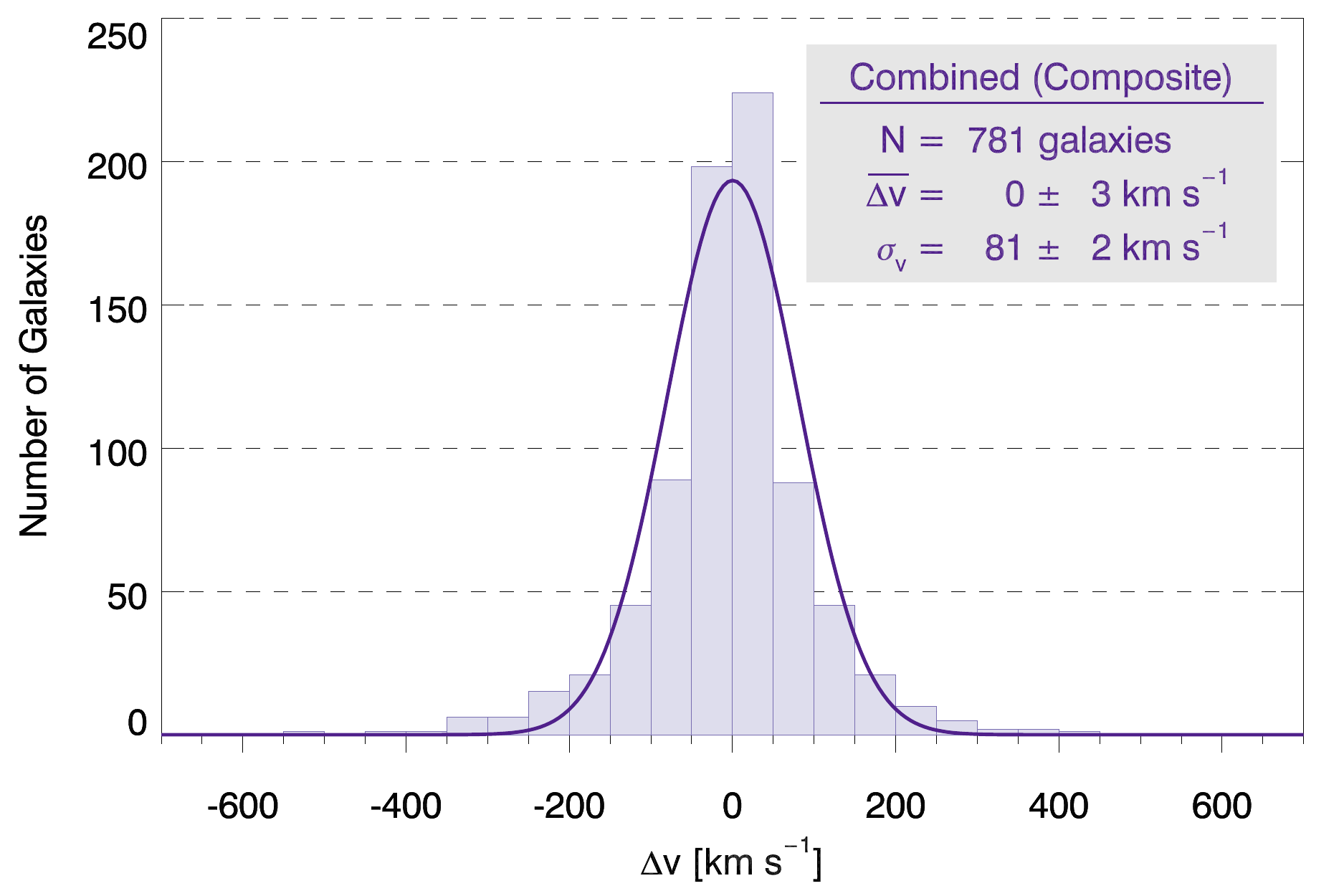}
  \centering\plotone{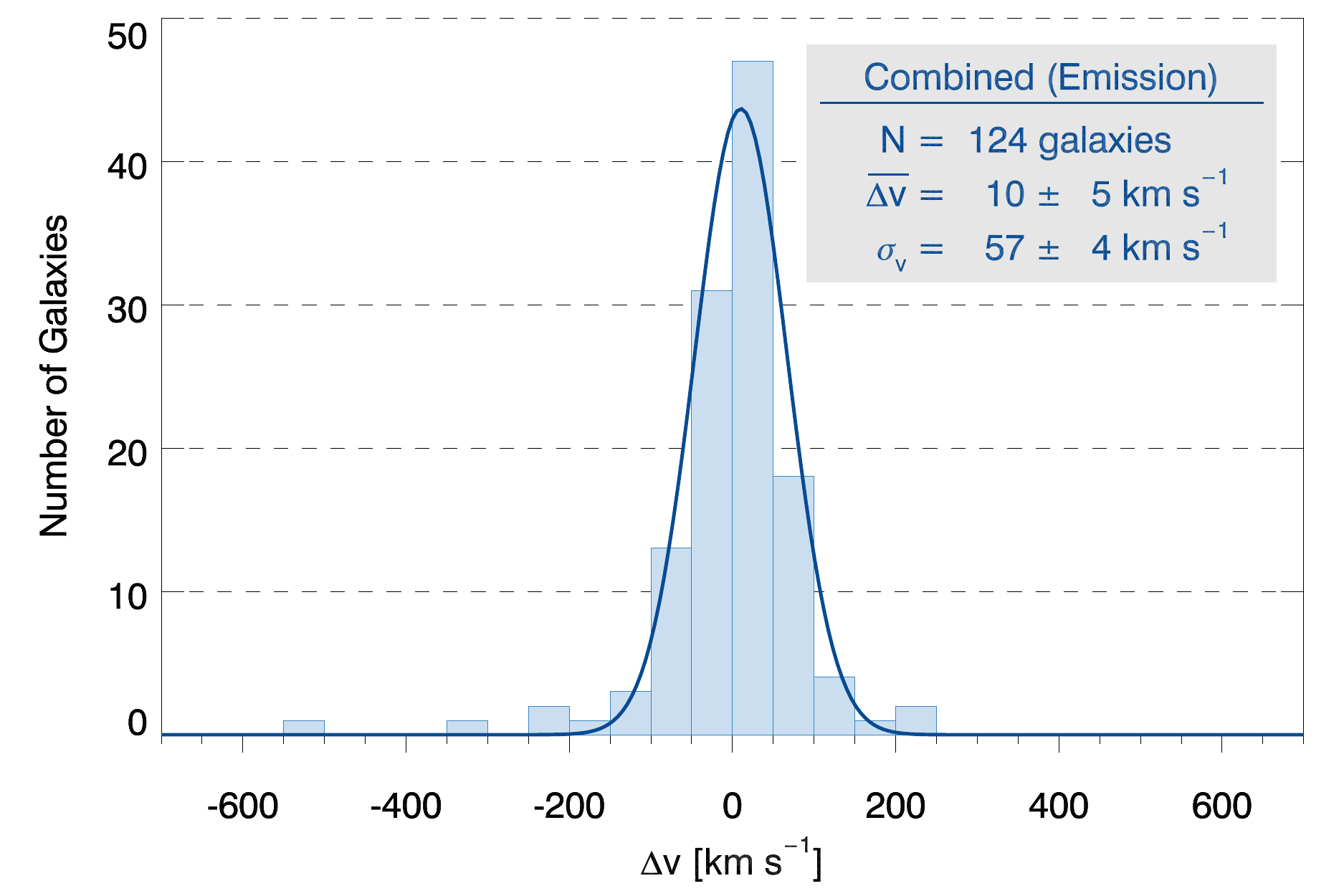}
  \vspace{-1em}
  \caption{Velocity offsets for galaxies with multiple observations from a single spectrograph. Emission-line galaxies have a narrower distribution than composite and absorption-line galaxies, and no significant systematic offsets are found.
  \label{fig:accuracy}}
\end{figure}

The fourth row of \autoref{tab:accuracy} shows the results of combining the previous three rows to maximize the sample size for any galaxy type. The distribution of velocity offsets for this combined sample is shown in \autoref{fig:accuracy} for each galaxy type. The overlaid Gaussians are not fits to the data, but rather expectations based on the measured means and standard deviations if the data are normally distributed (i.e., the only free parameters are mean, $\overline{\Delta v}$, and standard deviation, $\sigma_v$; the amplitude is proportional to $\sigma_v^{-1}$). In all cases the Gaussian expectations from \autoref{tab:accuracy} are a reasonable description of the data.

\begin{figure}
  \epsscale{1.0}
  \centering\plotone{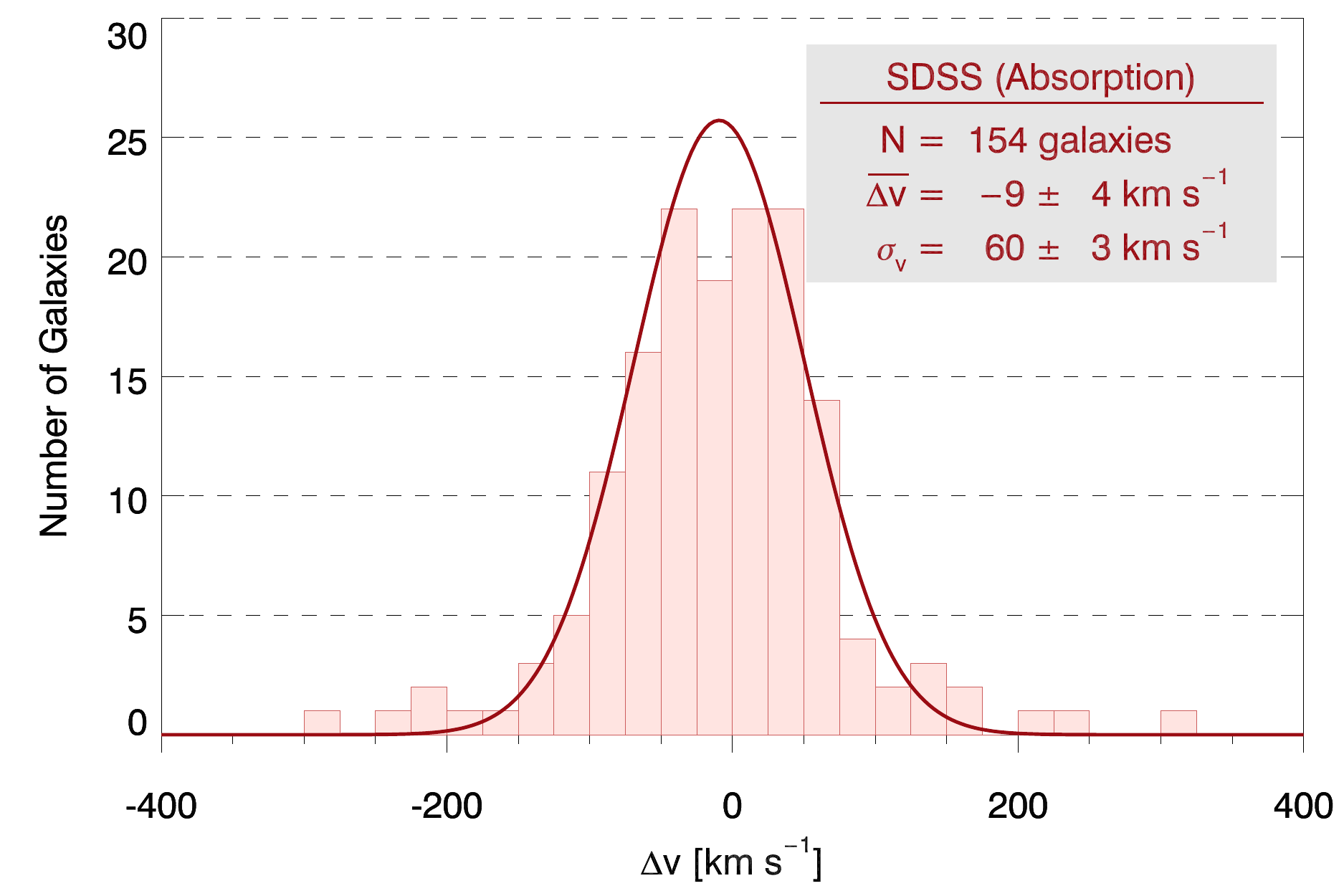}
  \centering\plotone{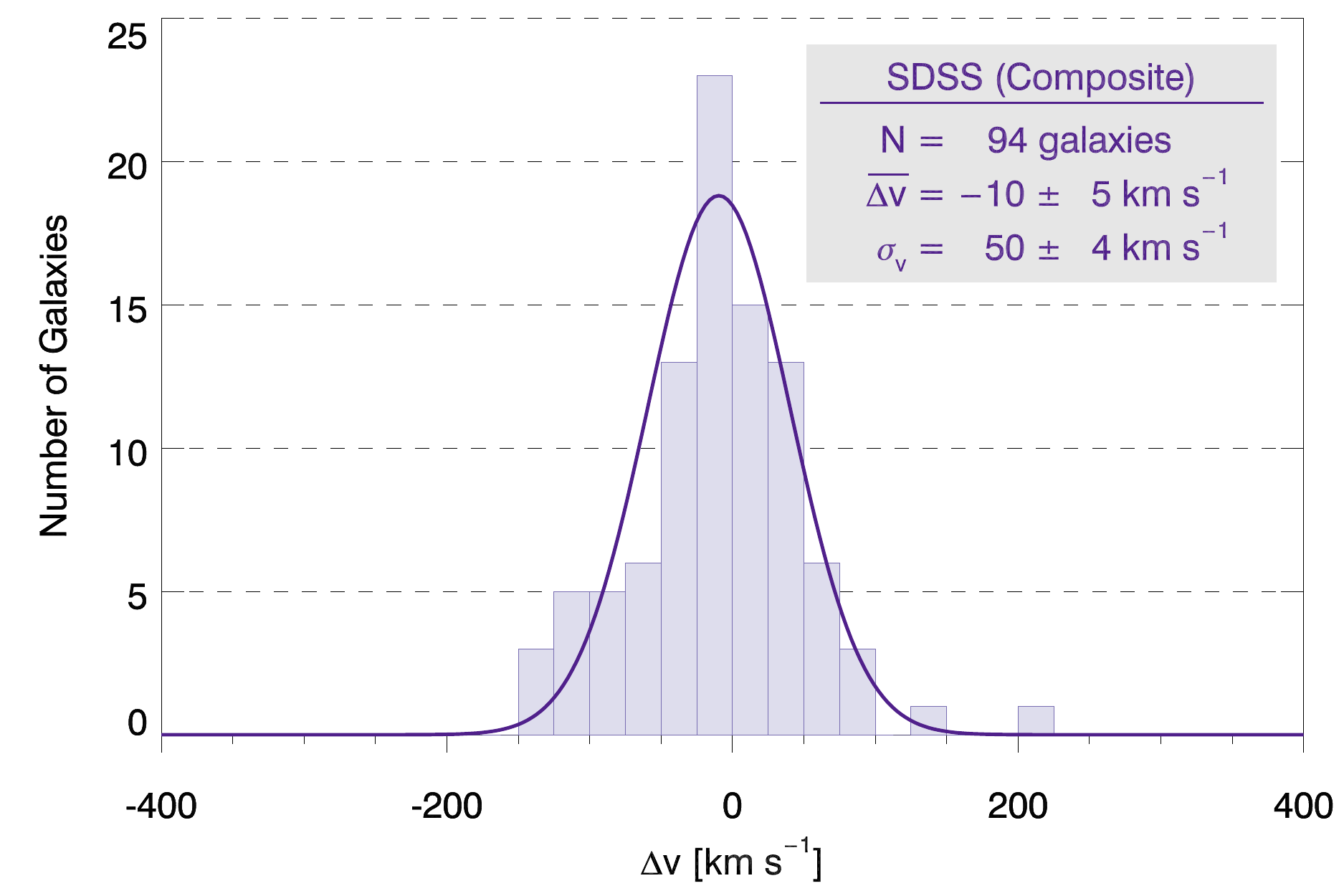}
  \centering\plotone{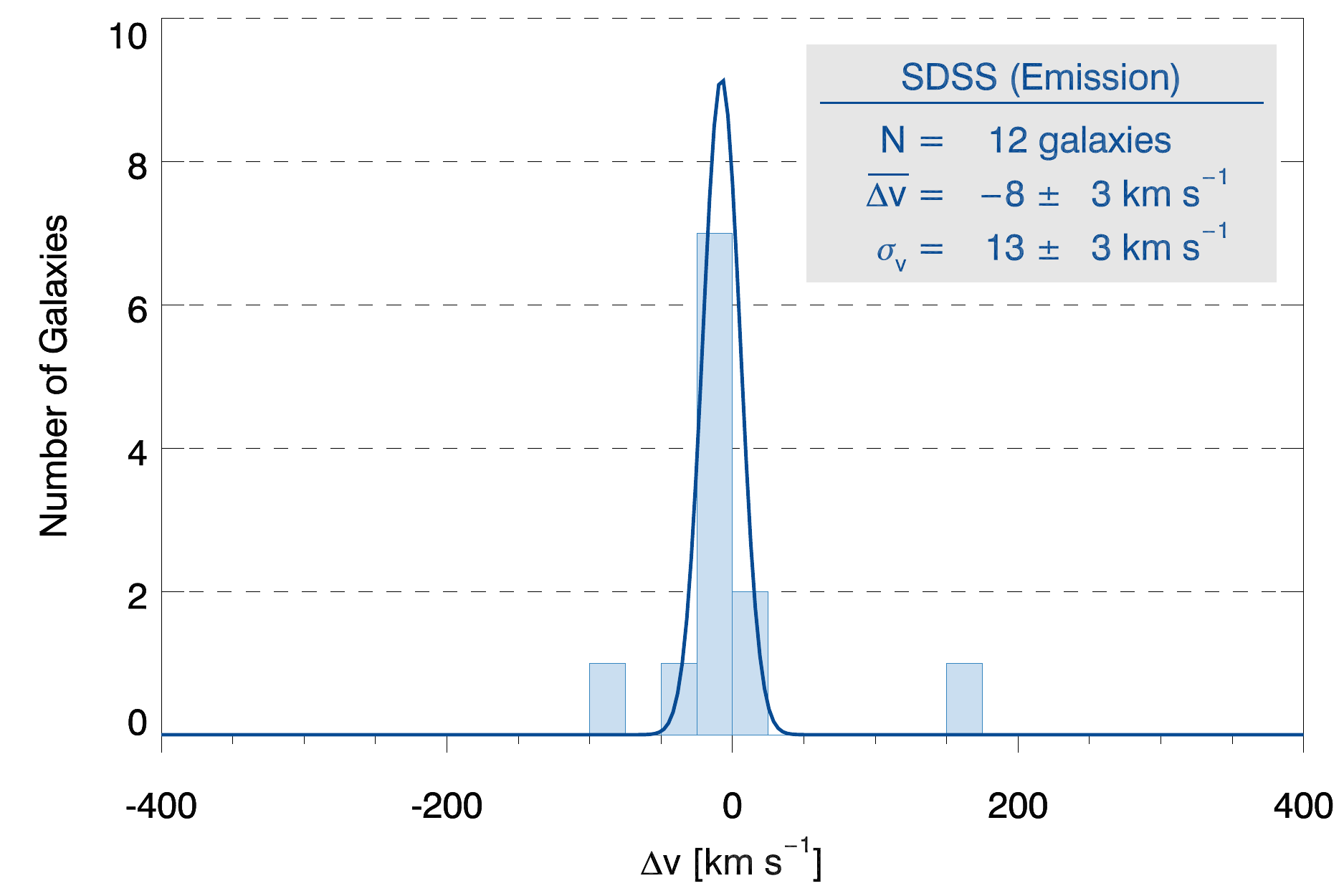}
  \vspace{-1em}
  \caption{Velocity offsets for galaxies with SDSS spectra. As with \autoref{fig:accuracy}, emission-line galaxies have a narrower distribution than composite and absorption-line galaxies, and no significant systematic offsets are found.
  \label{fig:sdss}}
\end{figure}

Some of our galaxy group sight lines were observed with both WIYN and MMT (\autoref{tab:groups_mos}), allowing us to search for systematic offsets in redshifts derived from HYDRA and Hectospec spectra that may arise due to varying wavelength coverage and S/N. The fifth row of \autoref{tab:accuracy} shows the results for $\sim120$ galaxies that have redshift determinations from both WIYN and MMT. The widths of the velocity distributions for each type of galaxy are very similar to what we found in our single-instrument analyses, but there is a 20-50~$\mathrm{km\,s^{-1}}$ characteristic difference (2-$4\sigma$ significance) between the two measurements. There is no clear cause for this discrepancy, but the instrumental setups do differ; with our chosen gratings (\autoref{sec:obs}) WIYN/HYDRA has a wavelength coverage of 3800-6600~\AA\ (notably, this means that we rarely observe H$\alpha$ emission) and resolution $\mathcal{R}\approx1200$, while MMT/Hectospec covers 3700-9100~\AA\ at somewhat lower resolution ($\mathcal{R}\approx1000$).

\newpage
\subsubsection{External Validation}
\label{sec:results:accuracy:external}

As described in \autoref{sec:obs}, we allowed galaxies with known SDSS redshifts to be observed as ``extra'' targets in our configurations if no additional objects with unknown redshift could be accommodated in the configuration design. This choice now gives us the opportunity to externally validate our redshift measurements with SDSS. To gather the largest possible sample we entered the galaxy positions from \autoref{tab:redshifts} into the SDSS CrossID tool\footnote{\url{http://skyserver.sdss.org/dr14/en/tools/crossid/crossid.aspx}}, then cleaned the results to remove spurious matches.

This resulted in 260~galaxies with measurements in both SDSS and \autoref{tab:redshifts}. The last row of \autoref{tab:accuracy} shows how our redshift measurements compare to the SDSS values for absorption-line, emission-line, and composite galaxies. The galaxy classification was performed exactly as before (i.e., from visual inspection of our spectra) to minimize systematic biases. Approximately 1/3 of the galaxies with SDSS redshifts are in regions surveyed only by WIYN, $\sim1/4$ are in regions surveyed only with MMT, and $\sim40$\% are near sight lines surveyed with both telescopes.

Our redshift measurements show excellent agreement with the SDSS values, as shown in \autoref{fig:sdss}. The average velocity offset between them is $\sim10~\mathrm{km\,s^{-1}}$ for all three galaxy types, which is small enough to be consistent with zero in all cases. As with our internal comparison, we find that the widths of the velocity distributions are narrower for emission-line galaxies than absorption-line galaxies, with composite galaxies having intermediate values.

Interestingly, the standard deviations between our redshifts and the SDSS values are smaller than the standard deviations from our internal consistency checks for all galaxy types. This is probably because galaxies with SDSS spectra are brighter \citep[$r<17.8$;][]{strauss02} than the typical galaxies we surveyed. Thus, their redshift estimates are more precise by virtue of having high S/N. Conversely, our internal consistency comparisons (\autoref{sec:results:accuracy:internal}) include fainter objects with lower S/N and less precise redshift measurements.

\subsection{Completeness}
\label{sec:results:completeness}

{In this Section, the completeness of these observations and this survey are discussed using several different measures. The first two measures (``observational completeness'' and ``targeting completeness'') describe how efficiently redshifts were determined from the spectra obtained and how efficiently each sight line was observed for galaxies, respectively. Many observational variables affect these two completeness measures, including transparency, seeing, total integration time obtained, galaxy central surface brightness and galaxy spectral classification (emission, absorption or composite; see \autoref{sec:results:accuracy}). These values provide an indication about whether re-observing fields with low completeness values would be important to do or not. The third measure, ``overall completeness'' is the important number for users of this survey as it provides a quantitative measurement of the likelihood that all or nearly all galaxies which are potentially associated with an absorber have been identified. These values are generally quite high ($>80$\%), excepting for the sight lines which targeted nearby galaxy groups where the overall completeness typically exceeds 60\% by design.}

\clearpage
\startlongtable
\begin{deluxetable*}{lDDDDDD}
\tablecaption{Survey Completeness Measures for Individual Sight Lines
\label{tab:completeness}}

\tablehead{ \colhead{Sight Line} & \twocolhead{Obs. Comp.}  & \twocolhead{Targ. Comp.} & \twocolhead{Overall Comp.} & \twocolhead{$g_{\rm lim}$} & \twocolhead{$\theta_{\rm lim}$} & \twocolhead{Lim. Comp.} \\[-1ex]
                                 & \twocolhead{(\%)}        & \twocolhead{(\%)}        & \twocolhead{(\%)}          & \twocolhead{(mag)}        & \twocolhead{(arcmin)}          & \twocolhead{(\%)} }

\decimalcolnumbers
\startdata
1ES~1028+511      & 100.0 &  98.5 &  89.0 & 20.0 & 18.9 &  90.9 \\
1ES~1553+113      &  99.2 &  64.8 &  94.7 & 20.0 & 20.0 &  94.7 \\
1SAX~J1032.3+5051 & 100.0 &  95.5 &  93.7 & 20.0 & 20.0 &  93.7 \\
3C~57             &  13.2 &  76.5 &  21.8 &   .  &   .  &    .  \\
3C~263            &  97.3 &  94.7 &  85.3 & 20.0 & 12.5 &  91.1 \\
B~1612+266        &  98.7 &  72.8 &  61.3 & 18.1 & 19.6 &  92.0 \\
CSO~1022          &  94.6 &  49.0 &  62.6 & 19.3 & 11.7 &  90.6 \\
CSO~1080          &  91.4 &  83.8 &  74.2 & 19.5 & 10.7 &  91.7 \\
FBQS~J1010+3003   & 100.0 &  97.1 &  87.0 & 20.0 & 12.9 &  90.1 \\
FBQS~J1030+3102   &  84.6 &  85.7 &  76.3 & 19.6 &  5.5 &  90.9 \\
FBQS~J1519+2838   &  93.8 &  56.6 &  73.2 & 20.0 &  6.5 &  91.3 \\
H~1821+643        &  96.6 &  96.0 &  87.0 &   .  &   .  &    .  \\
HE~0153--4520     &  84.8 &  98.9 &  95.6 & 20.0 & 20.0 &  95.6 \\
HE~0226--4110     &  80.7 &  97.3 &  90.5 & 20.0 & 20.0 &  90.5 \\
HE~0435--5304     &  71.6 & 100.0 &  72.9 & 19.4 &  7.7 &  92.3 \\
HE~0439--5254     &  64.2 &  94.4 &  71.6 & 18.7 &  9.5 &  93.8 \\
HS~1102+3441      & 100.0 &  88.7 &  87.9 & 20.0 & 16.9 &  90.5 \\
Mrk~421           &  97.4 &  56.7 &  84.6 &   .  &   .  &    .  \\
PG~0832+251       &  98.8 &  88.8 &  85.4 & 20.0 &  8.6 &  90.5 \\
PG~0953+414       &  95.2 &  91.7 &  81.2 & 20.0 &  9.3 &  90.2 \\
PG~1001+291       &  99.1 &  88.1 &  85.3 & 20.0 & 14.3 &  90.5 \\
PG~1048+342       & 100.0 &  92.2 &  87.1 & 20.0 & 17.1 &  90.3 \\
PG~1115+407       & 100.0 &  93.3 &  95.9 & 20.0 & 20.0 &  95.9 \\
PG~1116+215       &  99.1 &  59.7 &  83.9 & 20.0 & 15.0 &  90.7 \\
PG~1121+422       &  99.2 &  98.5 &  92.8 & 20.0 & 20.0 &  92.8 \\
PG~1216+069       &  95.1 &  91.1 &  90.3 & 20.0 & 20.0 &  90.3 \\
PG~1222+216       &  95.0 &  45.4 &  52.3 & 19.2 & 13.1 &  90.5 \\
PG~1259+593       & 100.0 &  95.4 &  91.8 & 20.0 & 20.0 &  91.8 \\
PG~1424+240       &  98.4 &  57.6 &  70.0 & 20.0 &  8.6 &  92.3 \\
PG~1626+554       &  96.4 &  83.4 &  79.6 & 20.0 &  9.3 &  91.9 \\
PHL~1811          &  69.5 &  98.2 &  83.9 & 20.0 & 13.9 &  90.4 \\
PKS~2005--489     &  72.1 &  88.9 &  81.5 &   .  &   .  &    .  \\
Q~1230+0115       & 100.0 &  86.9 &  83.5 & 19.6 &  8.7 &  90.9 \\
RBS~711           &  90.2 &  88.6 &  77.3 & 20.0 &  9.4 &  90.9 \\
RX~J0439.6--5311  &  60.0 &  91.1 &  64.0 & 17.8 & 20.0 &  91.2 \\
RX~J2154.1--4414  &  55.4 & 100.0 &  75.2 & 19.6 &  6.4 &  93.3 \\
S5~0716+714       &  95.4 &  57.0 &  53.5 & 18.9 & 15.5 &  90.6 \\
SBS~0956+509      &  88.9 &  72.3 &  78.6 & 20.0 &  4.9 &  90.5 \\
SBS~1108+560      & 100.0 &  93.8 &  92.0 & 20.0 & 20.0 &  92.0 \\
SBS~1122+594      &  99.3 &  92.5 &  92.2 & 20.0 & 20.0 &  92.2 \\
SDSS~J1028+2119   &  96.0 &  71.1 &  57.1 & 20.0 &  4.2 &  92.3 \\
SDSS~J1333+4518   &  99.1 &  99.2 &  78.0 & 20.0 &  3.8 &  90.9 \\
SDSS~J1439+3932   &  93.8 &  98.6 &  89.9 & 20.0 & 17.8 &  90.3 \\
SDSS~J1540--0205  &  95.7 &  63.2 &  70.7 & 20.0 &  4.7 &  93.8 \\
Ton~236           &  98.8 &  69.6 &  89.6 & 20.0 & 19.2 &  90.3 \\
Ton~580           & 100.0 & 100.0 &  93.8 & 20.0 & 20.0 &  93.8 \\
Ton~1187          & 100.0 &  86.0 &  87.7 & 19.7 & 20.0 &  90.2 \\
VII~Zw~244        &  99.6 &  84.6 &  84.6 & 20.0 & 18.8 &  90.1 \\
\enddata

\tablecomments{Observational completeness is the fraction of objects observed for which we could retrieve a redshift. Targeting completeness is the fraction of available targets that were observed. Overall completeness is the fraction of all targets with $g<20$ within $20\arcmin$ of the sight line for which redshifts are available from this survey or other sources. The quantities $g_{\rm lim}$ and $\theta_{\rm lim}$ define a volume around the sight line for which the overall completeness is at least 90\%, and the limiting completeness is the actual completeness level within that volume. See \autoref{sec:results:completeness} for details.}

\end{deluxetable*}

\autoref{tab:completeness} lists several measures of completeness for each sight line surveyed. The first is the observational completeness, or the fraction of all observed galaxies for which we were able to determine a redshift, regardless of their apparent brightness or location with respect to the QSO sight line. These values are generally quite high, with $\sim25$\% of the sight lines being 100\% complete and $\sim75$\% having observational completenesses $>90$\%. Only two sight lines (4\%) have observational completenesses $<60$\%. Northern-hemisphere sight lines tend to have higher observational completeness than southern-hemisphere ones. We attribute this to the fact that we were granted more observing time in the northern hemisphere, which enabled us to re-observe sight lines that experienced poor weather or instrumental problems in their initial observations. {Thus, the root cause of the lower observational completeness in southern hemisphere sight lines is lower S/N in the data (see \autoref{sec:obs}).}

The second measure of completeness in \autoref{tab:completeness} is the targeting completeness, or the fraction of all targeted galaxies located within $20\arcmin$ of the QSO sight line with $g<20$ that were observed. Recall that our target lists exclude galaxies with known redshifts from SDSS or other sources (\autoref{sec:obs}). Approximately 2/3 of the sight lines have targeting completeness $>80$\%, and no sight line has $<40$\%. The ten sight lines that were observed as part of the galaxy group survey (\autoref{sec:obs:group}) have lower targeting completenesses than those observed as part of the COS GTO survey (\autoref{sec:obs:gto}). This difference is by design (see \autoref{sec:obs:group} for details), but it is clear that the targeting completeness for the galaxy group sight lines is large enough to achieve their goal of a fair sampling of potential group members.

The third completeness measure in \autoref{tab:completeness} is the overall completeness, which is defined as the fraction of all galaxies with $g<20$ located within $20\arcmin$ of the sight line that have known redshifts from \autoref{tab:redshifts}, SDSS, or other published sources. To be consistent with our configuration design requirements (\autoref{sec:obs}), we also require that galaxies have photometric redshifts {(when available) no more than $\sigma_{\rm phot}$ larger than the AGN redshift} for inclusion in the overall completeness calculation. {In non-SDSS fields where photometric redshifts were unavailable, the overall completeness is based solely on apparent $g$-band magnitude and proximity to the QSO sight line.} The overall completeness has a larger spread of values than the observational and targeting completenesses; nevertheless, $>85$\% (41/48) of the sight lines have overall completeness $>70$\% and only one sight line has a value $<50$\%.

{These overall completeness values are conservative in the sense that they are likely lower limits on the actual completeness in any one sight line, or in the survey overall. This is because we have used photometric redshifts for some galaxies proximate to the sight lines. We include galaxies without spectroscopic redshifts but whose photometric redshifts are consistent with being foreground to the AGN (see above) as adding to the incompleteness of the survey. In non-SDSS fields no photometric redshifts are used so that the incompleteness is overestimated even more.}

Any single number is a crude measure of the completeness of a survey. Thus, we provide two additional estimates of the survey completeness around our sight lines. One measure is found in \autoref{sec:appendix}, which provides tables for individual sight lines listing overall completeness values in bins of magnitude and impact parameter from the QSO sight line. The other measure searches for regions surrounding each sight line where the overall completeness is $>90$\% for angular separations of $\theta \leq \theta_{\rm lim}$ and apparent $g$-band magnitudes of $g \leq g_{\rm lim}$. When performing this search we prioritized being complete to fainter magnitudes (larger $g_{\rm lim}$) over being complete to larger impact parameters, and we required the search volume to contain at least 10~galaxies. The values of $g_{\rm lim}$, $\theta_{\rm lim}$, and the limiting (overall) completeness in this region are listed in Columns~5-7 of \autoref{tab:completeness}, respectively.

Unfortunately, four of our sight lines (3C~57, H~1821+643, Mrk~421, and PKS~2005--489) do not reach the limiting completeness threshold for any combination of $g_{\rm lim}$ and $\theta_{\rm lim}$ in our data. On the other hand, there are 11~sight lines (23\%) that have $>90$\% completeness all the way to the edge of the survey region (i.e., $g_{\rm lim}=20.0$ and $\theta_{\rm lim}=20\arcmin$), and an additional 21~sight lines (44\%) that are complete to $g_{\rm lim}=20.0$ with $\theta_{\rm lim}$ ranging from $3.8\farcm$-$18\farcm9$. Thus, 2/3 of the sight lines are $>90$\% complete to galaxies with $g<20$ over some region of the survey volume.

The COS GTO survey was designed to be complete to $L \la 0.1\,L^*$ galaxies at $z\leq0.1$ located within 1~Mpc of the sight line. A completeness limit of $g_{\rm lim} = 20.0$ corresponds to $L\approx0.15\,L^*$ at $z=0.1$ \citep{montero-dorta09}, and 1~Mpc is equivalent to $\theta_{\rm lim} = 9\farcm0$ at $z=0.1$. Twenty-four of the 38 sight lines in the COS GTO survey (63\%) are $>90$\% complete to at least these thresholds; if we restrict our analysis to the northern hemisphere, 20/29 WIYN sight lines (69\%) meet the design goals.

\section{Discussion}
\label{sec:disc}

\begin{figure}
  \epsscale{1.0}
  \centering\plotone{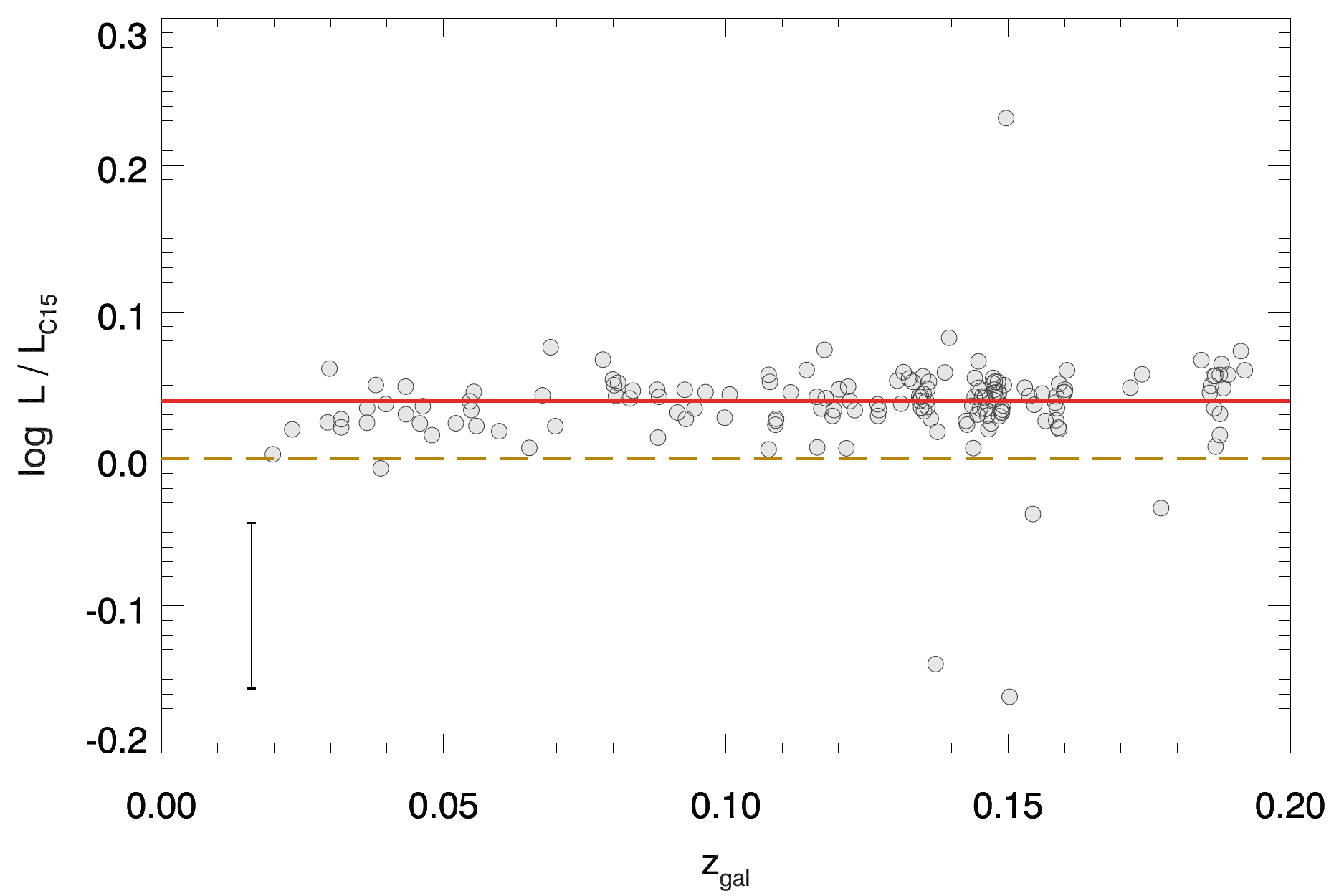}
  \centering\plotone{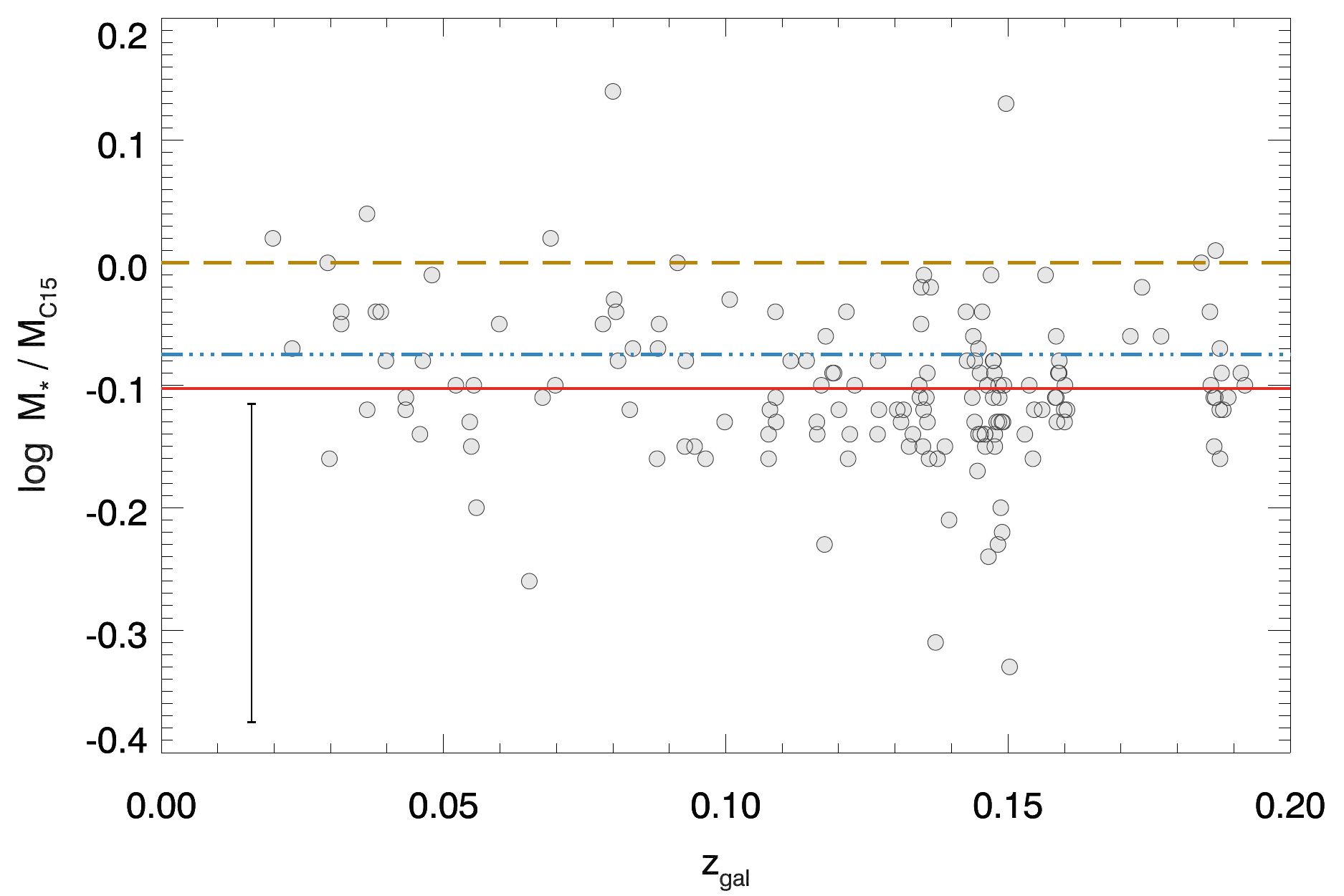}
  \centering\plotone{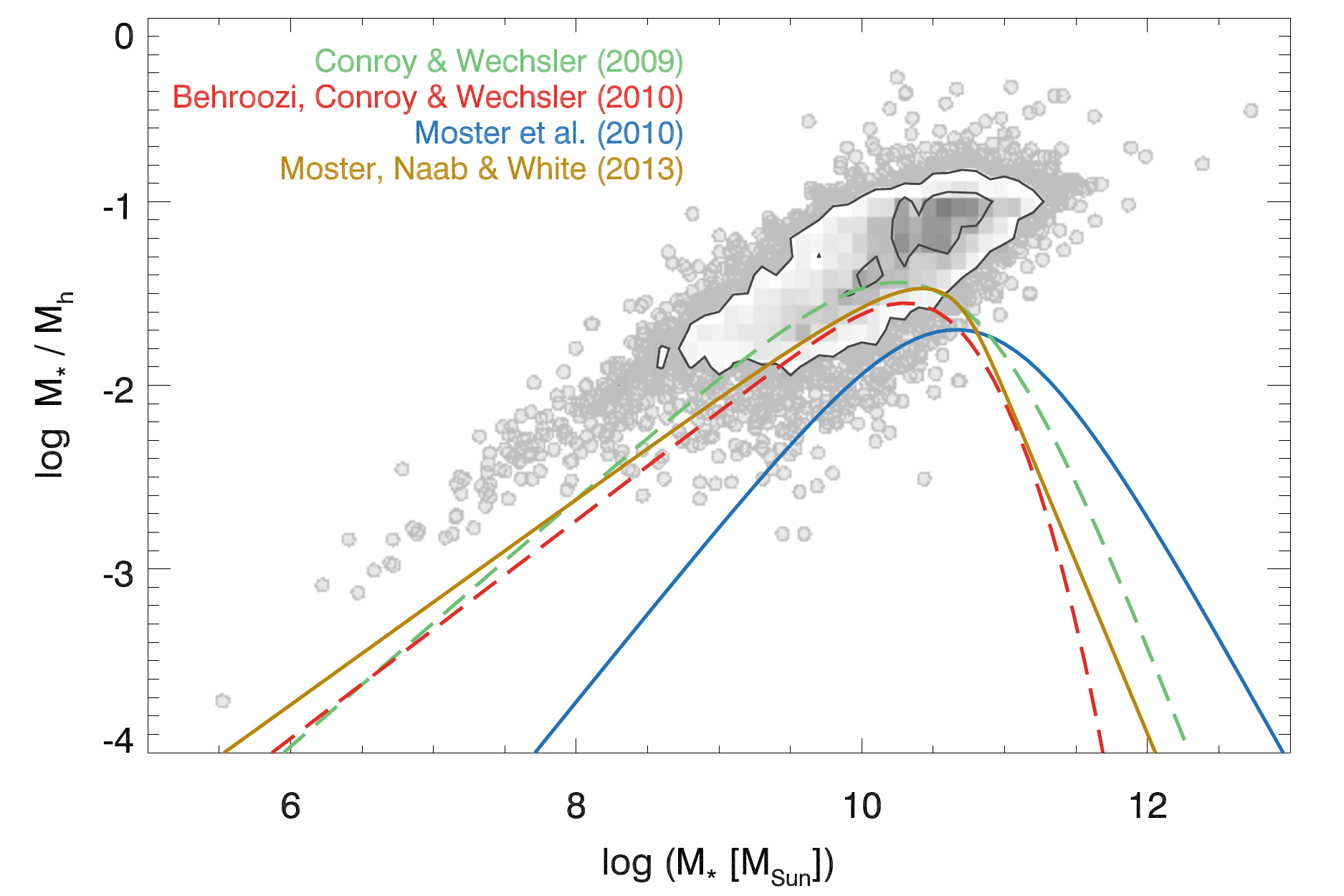}
  \vspace{-1ex}
  \caption{Comparison of the luminosities (top) and stellar masses (middle) from \autoref{tab:redshifts} with the measurements of \citet[C15]{chang15} as a function of redshift. Both exhibit slight systematic offsets (solid red line) compared to the \citet{chang15} values but have no clear trend with redshift. The bottom panel shows the distribution of $M_*/M_{\rm h}$ from \autoref{tab:redshifts} as a function of {$M_*$}. For a given stellar mass, our halo masses are systematically lower than those predicted from halo abundance matching algorithms (see text for details).
  \label{fig:mass}}
\end{figure}

\autoref{tab:redshifts} includes several columns intended to add value beyond the basic observables (position, redshift, apparent magnitudes) for each galaxy. Among them are the rest-frame $g$-band luminosity, stellar mass, and halo mass for each galaxy. \autoref{fig:mass} compares these values to those of \citet{chang15}, who used photometry from SDSS and the {\sl Wide-field Infrared Survey Explorer} ({\sl WISE}) to measure the stellar mass and luminosity of galaxies at $z\leq0.2$. 

The top panel of \autoref{fig:mass} compares the rest-frame $g$-band luminosities of the 158~galaxies that appear both in \autoref{tab:redshifts} and \citet{chang15}. The values in \autoref{tab:redshifts} are systematically larger by 0.04~dex (red line) with an rms scatter of $\sim0.02$~dex, but there is no clear trend with redshift, which we take as evidence that there are no systematic differences in the $K$-corrections employed. {A representative uncertainty range is shown in the bottom-left corner of the panel by assuming 0.1~mag uncertainty for each measurement once uncertainties in the Galactic foreground extinction removal and $K$-corrections are accounted for. The $0.04\pm0.02$~dex systematic shift between the two measurements is less than this representative uncertainty ($\approx0.06$~dex).}

The middle panel of \autoref{fig:mass} compares the stellar mass values for the overlapping galaxies. There is considerably more dispersion in this relationship because each of the stellar mass determinations has a $1\sigma$ uncertainty of $\pm0.1$~dex \citep[see representative uncertainty in bottom-left corner;][]{taylor11,chang15}. Nonetheless, it is clear that the values in \autoref{tab:redshifts} are smaller than those of \citet{chang15} by 0.1~dex (red line) with an rms scatter of $\sim0.06$~dex. Again we find no clear trend with redshift. We have also compared the stellar masses from \autoref{tab:redshifts} with the MPA-JHU measurements of SDSS galaxies \citep{kauffmann03,salim07} for the 172~galaxies in both samples and find very similar results (systematic offset of $-0.07$~dex with rms scatter of $\sim0.05$~dex; blue dash-dot line).

The bottom panel of \autoref{fig:mass} shows the ratio of $M_*/M_{\rm h}$ as a function of {stellar} mass for 8288 galaxies from \autoref{tab:redshifts} that have robust estimates of both quantities (i.e., they have $0.001 \leq z_{\rm gal} < 0.5$ and observed colors in the ranges specified in \autoref{sec:results:redshifts}.). Contours represent the density of the data points and are drawn at the $1\sigma$ and $2\sigma$ levels. Inside the contours we show the density of the data points in gray-scale, where darker colors indicate higher density. Outside of the $2\sigma$ contour the locations of individual data points are shown. Most of our galaxies have $\log{M_{\rm *}/M_{\Sun}}\sim10.5$ and $\log{M_*/M_{\rm h}}\sim-1$.

For comparison, we also show the predictions of several recent abundance-matching analyses \citep{conroy09,behroozi10,moster10,moster13} evaluated at $z=0.15$ (the median redshift of the plotted galaxies). While there is considerable variation in these relations, the largest $M_*/M_{\rm h}$ values they predict are all $\sim0.5$~dex below the highest density region for our galaxies. Since we have shown in the middle panel of \autoref{fig:mass} that there are only modest offsets between our stellar masses and those measured for the same galaxies by other methods, the culprit must be our halo mass estimates. In particular, our halo masses are systematically smaller than those inferred from abundance-matching prescriptions, causing our galaxies to cluster above their predictions in \autoref{fig:mass}.

Neither the existence nor the direction of this offset is surprising because the halo masses are estimated in very different ways. In abundance matching, a one-to-one correspondence is assumed between halo mass distributions from dark matter simulations and observed stellar mass functions. Thus, at a given redshift the percentiles of the stellar mass function are matched with those of the halo mass distribution to assign a stellar mass to a halo mass. However, at a certain point dark matter halos contain more than one galaxy and the correspondence predicted by abundance matching is between the stellar mass of the central (i.e., brightest or most massive) galaxy in the halo and the mass of the dark matter halo in which it resides, which may encapsulate an entire group or cluster of galaxies. Since we are interested in estimating the mass and extent of individual galaxies, we depart from an abundance matching formalism at $L\ga0.2\,L^*$, at which point we assume a constant mass-to-light ratio (see \citealp{stocke13} for details).

Figure~1 of \citet{stocke13} illustrates how our halo-mass prescription compares to pure abundance-matching results. While the differences are modest for $L \la L^*$ galaxies, our procedure yields halo masses that can be an order of magnitude smaller than abundance-matching predictions for brighter galaxies. This is largely due to the most massive halos including more than one galaxy, a central galaxy and many satellites, as in a group of galaxies.

\subsection{Galaxy-Absorber Connections}
\label{sec:disc:galabs}

Of the remaining value-added columns in \autoref{tab:redshifts}, the ``absorption flag'' in the final column is particularly interesting. It encodes the results of comparing the positions and redshifts of 8230~galaxies (all of the galaxies in \autoref{tab:redshifts} with $z_{\rm gal}>0.001$ located near one of the 47~sight lines with \hst/COS observations) with the redshifts of 1565~\ion{H}{1} absorbers from \hst/COS \citep{danforth16,stocke18}. The galaxy redshifts are in the range $0.00113 \leq z_{\rm gal} \leq 0.90909$, while \ion{H}{1} absorbers have redshifts in the range $0.00165 \leq z_{\rm abs} \leq 0.68278$ and column densities in the range $12.10 \leq \log{N_{\rm H\,I}} \leq 19.23$.

Absorption flags equal to zero (3781~galaxies) indicate that a galaxy is not within $1000~\mathrm{km\,s^{-1}}$ of an absorber, while a value of one (3699~galaxies) indicates that a galaxy is within $1000~\mathrm{km\,s^{-1}}$ of an absorber but is not the closest galaxy to that absorber. All velocity comparisons are done in the rest frame of the absorber (i.e., $\Delta v = c(z_{\rm gal}-z_{\rm abs})/(1+z_{\rm abs})$) and the ``closest'' galaxy to an absorber is defined to be the one with the smallest value of $D/R_{\rm vir}$, where $D$ is the three-dimensional distance between the galaxy and the absorber. The galaxy-absorber distance along the line of sight, $D_z$, is calculated using a reduced-Hubble-flow model where $D_z=0$ if $|\Delta v| \leq 400~\mathrm{km\,s^{-1}}$ and $D_z = (|\Delta v|-400~\mathrm{km\,s^{-1}})/H(z)$ otherwise. The three-dimensional distance is then found by summing $D_z$ and the impact parameter, $\rho$, in quadrature.

Absorption flags larger than one imply that a galaxy is, at a minimum, the closest galaxy in \autoref{tab:redshifts} to an absorber, in the sense described above. A value of two (179~galaxies) indicates that a closer galaxy is known from SDSS or other sources \citep[e.g.,][]{prochaska11}. A value of three (571~galaxies) indicates that a galaxy is the closest known galaxy within $20\arcmin$ and $1000~\mathrm{km\,s^{-1}}$ of an absorber.

However, this closest galaxy determination is not always unique; i.e., the closest galaxy to an absorber in \autoref{tab:redshifts} may also have a redshift from SDSS or elsewhere. We find 67 of these coincidences, such that 504/571 (88\%) of the galaxies with absorption flag values of three were not known to be the closest galaxy to an absorber before this work. Similarly, of all the galaxies with absorption flags of two or three, indicating that they are the closest galaxies in \autoref{tab:redshifts} to an absorber, 504/750 (67\%) represent newly discovered galaxy-absorber associations. This finding, in conjunction with the completeness analysis in \autoref{sec:results:completeness} and \autoref{tab:completeness}, suggests that the GTO and groups surveys presented in \autoref{sec:obs} and \autoref{sec:results} are deeper than other galaxy redshift surveys near the majority of these QSO sight lines.

The absorption flags in \autoref{tab:redshifts} do not address ambiguity in the closest galaxy associations. A choice must often be made between associating an absorber with a low-luminosity galaxy with a small impact parameter or a higher luminosity galaxy that is somewhat further away. This motivated our decision to divide by the estimated virial radius of each galaxy and associate the absorber with the galaxy with the smallest $D/R_{\rm vir}$ in this and previous works \citep[e.g.,][]{stocke13,keeney17}. While this does not remove the ambiguity altogether it at least provides an objective discriminator.

\citet{keeney17} studied the properties of 45~galaxies located within $2\,R_{\rm vir}$ of low-$z$ \ion{H}{1} absorbers, including measurements of properties (inclination, star formation rate, gas-phase metallicity) that we cannot measure for all galaxies in \autoref{tab:redshifts}. They also provided indicators of how unique each absorber-galaxy association is by examining the ratios of the impact parameters and galaxy-absorber velocity differences for the nearest and next-nearest galaxies \citep[see Figures~16 and 17 of][]{keeney17}. This analysis found that the galaxy-absorber associations are robust at $D\la1.4\,R_{\rm vir}$ and more questionable, particularly in velocity, at larger values \citep[see Section 7.2 in][]{keeney17}.

{When calculating three-dimensional galaxy-absorber distances, we adopt a reduced Hubble-flow velocity of $400~\mathrm{km\,s^{-1}}$ because it corresponds to the maximum rotation velocities in the most massive star-forming galaxies, and the largest velocity dispersions in passive galaxies. However, this value is assumed, not derived, so it creates some ambiguities in assigning individual galaxy-absorber associations. For example, if galaxy redshifts are varied by $\pm1\sigma_z$ from their nominal values, then 6/96 galaxy-absorber associations with $D \leq 1.4\,R_{\rm vir}$ change. Because a single galaxy can be associated with more than one absorber, only five galaxies are responsible for these changes; three of these galaxies had initial galaxy-absorber velocity differences $>300~\mathrm{km\,s^{-1}}$ and the other two had large redshift uncertainties of $c\sigma_z/(1+z_{\rm gal})>400~\mathrm{km\,s^{-1}}$. At larger distances ($D > 1.4\,R_{\rm vir}$) where \citet{keeney17} found that associations with an individual galaxy are not conclusive, $\approx95/1075$ (9\%) of all galaxy-absorber associations change when the galaxy redshifts are varied by $\pm1\sigma_z$.}

\begin{figure}
  \epsscale{1.0}
  \centering\plotone{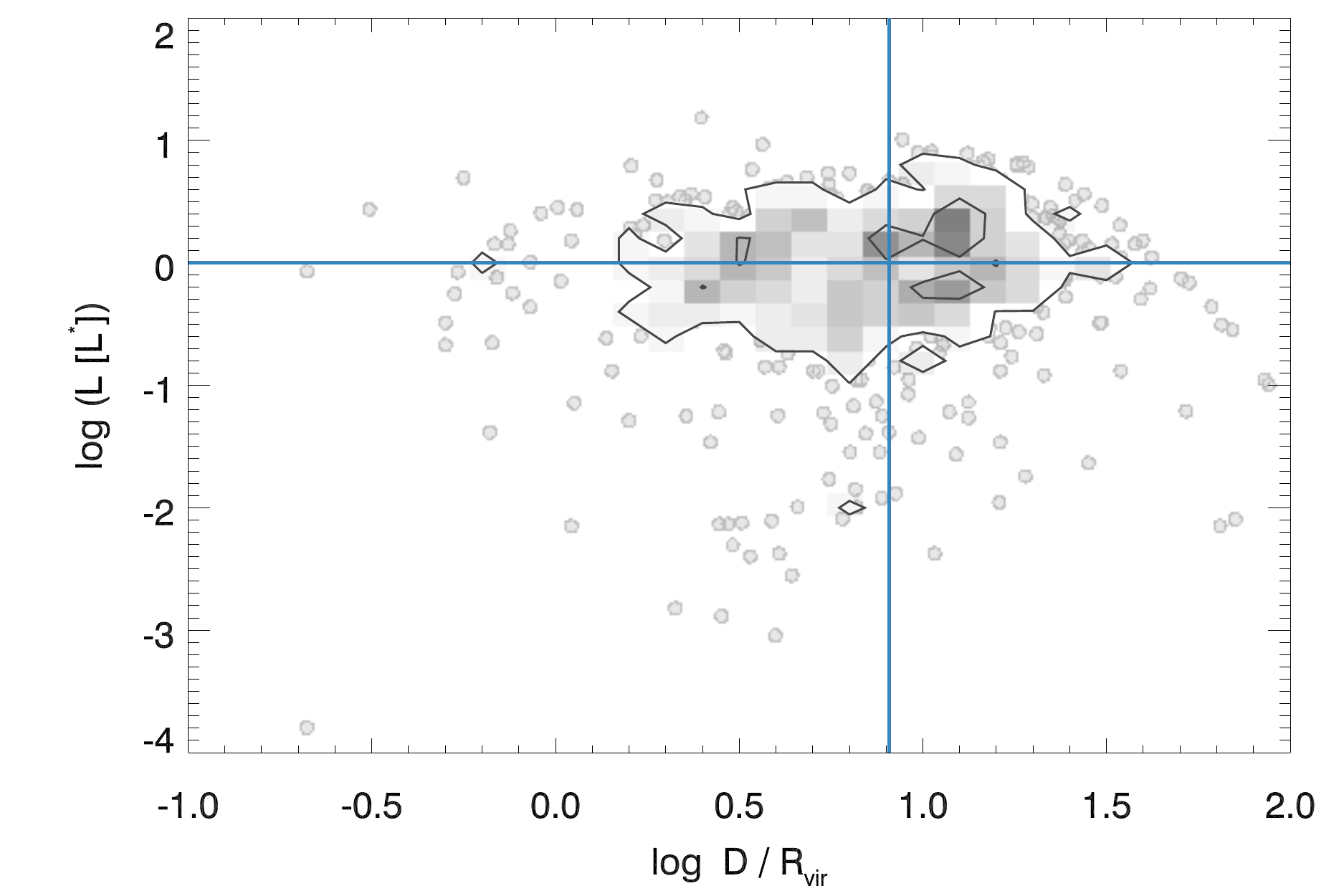}
  \centering\plotone{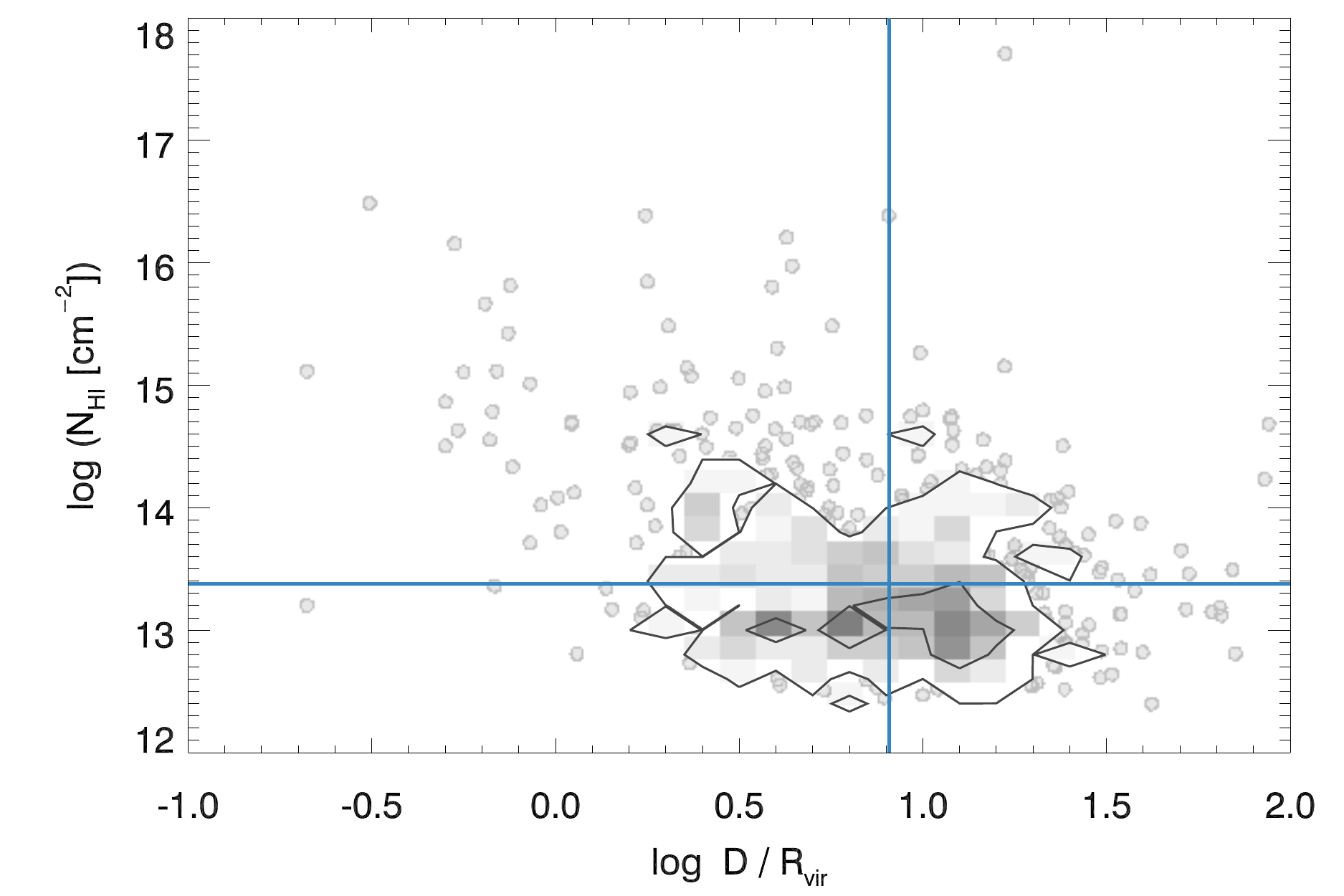}
  \centering\plotone{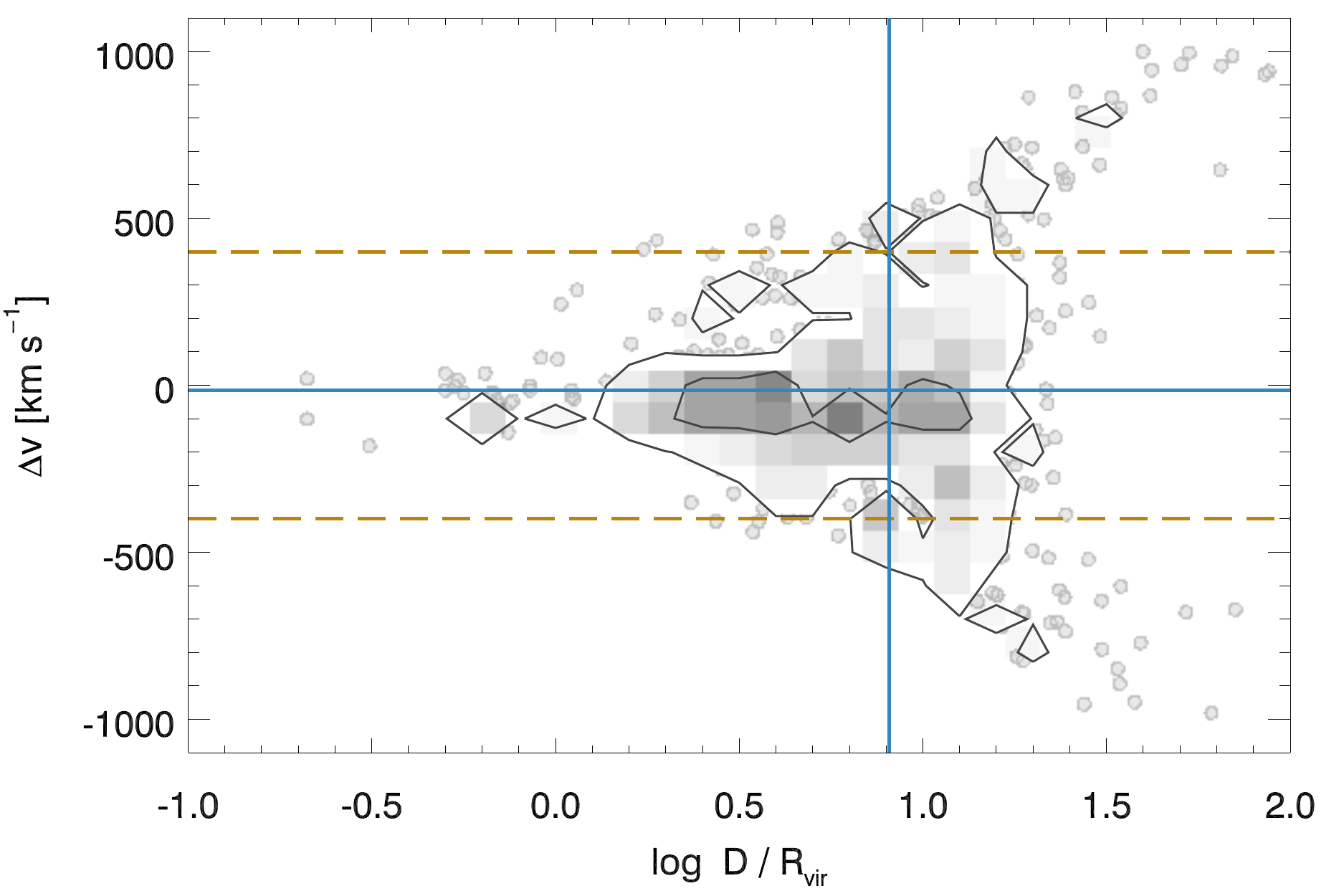}
  \vspace{-1ex}
  \caption{Nearest-galaxy luminosity (top), \ion{H}{1} column density (middle), and galaxy-absorber velocity difference (bottom) as a function of $D/R_{\rm vir}$ for galaxy-absorber associations in \autoref{tab:redshifts}. The solid blue lines indicate the median values of each property.
  \label{fig:norm}}
\end{figure}

\autoref{fig:norm} displays galaxy-absorber properties for the 571~galaxies in \autoref{tab:redshifts} identified as the closest known galaxy to an absorber. The data are displayed as in the bottom panel of \autoref{fig:mass}, where the contour levels are drawn at the $1\sigma$ and $2\sigma$ levels with respect to the galaxy density, data inside the $2\sigma$ contour are galaxy density, and data outside the $2\sigma$ contour are individual data points. Solid blue lines indicate the median value of each property, and the dashed lines in the bottom panel indicate our reduced-Hubble-flow velocity.

The typical galaxy-absorber association in \autoref{tab:redshifts} is between an $\sim L^*$ galaxy located $\sim8\,R_{\rm vir}$ from an absorber with $\log{N_{\rm H\,I}}\sim13.4$. No trend of galaxy luminosity as a function of $D/R_{\rm vir}$ is evident. The same can be said for the \ion{H}{1} column density for the data inside the contours; however, at $D\la3\,R_{\rm vir}$ there is a correlation between higher \ion{H}{1} column densities and closer galaxy-absorber separations. This trend is supported by the analysis of \citet[see their Figure~5]{stocke13}. All of the closest galaxy-absorber associations have $|\Delta v|<400~\mathrm{km\,s^{-1}}$ because we do not impose any line-of-sight corrections to the three-dimensional galaxy-absorber distance, $D$, until $|\Delta v|$ exceeds this threshold. A more complete analysis of nearest-neighbor statistics for close galaxy-absorber separations, particularly as it relates to the spread of metals in the IGM, can be found in \citet{pratt18}.

\subsection{Ly$\alpha$ Covering Fractions}
\label{sec:disc:covering}

Using the current survey it is also possible to calculate \ion{H}{1} Ly$\alpha$ covering fractions for galaxy-absorber associations. To do so, we searched for galaxies within $4\,R_{\rm vir}$ of absorbers with $0.042 \leq z_{\rm abs} \leq 0.135$. The maximum redshift is chosen such that $g=20$ corresponds to $L\leq0.3\,L^*$, and the minimum redshift is chosen such that $20\arcmin$ corresponds to $\rho\geq1$~Mpc. We also set a luminosity limit of $L\leq3\,L^*$ (i.e., $R_{\rm vir}<250$~kpc) to ensure that we are sensitive to all galaxies with $D<4\,R_{\rm vir}$ that are within 1~Mpc of the QSO sight line.

{To investigate the \ion{H}{1} Ly$\alpha$ covering fractions in this galaxy sample, we have employed the procedure used in \citet{stocke13} by which the redshift of each galaxy within $4\,R_{\rm vir}$ of the sight line is used as a marker to search for \ion{H}{1} absorption at $N_{\rm H\,I} \geq 10^{13}~\mathrm{cm^{-2}}$. If \ion{H}{1} absorption is present above this threshold, this constitutes a ``hit'' ($H$); if no absorption is present, this constitutes a ``miss'' ($M$). The covering fraction is simply $C=H/(H+M)$. Spectral regions not sensitive enough for the above limit to be detectable are not used in this analysis; e.g., locations of Galactic metal-line absorption or strong absorption at another redshift \citep{stocke13}.}

\begin{figure}
  \epsscale{1.0}
  \centering\plotone{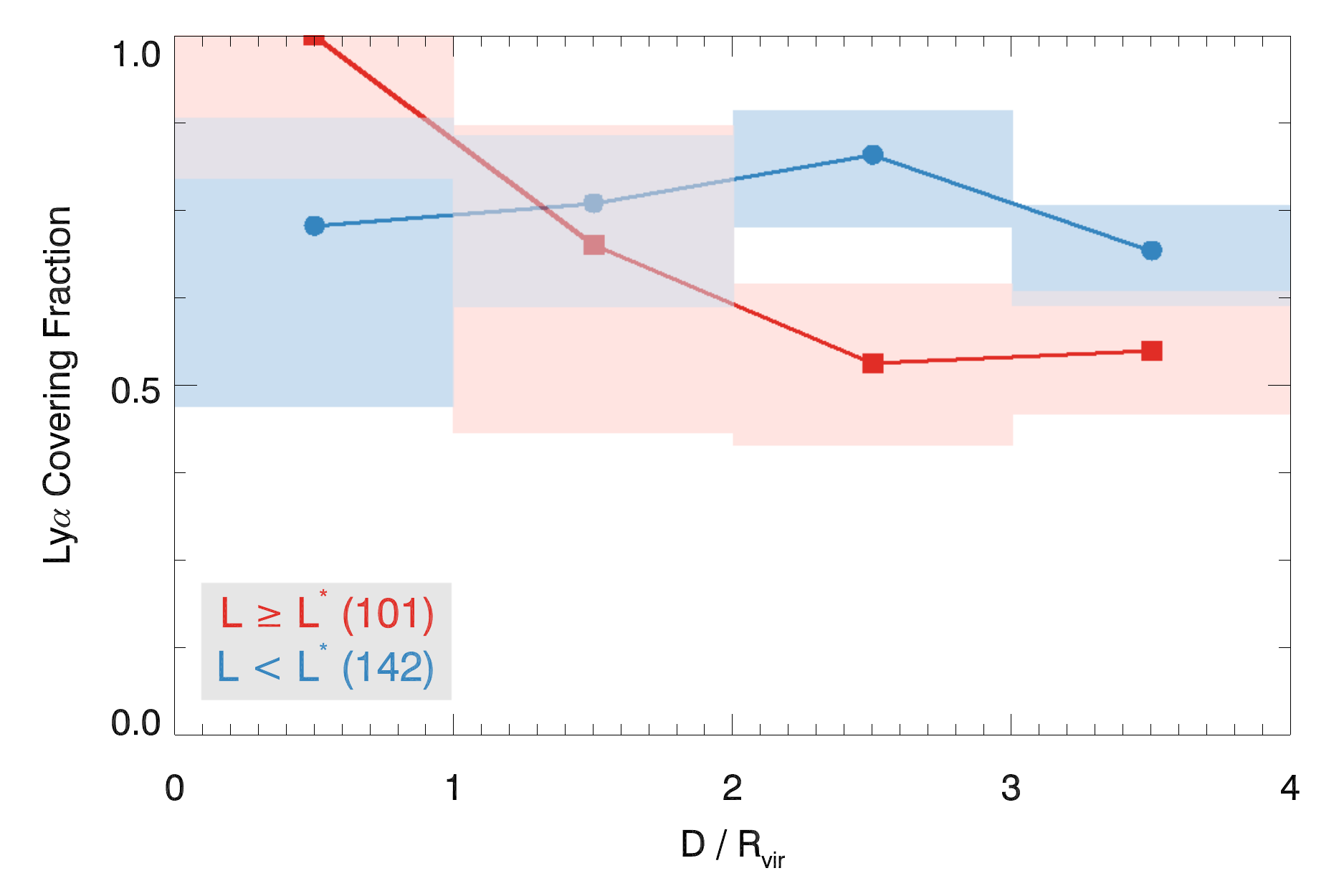}
  \centering\plotone{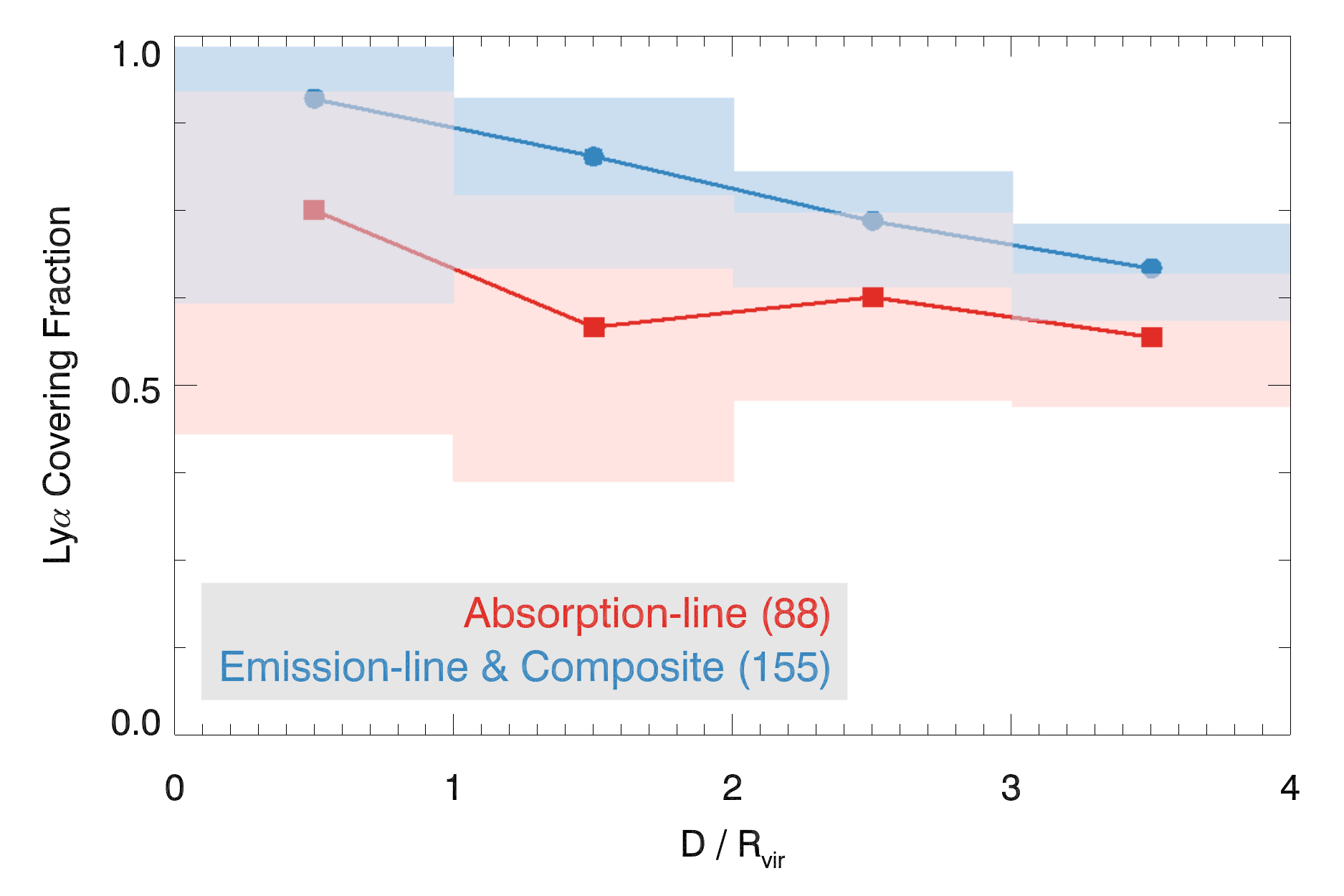}
  \vspace{-1ex}
  \caption{Covering fraction of \ion{H}{1} Ly$\alpha$ as a function of $D/R_{\rm vir}$ for galaxies of different luminosities (top) and spectral classifications (bottom). {Shaded regions indicate the 68\% confidence interval for the covering fraction \citep{gehrels86}.}
  \label{fig:fcov}}
\end{figure}

\autoref{fig:fcov} displays the results of this analysis. The top panel compares the Ly$\alpha$ covering fractions as a function of $D/R_{\rm vir}$ for galaxies with $L \geq L^*$ (red) and $L<L^*$ (blue). The number of galaxies in each subsample is listed in parentheses, and the shaded regions show the 68\% confidence Poisson uncertainties for each bin \citep{gehrels86}. While the uncertainties overlap in most of the radial bins, the covering fraction for $L<L^*$ galaxies are flatter than for more luminous galaxies.

{These results differ slightly} from the finding of \citet[see their Figure~7]{stocke13}, {with the largest difference being a ($\sim2\sigma$) larger covering fraction ($C=0.7$ compared to 0.5) for sub-L$^*$ galaxies}, but it should be noted that the absorber and galaxy samples used for the covering fractions also differ from what \citet{stocke13} used. Nevertheless, both \autoref{fig:fcov} and \citet{stocke13} find that galaxies with $L\gtrsim0.1\,L^*$ have \ion{H}{1} covering fractions $>50$\% out to $4\,R_{\rm vir}$ and that the covering fraction for $L>L^*$ galaxies is very high, consistent with unity inside $R_{\rm vir}$.

{The COS-Halos project \citet{werk13} finds a similarly high covering fraction for star-forming galaxies inside an impact parameter of $0.5\,R_{\rm vir}$, while \citet{wakker09} found unity \ion{H}{1} covering fractions out to $\sim R_{\rm vir}$ for massive galaxies. Similarly, covering factors close to unity are found at $z=2$-3 by \citet[][see their Table 6]{rudie12} with these high covering factors extending to much greater impact parameters at comparable column densities ($\log{N_{\rm H\,I}} \geq 13$) to this study and \citet{wakker09}. The \citet{rudie12} covering fractions for $\log{N_{\rm H\,I}} \geq 14$, which may be a more appropriate comparison between high- and low-$z$, are nearly identical to what is shown in \autoref{fig:fcov} and in \citet{wakker09} in showing a very slow decline out to several virial radii.}

The bottom panel of \autoref{fig:fcov} compares the covering fractions for absorption-line galaxies (red) and emission-line and composite galaxies (blue). We use these observationally defined samples as proxies for the star-forming and passive galaxy populations since it is not always possible to determine specific star formation rate using the current spectra because: (1) for most WIYN/HYDRA spectra, H$\alpha$ is not included in the observing band; and (2) more distant galaxies are not well-resolved in our images and so have only poorly determined sizes. We use the absorption-line galaxies, which show no sign of emission lines from star formation, as proxies for passive galaxies; all galaxies with emission-line and composite spectra are assumed to be star-forming.

\begin{figure}
  \epsscale{1.0}
  \centering\plotone{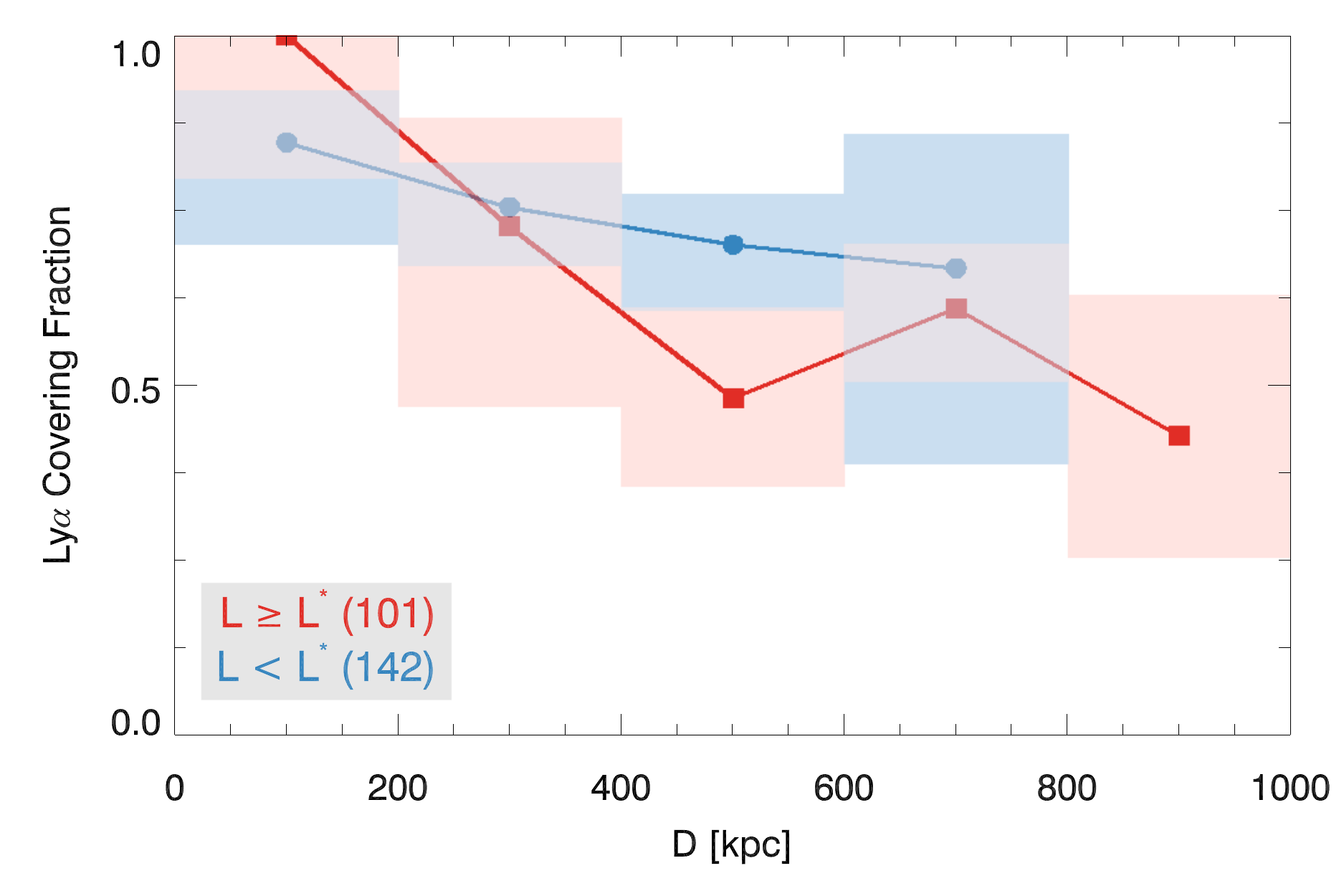}
  \centering\plotone{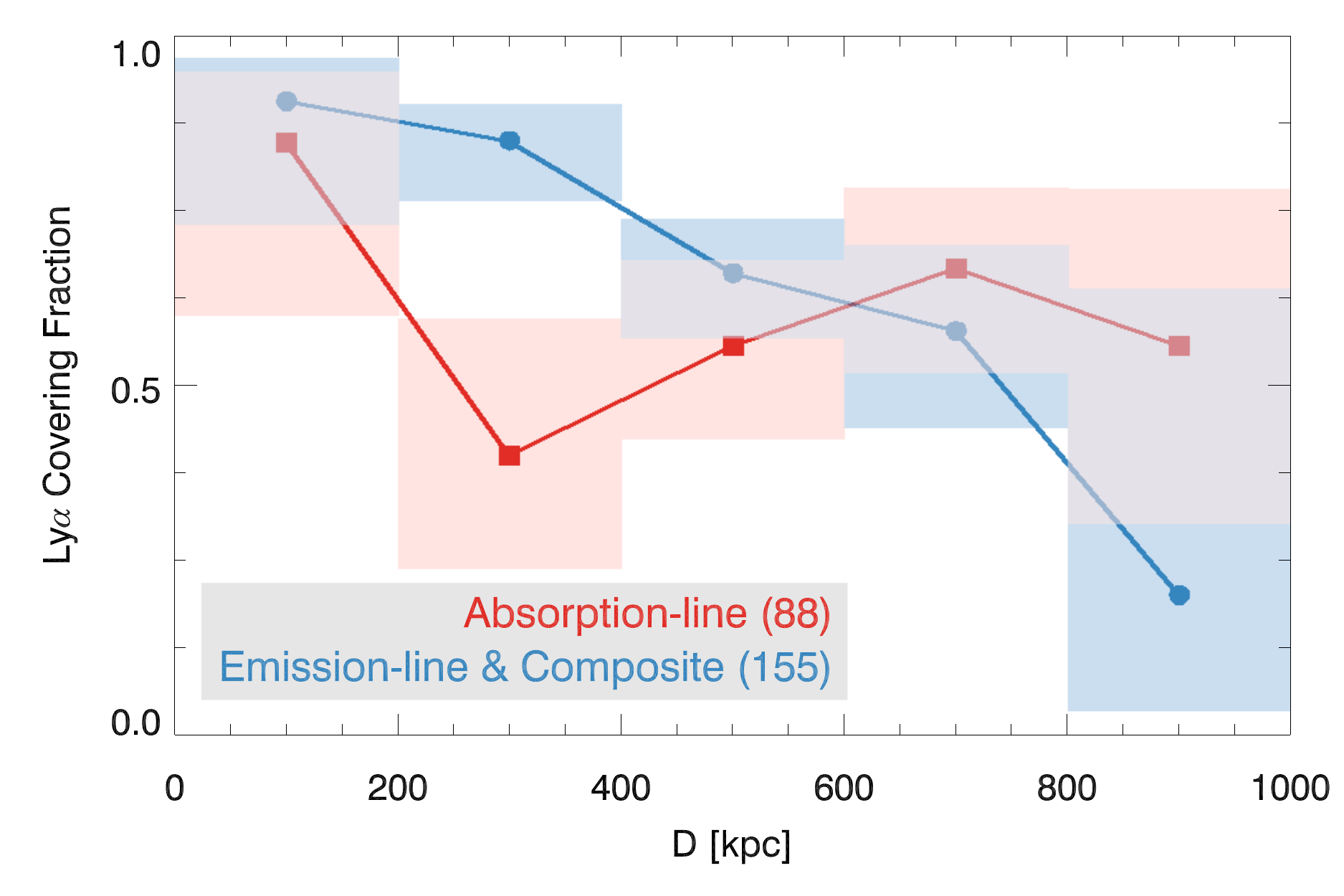}
  \vspace{-1ex}
  \caption{{Covering fraction of \ion{H}{1} Ly$\alpha$ as a function of $D$ for galaxies of different luminosities (top) and spectral classifications (bottom). Shaded regions indicate the 68\% confidence interval for the covering fraction \citep{gehrels86}.}
  \label{fig:fcov_rho}}
\end{figure}

The covering fraction for emission-line and composite (star-forming) galaxies is slightly higher than that of absorption-line (passive) galaxies at all radii, although the uncertainties in the radial bins overlap. {Despite showing no significant difference in covering fraction in the binned data shown in \autoref{fig:fcov}, taken altogether the emission-line and composite galaxies have $C=113/155=0.73^{+0.04}_{-0.05}$ within $4\,R_{\rm vir}$, while the absorption-line galaxies have $C=53/88=0.60^{+0.06}_{-0.07}$. The covering fraction for passive galaxies in our sample is therefore $\sim2\sigma$ lower than that of star-forming galaxies.} Since the data in the top panel of \autoref{fig:fcov} include some absorption-line galaxies, it is probable that all star-forming galaxies with $L\geq0.3\,L^*$ have a near-unity covering fraction inside their virial radius.

The other interesting new result is that the absorption-line galaxies show a rather flat covering fraction of $\sim60$\% out to $4\,R_{\rm vir}$ ($\sim750$~kpc for an $L^*$ galaxy). Previous hints of a flat covering fraction for passive galaxies come from \citet{keeney17}, whose sample included a few passive galaxies at impact parameters $>R_{\rm vir}$, and from \citet{stocke18}, who find Ly$\alpha$ and \ion{O}{6} absorption far enough away from individual bright passive galaxies that these authors ascribe the absorption to an intra-group medium, not to an individual galaxy. Lower overall covering fractions that are rather constant with impact parameter argue in favor of a more extensive, but clumpy, medium in systems of passive galaxies like dense groups of galaxies.

{\autoref{fig:fcov_rho} is analogous to \autoref{fig:fcov}, except that it shows the Ly$\alpha$ covering fractions as a function of physical distance, $D$, rather than normalized distance, $D/R_{\rm vir}$. The top panel shows covering fractions for galaxies of different luminosities, and has no information for $L<L^*$ galaxies in the outermost bin because those distances correspond to $D>4\,R_{\rm vir}$. To within the (largely overlapping) uncertainties, this plot shows a smooth, shallow decline in covering fraction with increasing distance for galaxies of all luminosities.}

{The bottom panel of \autoref{fig:fcov_rho} shows the covering fractions for star-forming and passive galaxies, defined by their spectral classification as above. The uncertainties are once again large, but there are hints of an intriguing trend whereby star-forming galaxies have higher covering fractions than passive galaxies at $D\lesssim400$~kpc, the two samples have comparable covering fractions when $D\sim400$-800~kpc, and then passive galaxies have larger covering fractions at $D\gtrsim800$~kpc. Larger sample sizes are required to determine whether this trend is robust.}

\section{Conclusions}
\label{sec:conc}

We present the basic results of a deep ($g \leq 20$) and wide ($\sim20\arcmin$ radius) galaxy redshift survey around 47~COS sight lines. These sight lines were selected by the COS science team and the UV-bright targets were well-observed ($\mathrm{S/N} > 15$) to provide the best available far-UV absorption line data for understanding the local CGM and IGM. Several studies have already utilized this redshift survey to characterize the CGM of star-forming and passive galaxies \citep{stocke13,keeney17}, to determine the maximum extent that metals spread away from their potential galaxy of origin \citep{stocke06,pratt18}, to probe the warm-hot intergalactic medium (WHIM) in nearby groups of galaxies \citep{stocke14, stocke17}, and to place an upper limit on the metallicity of gas clouds in cosmic voids \citep{stocke07}. This survey also provided galaxy environments for individual absorbers of interest including two cases of demonstrably warm-hot, \ion{O}{6}-only absorbers (i.e., no detectable \ion{H}{1} Ly$\alpha$) in the PKS~0405--123 and FBQS~1010+3003 sight lines \citep{savage10,stocke17} and a strong, multi-phase absorber in the 3C~263 sight line \citep{savage12}.

As part of the technical design for the majority of this survey (38 of 48 sight lines, including one that was not ultimately observed) we obtained high completeness ($>90$\%) over a $>1$~Mpc region down to $\la0.1\,L^*$ luminosities at $z \leq 0.1$ for most of the target sight lines (see \autoref{tab:completeness}). The completeness levels of individual sight lines as a function of both magnitude and impact parameter are provided in \autoref{sec:appendix}. The limiting luminosities obtained around $z\leq0.1$ absorbers are not so faint as to suggest that virtually all galaxies have been surveyed. Galaxies with $L\geq0.1\,L^*$ are thought to be the dominant source for metal-enriched gas expelled into the CGM and IGM \citep{prochaska11,stocke13,burchett15,keeney17,pratt18}, so these survey limits are sufficient for many local universe investigations.

For the remaining 10~sight lines probing nearby SDSS-selected galaxy groups, the basic design goal was to obtain good quality spectra for at least 20~members (including the original SDSS members) to constrain basic observables (group position on the sky, redshift, velocity dispersion) and derived quantities (total stellar mass, halo mass, group virial radius). While this goal was met for all ten sight lines and 12~galaxy groups probed \citep{stocke18}, the overall completeness percentages are not as high as for the 38~GTO sight lines (see \autoref{tab:completeness} and \autoref{sec:appendix}). Otherwise, all of the data products for these ten sight lines are the same as for the original 38 and are included in the general statistics for this survey.

We have used multiple observations for some galaxies to assess the overall redshift uncertainties for three classes of spectra: emission-line, absorption-line, and composite spectra with both emission and absorption lines present. As expected, redshifts obtained from emission lines alone have a small standard deviation and small offset when compared to the SDSS redshifts (\autoref{tab:accuracy}); this redshift uncertainty is comparable to the uncertainties for the COS UV absorption-line data \citep{danforth16,keeney17}. Redshifts for absorption-line and composite galaxies have larger standard deviations than emission-line galaxies but comparably small velocity offsets when comparison to SDSS measurements (\autoref{tab:accuracy}).

Regardless of spectral class, the error budget for galaxy redshifts does not add a significant uncertainty to making a case for or against an absorber-galaxy association because the redshift uncertainty is still less than the velocity dispersion in luminous, early-type galaxies. Further, it has become standard in the CGM field to reduce the Hubble-flow radial distance between absorbers and galaxies by a peculiar velocity of $400~\mathrm{km\,s^{-1}}$ (as we have done here), placing an absorber at the same radial distance as a potential associated galaxy if their recessional velocities match to within that margin \citep{morris93,penton02,prochaska11,stocke13,keeney17}. The redshift uncertainties reported herein for all galaxy types are significantly less than this canonical peculiar velocity.

The observational data presented here for $\sim9000$ individual galaxies around these sight lines provide a wealth of material with which to investigate the connections between gas and galaxies in the local universe. In addition to the basic observables, we also provide rest-frame $g$-band luminosities and impact parameters, as well as stellar mass, halo mass, and virial radius estimates for each galaxy. We compare our estimates to previously-published values in \autoref{sec:disc}. We also assess each galaxy's proximity to \ion{H}{1} absorbers from \citet{danforth16} and \citet{stocke18}, and use this information to determine Ly$\alpha$ covering fractions for galaxies of different luminosities and spectral classifications.

Covering fractions for emission-line and composite spectrum galaxies are high, nearly 90\% at $\rho \leq 2\,R_{\rm vir}$, as has been seen earlier by \citet{stocke13} and the COS-Halos team \citep{werk14}. These data are consistent with covering fractions at or very close to unity for all star-forming galaxies with $L\geq0.3\,L^*$. However, galaxies with only absorption-line spectra show a consistently lower covering fraction of $\approx60$\% out to at least $4\,R_{\rm vir}$, which indicates that the gaseous environment of passive galaxies may be patchier than for star-forming galaxies. The rather constant covering fraction for the absorption-line galaxies means that their gaseous structures have larger scale lengths than for the star-forming galaxies, suggesting that this gas is related to entire groups of galaxies (see \citealp{keeney17} and \citealp{stocke18}). 

Although the UV-bright targets often have substantial redshifts allowing for a long pathlength with which to probe the CGM and IGM, the connections with individual galaxies and large-scale galaxy structures is still best done at the lowest redshifts due to galaxy redshift survey limitations in any apparent-magnitude-limited sample. While use of the SDSS spectroscopic survey has proven essential for some studies of metal-rich absorbers and galaxies \citep[i.e.,][]{burchett16}, SDSS is quite shallow allowing the gas-galaxy connection to be studied only very nearby. The \citet{burchett16} survey, based as it is on \ion{C}{4} absorption, is also confined to very low redshfits due to the rapidly declining sensitivity of the COS detector at the long wavelength end of the G160M grating mode (i.e., $\lambda >1700$~\AA, or $z_{\rm C\,IV}\ga0.1$).

However, studies of the extent to which various metal ions extend away from galaxies \citep{chen98,penton04,stocke06,stocke13,pratt18} show that the combination of line strength and absorber ionization parameter \citep[see][]{stocke07} allows \ion{O}{6} absorption to trace metals more sensitively and so to much larger distances away from galaxies. The \ion{O}{6} doublet is also one of our very best ways to detect the ``warm-hot'' intergalactic medium (WHIM). For these reasons, it is imperative to have in-hand a galaxy redshift survey significantly deeper than SDSS since the \ion{O}{6} doublet enters the COS standard FUV band at $z\ga0.1$ (i.e., at the same redshift that \ion{C}{4} leaves the COS bandpass), where SDSS provides close to complete redshifts for galaxies of $L>0.4\,L^*$ at $z=0.1$ \citep{strauss02,montero-dorta09}. The results of the \citet{burchett16} survey suggest that galaxies associated with low-$z$ absorbers invariably are at $L>0.3\,L^*$, implying that the SDSS alone is insufficient to study the distribution of metal-enriched gas around galaxies at $z>0.1$.

The COS G130M/1055, G130M/1096, and G130M/1222 modes push the discovery of \ion{O}{6} absorbers to $z<0.1$, but do so at decreased resolution and sensitivity. Instead, the most important redshift range is $z=0.1$-0.17, bounded by the stronger line of the \ion{O}{6} doublet entering the COS G130M/1291 band at 1136~\AA\ and the weaker line of the doublet becoming obscured by the blue-side damping wing of Galactic Ly$\alpha$ absorption at $\sim1212$~\AA. This range is especially important for detecting \ion{O}{6} because there is no confusion with intervening Ly$\alpha$ absorption at these wavelengths, which are blueward of the Ly$\alpha$ rest wavelength. Using the nominal $g=20$ limit for this survey, \ion{O}{6} absorbers have galaxy redshifts determined completely down to $L>0.5\,L^*$ or lower throughout this range, whereas SDSS is only complete to $L>1.3\,L^*$ at $z=0.17$. Thus, this redshift survey is almost completely sufficient for galaxy associations with metal-enriched absorbers. We suggest that future studies of the WHIM and of metal distibution in general concentrate on this redshift interval.

However, if individual absorbers of interest are found in \hst/COS spectra at higher redshifts, the current galaxy survey may not be deep enough to characterize completely the absorber environment.  For example, the very unusual \ion{H}{1} + \ion{O}{6} absorption-line system in the HE~0153--4520 sight line at $z=0.22601$ contains both Lyman limit system (LLS) and WHIM absorbers \citep{savage11} and has a nearest galaxy in this survey at $\rho=80$~kpc (galaxy he0153\_192\_22 in \autoref{tab:redshifts}) that can account for the LLS. And, while a rich galaxy group appears to be present as well (roughly a dozen $L \ga L^*$ galaxies with $\rho\la4$~Mpc are found within $1000~\mathrm{km\,s^{-1}}$ of the absorber redshift in \autoref{tab:redshifts}) that could account for the WHIM absorber, a deeper galaxy survey in this region is required to characterize this group completely \citep[see group characterization discussions in][]{stocke14,stocke18}. However, the current survey provides a good basis for a further investigation of this WHIM absorber.

We provide electronic versions of the reduced, wavelength-calibrated spectra of all galaxies included in this survey to aid future, low-$z$ CGM and IGM studies, as well as new research not yet conceived.

\acknowledgments
We gratefully acknowledge access to high-quality telescopes and multi-object spectrographs from Emma Ryan-Weber, Bart Wakker, Chris Impey, and Buell Jannuzi, and observing assistance from Lisa Winter, Kevin France, Eric Burgh, Emma Ryan-Weber, and Glenn Kacprzak. We would also like to thank Helen Yamamoto and Ben Weiner for their invaluable help with data reduction. This work was supported by NASA grants NNX08AC146 and NAS5-98043 to the University of Colorado at Boulder for the \hst/COS project. B.A.K., J.T.S., C.T.P., and J.D.D. gratefully acknowledge support from NSF grant AST1109117. This research has made use of the NASA/IPAC Extragalactic Database (NED) and the NASA/IPAC Infrared Science Archive (IRSA), which are operated by the Jet Propulsion Laboratory, California Institute of Technology, under contract with the National Aeronautics and Space Administration.

\facilities{AAT (AA$\Omega$), Blanco (MOSAIC), IRSA, MMT (Hectospec), NED, Sloan, WIYN (HYDRA), WIYN:0.9m (MOSAIC)} 
\software{IRAF, IDL, 2dfdr (\url{http://www.aao.gov.au/science/software/2dfdr}), HSRed (\url{http://github.com/richardjcool/HSRed}), PPP \citep{yee91}}

\bibliographystyle{yahapj}
\bibliography{references}

\begin{thebibliography}{}
\providecommand\natexlab[1]{#1}
\providecommand\JournalTitle[1]{#1}

\bibitem[{{Alam} {et~al.}(2015){Alam}, {Albareti}, {Allende Prieto}, {Anders},
  {Anderson}, {Anderton}, {Andrews}, {Armengaud}, {Aubourg}, {Bailey}, \&
  et~al.}]{alam15}
{Alam}, S., {Albareti}, F.~D., {Allende Prieto}, C., {et~al.} 2015,
  \href{http://dx.doi.org/10.1088/0067-0049/219/1/12}{\JournalTitle{\apjs},
  219, 12}

\bibitem[{{Barden} {et~al.}(1993){Barden}, {Armandroff}, {Massey}, {Groves},
  {Rudeen}, {Vaughnn}, \& {Muller}}]{barden93}
{Barden}, S.~C., {Armandroff}, T., {Massey}, P., {et~al.} 1993, in Astronomical
  Society of the Pacific Conference Series, Vol.~37, Fiber Optics in Astronomy
  II, ed. P.~M. {Gray}, 185

\bibitem[{{Beck} {et~al.}(2017){Beck}, {Lin}, {Ishida}, {Gieseke}, {de Souza},
  {Costa-Duarte}, {Hattab}, \& {Krone-Martins}}]{beck17}
{Beck}, R., {Lin}, C.-A., {Ishida}, E.~E.~O., {et~al.} 2017,
  \href{http://dx.doi.org/10.1093/mnras/stx687}{\JournalTitle{\mnras}, 468,
  4323}

\bibitem[{{Behroozi} {et~al.}(2010){Behroozi}, {Conroy}, \&
  {Wechsler}}]{behroozi10}
{Behroozi}, P.~S., {Conroy}, C., \& {Wechsler}, R.~H. 2010,
  \href{http://dx.doi.org/10.1088/0004-637X/717/1/379}{\JournalTitle{\apj},
  717, 379}

\bibitem[{{Bershady} {et~al.}(2008){Bershady}, {Barden}, {Blanche}, {Blanco},
  {Corson}, {Crawford}, {Glaspey}, {Habraken}, {Jacoby}, {Keyes}, {Knezek},
  {Lemaire}, {Liang}, {McDougall}, {Poczulp}, {Sawyer}, {Westfall}, \&
  {Willmarth}}]{bershady08}
{Bershady}, M., {Barden}, S., {Blanche}, P.-A., {et~al.} 2008,
  \href{http://dx.doi.org/10.1117/12.789112}{in \procspie, Vol. 7014,
  Ground-based and Airborne Instrumentation for Astronomy II}, 70140H

\bibitem[{{Binney} \& {Tremaine}(1987)}]{binney87}
{Binney}, J., \& {Tremaine}, S. 1987, {Galactic Dynamics}

\bibitem[{{Burchett} {et~al.}(2015){Burchett}, {Tripp}, {Prochaska}, {Werk},
  {Tumlinson}, {O'Meara}, {Bordoloi}, {Katz}, \& {Willmer}}]{burchett15}
{Burchett}, J.~N., {Tripp}, T.~M., {Prochaska}, J.~X., {et~al.} 2015,
  \href{http://dx.doi.org/10.1088/0004-637X/815/2/91}{\JournalTitle{\apj}, 815,
  91}

\bibitem[{{Burchett} {et~al.}(2016){Burchett}, {Tripp}, {Bordoloi}, {Werk},
  {Prochaska}, {Tumlinson}, {Willmer}, {O'Meara}, \& {Katz}}]{burchett16}
{Burchett}, J.~N., {Tripp}, T.~M., {Bordoloi}, R., {et~al.} 2016,
  \href{http://dx.doi.org/10.3847/0004-637X/832/2/124}{\JournalTitle{\apj},
  832, 124}

\bibitem[{{Chang} {et~al.}(2015){Chang}, {van der Wel}, {da Cunha}, \&
  {Rix}}]{chang15}
{Chang}, Y.-Y., {van der Wel}, A., {da Cunha}, E., \& {Rix}, H.-W. 2015,
  \href{http://dx.doi.org/10.1088/0067-0049/219/1/8}{\JournalTitle{\apjs}, 219,
  8}

\bibitem[{{Chen} {et~al.}(1998){Chen}, {Lanzetta}, {Webb}, \&
  {Barcons}}]{chen98}
{Chen}, H.-W., {Lanzetta}, K.~M., {Webb}, J.~K., \& {Barcons}, X. 1998,
  \href{http://dx.doi.org/10.1086/305554}{\JournalTitle{\apj}, 498, 77}

\bibitem[{{Chen} \& {Mulchaey}(2009)}]{chen09}
{Chen}, H.-W., \& {Mulchaey}, J.~S. 2009,
  \href{http://dx.doi.org/10.1088/0004-637X/701/2/1219}{\JournalTitle{\apj},
  701, 1219}

\bibitem[{{Chilingarian} {et~al.}(2010){Chilingarian}, {Melchior}, \&
  {Zolotukhin}}]{chilingarian10}
{Chilingarian}, I.~V., {Melchior}, A.-L., \& {Zolotukhin}, I.~Y. 2010,
  \href{http://dx.doi.org/10.1111/j.1365-2966.2010.16506.x}{\JournalTitle{\mnras},
  405, 1409}

\bibitem[{{Chilingarian} \& {Zolotukhin}(2012)}]{chilingarian12}
{Chilingarian}, I.~V., \& {Zolotukhin}, I.~Y. 2012,
  \href{http://dx.doi.org/10.1111/j.1365-2966.2011.19837.x}{\JournalTitle{\mnras},
  419, 1727}

\bibitem[{{Conroy} \& {Wechsler}(2009)}]{conroy09}
{Conroy}, C., \& {Wechsler}, R.~H. 2009,
  \href{http://dx.doi.org/10.1088/0004-637X/696/1/620}{\JournalTitle{\apj},
  696, 620}

\bibitem[{{Danforth} {et~al.}(2016){Danforth}, {Keeney}, {Tilton}, {Shull},
  {Stocke}, {Stevans}, {Pieri}, {Savage}, {France}, {Syphers}, {Smith},
  {Green}, {Froning}, {Penton}, \& {Osterman}}]{danforth16}
{Danforth}, C.~W., {Keeney}, B.~A., {Tilton}, E.~M., {et~al.} 2016,
  \href{http://dx.doi.org/10.3847/0004-637X/817/2/111}{\JournalTitle{\apj},
  817, 111}

\bibitem[{{Fabricant} {et~al.}(2005){Fabricant}, {Fata}, {Roll}, {Hertz},
  {Caldwell}, {Gauron}, {Geary}, {McLeod}, {Szentgyorgyi}, {Zajac}, {Kurtz},
  {Barberis}, {Bergner}, {Brown}, {Conroy}, {Eng}, {Geller}, {Goddard},
  {Honsa}, {Mueller}, {Mink}, {Ordway}, {Tokarz}, {Woods}, {Wyatt}, {Epps}, \&
  {Dell'Antonio}}]{fabricant05}
{Fabricant}, D., {Fata}, R., {Roll}, J., {et~al.} 2005,
  \href{http://dx.doi.org/10.1086/497385}{\JournalTitle{\pasp}, 117, 1411}

\bibitem[{{Fitzpatrick}(1999)}]{fitzpatrick99}
{Fitzpatrick}, E.~L. 1999,
  \href{http://dx.doi.org/10.1086/316293}{\JournalTitle{\pasp}, 111, 63}

\bibitem[{{Gehrels}(1986)}]{gehrels86}
{Gehrels}, N. 1986,
  \href{http://dx.doi.org/10.1086/164079}{\JournalTitle{\apj}, 303, 336}

\bibitem[{{Green} {et~al.}(2012){Green}, {Froning}, {Osterman}, {Ebbets},
  {Heap}, {Leitherer}, {Linsky}, {Savage}, {Sembach}, {Shull}, {Siegmund},
  {Snow}, {Spencer}, {Stern}, {Stocke}, {Welsh}, {B{\'e}land}, {Burgh},
  {Danforth}, {France}, {Keeney}, {McPhate}, {Penton}, {Andrews},
  {Brownsberger}, {Morse}, \& {Wilkinson}}]{green12}
{Green}, J.~C., {Froning}, C.~S., {Osterman}, S., {et~al.} 2012,
  \href{http://dx.doi.org/10.1088/0004-637X/744/1/60}{\JournalTitle{\apj}, 744,
  60}

\bibitem[{{Hayward} \& {Hopkins}(2017)}]{hayward17}
{Hayward}, C.~C., \& {Hopkins}, P.~F. 2017,
  \href{http://dx.doi.org/10.1093/mnras/stw2888}{\JournalTitle{\mnras}, 465,
  1682}

\bibitem[{{Hinshaw} {et~al.}(2013){Hinshaw}, {Larson}, {Komatsu}, {Spergel},
  {Bennett}, {Dunkley}, {Nolta}, {Halpern}, {Hill}, {Odegard}, {Page}, {Smith},
  {Weiland}, {Gold}, {Jarosik}, {Kogut}, {Limon}, {Meyer}, {Tucker}, {Wollack},
  \& {Wright}}]{hinshaw13}
{Hinshaw}, G., {Larson}, D., {Komatsu}, E., {et~al.} 2013,
  \href{http://dx.doi.org/10.1088/0067-0049/208/2/19}{\JournalTitle{\apjs},
  208, 19}

\bibitem[{{Johnson} {et~al.}(2013){Johnson}, {Chen}, \& {Mulchaey}}]{johnson13}
{Johnson}, S.~D., {Chen}, H.-W., \& {Mulchaey}, J.~S. 2013,
  \href{http://dx.doi.org/10.1093/mnras/stt1137}{\JournalTitle{\mnras}, 434,
  1765}

\bibitem[{{Johnson} {et~al.}(2017){Johnson}, {Chen}, {Mulchaey}, {Schaye}, \&
  {Straka}}]{johnson17}
{Johnson}, S.~D., {Chen}, H.-W., {Mulchaey}, J.~S., {Schaye}, J., \& {Straka},
  L.~A. 2017,
  \href{http://dx.doi.org/10.3847/2041-8213/aa9370}{\JournalTitle{\apjl}, 850,
  L10}

\bibitem[{{Kauffmann} {et~al.}(2003){Kauffmann}, {Heckman}, {White}, {Charlot},
  {Tremonti}, {Brinchmann}, {Bruzual}, {Peng}, {Seibert}, {Bernardi},
  {Blanton}, {Brinkmann}, {Castander}, {Cs{\'a}bai}, {Fukugita}, {Ivezic},
  {Munn}, {Nichol}, {Padmanabhan}, {Thakar}, {Weinberg}, \&
  {York}}]{kauffmann03}
{Kauffmann}, G., {Heckman}, T.~M., {White}, S.~D.~M., {et~al.} 2003,
  \href{http://dx.doi.org/10.1046/j.1365-8711.2003.06291.x}{\JournalTitle{\mnras},
  341, 33}

\bibitem[{{Keeney} {et~al.}(2013){Keeney}, {Stocke}, {Rosenberg}, {Danforth},
  {Ryan-Weber}, {Shull}, {Savage}, \& {Green}}]{keeney13}
{Keeney}, B.~A., {Stocke}, J.~T., {Rosenberg}, J.~L., {et~al.} 2013,
  \href{http://dx.doi.org/10.1088/0004-637X/765/1/27}{\JournalTitle{\apj}, 765,
  27}

\bibitem[{{Keeney} {et~al.}(2017){Keeney}, {Stocke}, {Danforth}, {Shull},
  {Pratt}, {Froning}, {Green}, {Penton}, \& {Savage}}]{keeney17}
{Keeney}, B.~A., {Stocke}, J.~T., {Danforth}, C.~W., {et~al.} 2017,
  \href{http://dx.doi.org/10.3847/1538-4365/aa6b59}{\JournalTitle{\apjs}, 230,
  6}

\bibitem[{{Kere{\v s}} \& {Hernquist}(2009)}]{keres09}
{Kere{\v s}}, D., \& {Hernquist}, L. 2009,
  \href{http://dx.doi.org/10.1088/0004-637X/700/1/L1}{\JournalTitle{\apjl},
  700, L1}

\bibitem[{{Montero-Dorta} \& {Prada}(2009)}]{montero-dorta09}
{Montero-Dorta}, A.~D., \& {Prada}, F. 2009,
  \href{http://dx.doi.org/10.1111/j.1365-2966.2009.15197.x}{\JournalTitle{\mnras},
  399, 1106}

\bibitem[{{Morris} {et~al.}(1993){Morris}, {Weymann}, {Dressler}, {McCarthy},
  {Smith}, {Terrile}, {Giovanelli}, \& {Irwin}}]{morris93}
{Morris}, S.~L., {Weymann}, R.~J., {Dressler}, A., {et~al.} 1993,
  \href{http://dx.doi.org/10.1086/173505}{\JournalTitle{\apj}, 419, 524}

\bibitem[{{Moster} {et~al.}(2013){Moster}, {Naab}, \& {White}}]{moster13}
{Moster}, B.~P., {Naab}, T., \& {White}, S.~D.~M. 2013,
  \href{http://dx.doi.org/10.1093/mnras/sts261}{\JournalTitle{\mnras}, 428,
  3121}

\bibitem[{{Moster} {et~al.}(2010){Moster}, {Somerville}, {Maulbetsch}, {van den
  Bosch}, {Macci{\`o}}, {Naab}, \& {Oser}}]{moster10}
{Moster}, B.~P., {Somerville}, R.~S., {Maulbetsch}, C., {et~al.} 2010,
  \href{http://dx.doi.org/10.1088/0004-637X/710/2/903}{\JournalTitle{\apj},
  710, 903}

\bibitem[{{Muratov} {et~al.}(2015){Muratov}, {Kere{\v s}},
  {Faucher-Gigu{\`e}re}, {Hopkins}, {Quataert}, \& {Murray}}]{muratov15}
{Muratov}, A.~L., {Kere{\v s}}, D., {Faucher-Gigu{\`e}re}, C.-A., {et~al.}
  2015, \href{http://dx.doi.org/10.1093/mnras/stv2126}{\JournalTitle{\mnras},
  454, 2691}

\bibitem[{{Muratov} {et~al.}(2017){Muratov}, {Kere{\v s}},
  {Faucher-Gigu{\`e}re}, {Hopkins}, {Ma}, {Angl{\'e}s-Alc{\'a}zar}, {Chan},
  {Torrey}, {Hafen}, {Quataert}, \& {Murray}}]{muratov17}
---. 2017, \href{http://dx.doi.org/10.1093/mnras/stx667}{\JournalTitle{\mnras},
  468, 4170}

\bibitem[{{Pagel}(2008)}]{pagel08}
{Pagel}, B.~E.~J. 2008, in Astronomical Society of the Pacific Conference
  Series, Vol. 390, Pathways Through an Eclectic Universe, ed. J.~H. {Knapen},
  T.~J. {Mahoney}, \& A.~{Vazdekis}, 483

\bibitem[{{Penton} {et~al.}(2002){Penton}, {Stocke}, \& {Shull}}]{penton02}
{Penton}, S.~V., {Stocke}, J.~T., \& {Shull}, J.~M. 2002,
  \href{http://dx.doi.org/10.1086/324483}{\JournalTitle{\apj}, 565, 720}

\bibitem[{{Penton} {et~al.}(2004){Penton}, {Stocke}, \& {Shull}}]{penton04}
---. 2004, \href{http://dx.doi.org/10.1086/382877}{\JournalTitle{\apjs}, 152,
  29}

\bibitem[{{Pratt} {et~al.}(2018){Pratt}, {Stocke}, {Keeney}, \&
  {Danforth}}]{pratt18}
{Pratt}, C.~T., {Stocke}, J.~T., {Keeney}, B.~A., \& {Danforth}, C.~W. 2018,
  \href{http://dx.doi.org/10.3847/1538-4357/aaaaac}{\JournalTitle{\apj}, 855,
  18}

\bibitem[{{Prochaska} {et~al.}(2011){Prochaska}, {Weiner}, {Chen}, {Cooksey},
  \& {Mulchaey}}]{prochaska11}
{Prochaska}, J.~X., {Weiner}, B., {Chen}, H.-W., {Cooksey}, K.~L., \&
  {Mulchaey}, J.~S. 2011,
  \href{http://dx.doi.org/10.1088/0067-0049/193/2/28}{\JournalTitle{\apjs},
  193, 28}

\bibitem[{{Prochaska} {et~al.}(2017){Prochaska}, {Werk}, {Worseck}, {Tripp},
  {Tumlinson}, {Burchett}, {Fox}, {Fumagalli}, {Lehner}, {Peeples}, \&
  {Tejos}}]{prochaska17}
{Prochaska}, J.~X., {Werk}, J.~K., {Worseck}, G., {et~al.} 2017,
  \href{http://dx.doi.org/10.3847/1538-4357/aa6007}{\JournalTitle{\apj}, 837,
  169}

\bibitem[{{Rudie} {et~al.}(2012){Rudie}, {Steidel}, {Trainor}, {Rakic},
  {Bogosavljevi{\'c}}, {Pettini}, {Reddy}, {Shapley}, {Erb}, \&
  {Law}}]{rudie12}
{Rudie}, G.~C., {Steidel}, C.~C., {Trainor}, R.~F., {et~al.} 2012,
  \href{http://dx.doi.org/10.1088/0004-637X/750/1/67}{\JournalTitle{\apj}, 750,
  67}

\bibitem[{{Salim} {et~al.}(2007){Salim}, {Rich}, {Charlot}, {Brinchmann},
  {Johnson}, {Schiminovich}, {Seibert}, {Mallery}, {Heckman}, {Forster},
  {Friedman}, {Martin}, {Morrissey}, {Neff}, {Small}, {Wyder}, {Bianchi},
  {Donas}, {Lee}, {Madore}, {Milliard}, {Szalay}, {Welsh}, \& {Yi}}]{salim07}
{Salim}, S., {Rich}, R.~M., {Charlot}, S., {et~al.} 2007,
  \href{http://dx.doi.org/10.1086/519218}{\JournalTitle{\apjs}, 173, 267}

\bibitem[{{Savage} {et~al.}(2012){Savage}, {Kim}, {Keeney}, {Narayanan},
  {Stocke}, {Syphers}, \& {Wakker}}]{savage12}
{Savage}, B.~D., {Kim}, T.-S., {Keeney}, B., {et~al.} 2012,
  \href{http://dx.doi.org/10.1088/0004-637X/753/1/80}{\JournalTitle{\apj}, 753,
  80}

\bibitem[{{Savage} {et~al.}(2014){Savage}, {Kim}, {Wakker}, {Keeney}, {Shull},
  {Stocke}, \& {Green}}]{savage14}
{Savage}, B.~D., {Kim}, T.-S., {Wakker}, B.~P., {et~al.} 2014,
  \href{http://dx.doi.org/10.1088/0067-0049/212/1/8}{\JournalTitle{\apjs}, 212,
  8}

\bibitem[{{Savage} {et~al.}(2011){Savage}, {Narayanan}, {Lehner}, \&
  {Wakker}}]{savage11}
{Savage}, B.~D., {Narayanan}, A., {Lehner}, N., \& {Wakker}, B.~P. 2011,
  \href{http://dx.doi.org/10.1088/0004-637X/731/1/14}{\JournalTitle{\apj}, 731,
  14}

\bibitem[{{Savage} {et~al.}(2010){Savage}, {Narayanan}, {Wakker}, {Stocke},
  {Keeney}, {Shull}, {Sembach}, {Yao}, \& {Green}}]{savage10}
{Savage}, B.~D., {Narayanan}, A., {Wakker}, B.~P., {et~al.} 2010,
  \href{http://dx.doi.org/10.1088/0004-637X/719/2/1526}{\JournalTitle{\apj},
  719, 1526}

\bibitem[{{Schlafly} \& {Finkbeiner}(2011)}]{schlafly11}
{Schlafly}, E.~F., \& {Finkbeiner}, D.~P. 2011,
  \href{http://dx.doi.org/10.1088/0004-637X/737/2/103}{\JournalTitle{\apj},
  737, 103}

\bibitem[{{Sharp} {et~al.}(2006){Sharp}, {Saunders}, {Smith}, {Churilov},
  {Correll}, {Dawson}, {Farrel}, {Frost}, {Haynes}, {Heald}, {Lankshear},
  {Mayfield}, {Waller}, \& {Whittard}}]{sharp06}
{Sharp}, R., {Saunders}, W., {Smith}, G., {et~al.} 2006,
  \href{http://dx.doi.org/10.1117/12.671022}{in \procspie, Vol. 6269, Society
  of Photo-Optical Instrumentation Engineers (SPIE) Conference Series}, 62690G

\bibitem[{{Starck} {et~al.}(1997){Starck}, {Siebenmorgen}, \&
  {Gredel}}]{starck97}
{Starck}, J.-L., {Siebenmorgen}, R., \& {Gredel}, R. 1997,
  \href{http://dx.doi.org/10.1086/304186}{\JournalTitle{\apj}, 482, 1011}

\bibitem[{{Stocke} {et~al.}(2018){Stocke}, {Danforth}, {Keeney}, {Oppenheimer},
  {Pratt}, {Berlind}, {Impey}, \& {Jannuzi}}]{stocke18}
{Stocke}, J.~T., {Danforth}, C.~W., {Keeney}, B.~A., {et~al.} 2018,
  \JournalTitle{\apj}, submitted

\bibitem[{{Stocke} {et~al.}(2007){Stocke}, {Danforth}, {Shull}, {Penton}, \&
  {Giroux}}]{stocke07}
{Stocke}, J.~T., {Danforth}, C.~W., {Shull}, J.~M., {Penton}, S.~V., \&
  {Giroux}, M.~L. 2007,
  \href{http://dx.doi.org/10.1086/522920}{\JournalTitle{\apj}, 671, 146}

\bibitem[{{Stocke} {et~al.}(2017){Stocke}, {Keeney}, {Danforth}, {Oppenheimer},
  {Pratt}, \& {Berlind}}]{stocke17}
{Stocke}, J.~T., {Keeney}, B.~A., {Danforth}, C.~W., {et~al.} 2017,
  \href{http://dx.doi.org/10.3847/1538-4357/aa64e2}{\JournalTitle{\apj}, 838,
  37}

\bibitem[{{Stocke} {et~al.}(2013){Stocke}, {Keeney}, {Danforth}, {Shull},
  {Froning}, {Green}, {Penton}, \& {Savage}}]{stocke13}
---. 2013,
  \href{http://dx.doi.org/10.1088/0004-637X/763/2/148}{\JournalTitle{\apj},
  763, 148}

\bibitem[{{Stocke} {et~al.}(2006){Stocke}, {Penton}, {Danforth}, {Shull},
  {Tumlinson}, \& {McLin}}]{stocke06}
{Stocke}, J.~T., {Penton}, S.~V., {Danforth}, C.~W., {et~al.} 2006,
  \href{http://dx.doi.org/10.1086/500386}{\JournalTitle{\apj}, 641, 217}

\bibitem[{{Stocke} {et~al.}(2014){Stocke}, {Keeney}, {Danforth}, {Syphers},
  {Yamamoto}, {Shull}, {Green}, {Froning}, {Savage}, {Wakker}, {Kim},
  {Ryan-Weber}, \& {Kacprzak}}]{stocke14}
{Stocke}, J.~T., {Keeney}, B.~A., {Danforth}, C.~W., {et~al.} 2014,
  \href{http://dx.doi.org/10.1088/0004-637X/791/2/128}{\JournalTitle{\apj},
  791, 128}

\bibitem[{{Strauss} {et~al.}(2002){Strauss}, {Weinberg}, {Lupton}, {Narayanan},
  {Annis}, {Bernardi}, {Blanton}, {Burles}, {Connolly}, {Dalcanton}, {Doi},
  {Eisenstein}, {Frieman}, {Fukugita}, {Gunn}, {Ivezi{\'c}}, {Kent}, {Kim},
  {Knapp}, {Kron}, {Munn}, {Newberg}, {Nichol}, {Okamura}, {Quinn}, {Richmond},
  {Schlegel}, {Shimasaku}, {SubbaRao}, {Szalay}, {Vanden Berk}, {Vogeley},
  {Yanny}, {Yasuda}, {York}, \& {Zehavi}}]{strauss02}
{Strauss}, M.~A., {Weinberg}, D.~H., {Lupton}, R.~H., {et~al.} 2002,
  \href{http://dx.doi.org/10.1086/342343}{\JournalTitle{\aj}, 124, 1810}

\bibitem[{{Taylor} {et~al.}(2011){Taylor}, {Hopkins}, {Baldry}, {Brown},
  {Driver}, {Kelvin}, {Hill}, {Robotham}, {Bland-Hawthorn}, {Jones}, {Sharp},
  {Thomas}, {Liske}, {Loveday}, {Norberg}, {Peacock}, {Bamford}, {Brough},
  {Colless}, {Cameron}, {Conselice}, {Croom}, {Frenk}, {Gunawardhana},
  {Kuijken}, {Nichol}, {Parkinson}, {Phillipps}, {Pimbblet}, {Popescu},
  {Prescott}, {Sutherland}, {Tuffs}, {van Kampen}, \& {Wijesinghe}}]{taylor11}
{Taylor}, E.~N., {Hopkins}, A.~M., {Baldry}, I.~K., {et~al.} 2011,
  \href{http://dx.doi.org/10.1111/j.1365-2966.2011.19536.x}{\JournalTitle{\mnras},
  418, 1587}

\bibitem[{{Thom} {et~al.}(2012){Thom}, {Tumlinson}, {Werk}, {Prochaska},
  {Oppenheimer}, {Peeples}, {Tripp}, {Katz}, {O'Meara}, {Ford}, {Dav{\'e}},
  {Sembach}, \& {Weinberg}}]{thom12}
{Thom}, C., {Tumlinson}, J., {Werk}, J.~K., {et~al.} 2012,
  \href{http://dx.doi.org/10.1088/2041-8205/758/2/L41}{\JournalTitle{\apjl},
  758, L41}

\bibitem[{{Tumlinson} \& {Fang}(2005)}]{tumlinson05}
{Tumlinson}, J., \& {Fang}, T. 2005,
  \href{http://dx.doi.org/10.1086/430142}{\JournalTitle{\apjl}, 623, L97}

\bibitem[{{Tumlinson} {et~al.}(2017){Tumlinson}, {Peeples}, \&
  {Werk}}]{tumlinson17}
{Tumlinson}, J., {Peeples}, M.~S., \& {Werk}, J.~K. 2017,
  \href{http://dx.doi.org/10.1146/annurev-astro-091916-055240}{\JournalTitle{\araa},
  55, 389}

\bibitem[{{Tumlinson} {et~al.}(2011){Tumlinson}, {Thom}, {Werk}, {Prochaska},
  {Tripp}, {Weinberg}, {Peeples}, {O'Meara}, {Oppenheimer}, {Meiring}, {Katz},
  {Dav{\'e}}, {Ford}, \& {Sembach}}]{tumlinson11}
{Tumlinson}, J., {Thom}, C., {Werk}, J.~K., {et~al.} 2011,
  \href{http://dx.doi.org/10.1126/science.1209840}{\JournalTitle{Science}, 334,
  948}

\bibitem[{{Valdes}(1998)}]{valdes98}
{Valdes}, F.~G. 1998, in Astronomical Society of the Pacific Conference Series,
  Vol. 145, Astronomical Data Analysis Software and Systems VII, ed.
  R.~{Albrecht}, R.~N. {Hook}, \& H.~A. {Bushouse}, 53

\bibitem[{{Veilleux} {et~al.}(2005){Veilleux}, {Cecil}, \&
  {Bland-Hawthorn}}]{veilleux05}
{Veilleux}, S., {Cecil}, G., \& {Bland-Hawthorn}, J. 2005,
  \href{http://dx.doi.org/10.1146/annurev.astro.43.072103.150610}{\JournalTitle{\araa},
  43, 769}

\bibitem[{{Wakker} \& {Savage}(2009)}]{wakker09}
{Wakker}, B.~P., \& {Savage}, B.~D. 2009,
  \href{http://dx.doi.org/10.1088/0067-0049/182/1/378}{\JournalTitle{\apjs},
  182, 378}

\bibitem[{{Werk} {et~al.}(2013){Werk}, {Prochaska}, {Thom}, {Tumlinson},
  {Tripp}, {O'Meara}, \& {Peeples}}]{werk13}
{Werk}, J.~K., {Prochaska}, J.~X., {Thom}, C., {et~al.} 2013,
  \href{http://dx.doi.org/10.1088/0067-0049/204/2/17}{\JournalTitle{\apjs},
  204, 17}

\bibitem[{{Werk} {et~al.}(2014){Werk}, {Prochaska}, {Tumlinson}, {Peeples},
  {Tripp}, {Fox}, {Lehner}, {Thom}, {O'Meara}, {Ford}, {Bordoloi}, {Katz},
  {Tejos}, {Oppenheimer}, {Dav{\'e}}, \& {Weinberg}}]{werk14}
{Werk}, J.~K., {Prochaska}, J.~X., {Tumlinson}, J., {et~al.} 2014,
  \href{http://dx.doi.org/10.1088/0004-637X/792/1/8}{\JournalTitle{\apj}, 792,
  8}

\bibitem[{{Yee}(1991)}]{yee91}
{Yee}, H.~K.~C. 1991,
  \href{http://dx.doi.org/10.1086/132834}{\JournalTitle{\pasp}, 103, 396}

\end{thebibliography}

\appendix
\section{Detailed Sight Line Completeness Tables}
\label{sec:appendix}

Here we provide tables detailing the overall completeness (\autoref{sec:results:completeness}) as a function of apparent $g$-band magnitude and angular separation between the galaxy and the QSO for each sight line. Unlike in \autoref{tab:completeness}, we show fractional completion (i.e., the number of galaxies in a given bin with redshift determinations divided by the number of targeted galaxies in that bin) instead of completion percentages. The number of targeted galaxies in a bin include objects of all priorities (\autoref{sec:obs}), and the redshift determinations come from our surveys (\autoref{tab:redshifts}), SDSS DR14, and published studies around individual sight lines \citep[e.g.,][]{prochaska11}.

\bigskip
\twocolumngrid
\clearpage
\begin{deluxetable}{c|CCCC}
\tablecaption{1ES~1028+511 Completeness
\label{tab:1es1028}}

\tablehead{ \colhead{$g$} & \colhead{$\leq5\arcmin$} & \colhead{$5\arcmin$-$10\arcmin$} & \colhead{$10\arcmin$-$15\arcmin$} & \colhead{$15\arcmin$-$20\arcmin$} }

\startdata
$\leq18$ & \nodata &   1/2   &   3/4   &   5/6   \\
18-18.5  &   1/1   &   2/2   &   1/1   &   2/2   \\
18.5-19  & \nodata &   3/3   &   4/5   &   4/6   \\
19-19.5  & \nodata &   3/3   &   3/3   &  11/11  \\
19.5-20  &   4/4   &   5/5   &  15/15  &  14/18  \\
\enddata

\end{deluxetable}
\begin{deluxetable}{c|CCCC}
\tablecaption{1ES~1553+113 Completeness
\label{tab:1es1553}}

\tablehead{ \colhead{$g$} & \colhead{$\leq5\arcmin$} & \colhead{$5\arcmin$-$10\arcmin$} & \colhead{$10\arcmin$-$15\arcmin$} & \colhead{$15\arcmin$-$20\arcmin$} }

\startdata
$\leq18$ &   2/3   &   3/3   &   8/9   &  14/14  \\
18-18.5  &   1/1   &   4/4   &   5/5   &   4/4   \\
18.5-19  &   3/3   &   6/7   &   6/6   &  15/16  \\
19-19.5  &   3/3   &   3/3   &   9/9   &  16/16  \\
19.5-20  &   2/2   &  10/10  &  25/26  &  23/27  \\
\enddata

\end{deluxetable}
\begin{deluxetable}{c|CCCC}
\tablecaption{1SAX~J1032.3+5051 Completeness
\label{tab:1sax1032}}

\tablehead{ \colhead{$g$} & \colhead{$\leq5\arcmin$} & \colhead{$5\arcmin$-$10\arcmin$} & \colhead{$10\arcmin$-$15\arcmin$} & \colhead{$15\arcmin$-$20\arcmin$} }

\startdata
$\leq18$ & \nodata &   2/2   &   5/6   & \nodata \\
18-18.5  & \nodata &   1/1   &   2/2   &   3/3   \\
18.5-19  & \nodata &   4/4   &   3/4   &   6/6   \\
19-19.5  &   1/1   &   2/2   &   1/1   &  12/12  \\
19.5-20  &   2/2   &   5/6   &   7/7   &  18/20  \\
\enddata

\end{deluxetable}
\begin{deluxetable}{c|CCCC}
\tablecaption{3C~57 Completeness
\label{tab:3c57}}

\tablehead{ \colhead{$g$} & \colhead{$\leq5\arcmin$} & \colhead{$5\arcmin$-$10\arcmin$} & \colhead{$10\arcmin$-$15\arcmin$} & \colhead{$15\arcmin$-$20\arcmin$} }

\startdata
$\leq18$ & \nodata &   2/2   &   5/8   &   8/10  \\
18-18.5  & \nodata & \nodata & \nodata &   1/3   \\
18.5-19  &   0/1   &   1/1   &   1/2   &   1/6   \\
19-19.5  &   0/3   &   0/3   &   2/8   &   3/16  \\
19.5-20  &   0/6   &   0/11  &   0/14  &   2/25  \\
\enddata

\end{deluxetable}
\begin{deluxetable}{c|CCCC}
\tablecaption{3C~263 Completeness
\label{tab:3c263}}

\tablehead{ \colhead{$g$} & \colhead{$\leq5\arcmin$} & \colhead{$5\arcmin$-$10\arcmin$} & \colhead{$10\arcmin$-$15\arcmin$} & \colhead{$15\arcmin$-$20\arcmin$} }

\startdata
$\leq18$ & \nodata &   4/5   &   1/2   &   1/1   \\
18-18.5  &   2/2   &   2/2   &   3/3   &   3/4   \\
18.5-19  & \nodata &   3/3   &   5/6   &  10/11  \\
19-19.5  & \nodata &  10/10  &   7/8   &   8/9   \\
19.5-20  &   4/4   &  11/12  &  14/19  &  22/28  \\
\enddata

\end{deluxetable}
\begin{deluxetable}{c|CCCC}
\tablecaption{B~1612+266 Completeness
\label{tab:b1612}}

\tablehead{ \colhead{$g$} & \colhead{$\leq5\arcmin$} & \colhead{$5\arcmin$-$10\arcmin$} & \colhead{$10\arcmin$-$15\arcmin$} & \colhead{$15\arcmin$-$20\arcmin$} }

\startdata
$\leq18$ & \nodata &   6/8   &   5/5   &   8/9   \\
18-18.5  &   1/1   &   2/4   &   2/2   &   3/4   \\
18.5-19  &   1/2   &   3/7   &   4/10  &   5/11  \\
19-19.5  &   4/6   &   9/14  &   3/8   &  13/21  \\
19.5-20  &   4/5   &  11/19  &  17/27  &  16/28  \\
\enddata

\end{deluxetable}

\clearpage
\begin{deluxetable}{c|CCCC}
\tablecaption{CSO~1022 Completeness
\label{tab:cso1022}}

\tablehead{ \colhead{$g$} & \colhead{$\leq5\arcmin$} & \colhead{$5\arcmin$-$10\arcmin$} & \colhead{$10\arcmin$-$15\arcmin$} & \colhead{$15\arcmin$-$20\arcmin$} }

\startdata
$\leq18$ &   3/3   &   5/6   &  13/15  &   6/6   \\
18-18.5  &   2/2   &   7/7   &  11/13  &   5/6   \\
18.5-19  &   6/6   &   3/4   &  11/15  &  11/18  \\
19-19.5  &   6/11  &   7/10  &  14/22  &  13/34  \\
19.5-20  &   1/2   &  18/25  &  17/37  &  15/36  \\
\enddata

\end{deluxetable}
\begin{deluxetable}{c|CCCC}
\tablecaption{CSO~1080 Completeness
\label{tab:cso1080}}

\tablehead{ \colhead{$g$} & \colhead{$\leq5\arcmin$} & \colhead{$5\arcmin$-$10\arcmin$} & \colhead{$10\arcmin$-$15\arcmin$} & \colhead{$15\arcmin$-$20\arcmin$} }

\startdata
$\leq18$ & \nodata &   0/1   &   5/6   &   2/2   \\
18-18.5  & \nodata &   4/4   &   2/2   &   7/8   \\
18.5-19  &   1/1   &   4/5   &   4/6   &  13/18  \\
19-19.5  &   1/1   &   8/8   &  13/19  &  16/24  \\
19.5-20  &   5/6   &  10/13  &  17/22  &  20/32  \\
\enddata

\end{deluxetable}
\begin{deluxetable}{c|CCCC}
\tablecaption{FBQS~J1010+3003 Completeness
\label{tab:fbqs1010}}

\tablehead{ \colhead{$g$} & \colhead{$\leq5\arcmin$} & \colhead{$5\arcmin$-$10\arcmin$} & \colhead{$10\arcmin$-$15\arcmin$} & \colhead{$15\arcmin$-$20\arcmin$} }

\startdata
$\leq18$ &   2/2   &   2/2   &   2/4   &   6/7   \\
18-18.5  &   2/2   &   3/3   &   3/3   &   3/3   \\
18.5-19  &   2/2   &   5/5   &   9/9   &  15/15  \\
19-19.5  &   2/2   &  10/11  &  12/14  &  16/18  \\
19.5-20  &   4/6   &  13/15  &  24/31  &  19/23  \\
\enddata

\end{deluxetable}
\begin{deluxetable}{c|CCCC}
\tablecaption{FBQS~J1030+3102 Completeness
\label{tab:fbqs1030}}

\tablehead{ \colhead{$g$} & \colhead{$\leq5\arcmin$} & \colhead{$5\arcmin$-$10\arcmin$} & \colhead{$10\arcmin$-$15\arcmin$} & \colhead{$15\arcmin$-$20\arcmin$} }

\startdata
$\leq18$ &   2/2   &   4/6   &   9/10  &   3/4   \\
18-18.5  &   2/2   &   2/2   &   8/8   &   2/2   \\
18.5-19  &   1/1   &   3/5   &   2/3   &  13/15  \\
19-19.5  &   1/2   &   4/5   &   8/10  &   7/10  \\
19.5-20  &   1/4   &   8/10  &  12/17  &  11/17  \\
\enddata

\end{deluxetable}
\begin{deluxetable}{c|CCCC}
\tablecaption{FBQS~J11519+2838 Completeness
\label{tab:fbqs1519}}

\tablehead{ \colhead{$g$} & \colhead{$\leq5\arcmin$} & \colhead{$5\arcmin$-$10\arcmin$} & \colhead{$10\arcmin$-$15\arcmin$} & \colhead{$15\arcmin$-$20\arcmin$} }

\startdata
$\leq18$ & \nodata &   8/9   &  19/19  &  25/27  \\
18-18.5  & \nodata &   8/8   &  11/11  &  20/21  \\
18.5-19  &   2/2   &   9/9   &  19/20  &  32/37  \\
19-19.5  &   2/2   &   5/9   &  14/25  &  16/43  \\
19.5-20  &   5/7   &  18/21  &  14/24  &  32/60  \\
\enddata

\end{deluxetable}
\begin{deluxetable}{c|CCCC}
\tablecaption{H~1821+643 Completeness
\label{tab:h1821}}

\tablehead{ \colhead{$g$} & \colhead{$\leq5\arcmin$} & \colhead{$5\arcmin$-$10\arcmin$} & \colhead{$10\arcmin$-$15\arcmin$} & \colhead{$15\arcmin$-$20\arcmin$} }

\startdata
$\leq18$ &   0/3   &   1/1   &   6/7   &  10/10  \\
18-18.5  &   1/1   &   2/2   &   3/4   &   5/5   \\
18.5-19  &   1/1   &   2/2   &   5/6   &  11/11  \\
19-19.5  &   3/3   &   5/7   &   5/6   &   7/9   \\
19.5-20  &   2/2   &  10/10  &   9/13  &  26/28  \\
\enddata

\end{deluxetable}

\clearpage
\begin{deluxetable}{c|CCCC}
\tablecaption{HE~0153--4520 Completeness
\label{tab:he0153}}

\tablehead{ \colhead{$g$} & \colhead{$\leq5\arcmin$} & \colhead{$5\arcmin$-$10\arcmin$} & \colhead{$10\arcmin$-$15\arcmin$} & \colhead{$15\arcmin$-$20\arcmin$} }

\startdata
$\leq18$ &   5/5   &   6/6   &   8/8   &   8/8   \\
18-18.5  & \nodata & \nodata &   2/2   &   4/4   \\
18.5-19  & \nodata &   1/1   &   2/2   &   2/2   \\
19-19.5  & \nodata &   2/2   &   5/5   &   7/8   \\
19.5-20  &   4/4   &   5/7   &   8/9   &  18/18  \\
\enddata

\end{deluxetable}
\begin{deluxetable}{c|CCCC}
\tablecaption{HE~0226--4110 Completeness
\label{tab:he0226}}

\tablehead{ \colhead{$g$} & \colhead{$\leq5\arcmin$} & \colhead{$5\arcmin$-$10\arcmin$} & \colhead{$10\arcmin$-$15\arcmin$} & \colhead{$15\arcmin$-$20\arcmin$} }

\startdata
$\leq18$ &   1/1   &   3/3   &   2/2   &   3/3   \\
18-18.5  & \nodata & \nodata & \nodata &   4/4   \\
18.5-19  & \nodata &   2/3   &   2/2   &   3/3   \\
19-19.5  &   1/1   &   2/2   &   6/8   &   7/8   \\
19.5-20  &   1/1   &   7/7   &  11/13  &  12/13  \\
\enddata

\end{deluxetable}
\begin{deluxetable}{c|CCCC}
\tablecaption{HE~0435--5304 Completeness
\label{tab:he0435}}

\tablehead{ \colhead{$g$} & \colhead{$\leq5\arcmin$} & \colhead{$5\arcmin$-$10\arcmin$} & \colhead{$10\arcmin$-$15\arcmin$} & \colhead{$15\arcmin$-$20\arcmin$} }

\startdata
$\leq18$ & \nodata &   2/4   &   5/6   &  17/17  \\
18-18.5  &   1/1   &   1/2   &   2/3   &   1/3   \\
18.5-19  & \nodata &   4/5   &   5/8   &  13/18  \\
19-19.5  &   3/4   &   3/6   &   8/11  &  17/23  \\
19.5-20  &   4/4   &  14/21  &  20/26  &  25/37  \\
\enddata

\end{deluxetable}
\begin{deluxetable}{c|CCCC}
\tablecaption{HE~0439--5254 Completeness
\label{tab:he0439}}

\tablehead{ \colhead{$g$} & \colhead{$\leq5\arcmin$} & \colhead{$5\arcmin$-$10\arcmin$} & \colhead{$10\arcmin$-$15\arcmin$} & \colhead{$15\arcmin$-$20\arcmin$} }

\startdata
$\leq18$ & \nodata &   8/9   &  16/19  &  12/13  \\
18-18.5  & \nodata &   5/5   &   1/3   &   2/3   \\
18.5-19  &   0/1   &   6/8   &  11/16  &  14/15  \\
19-19.5  & \nodata &   5/9   &  11/17  &  17/22  \\
19.5-20  &   1/2   &  10/17  &  13/30  &  27/33  \\
\enddata

\end{deluxetable}
\begin{deluxetable}{c|CCCC}
\tablecaption{HS~1102+3441 Completeness
\label{tab:hs1102}}

\tablehead{ \colhead{$g$} & \colhead{$\leq5\arcmin$} & \colhead{$5\arcmin$-$10\arcmin$} & \colhead{$10\arcmin$-$15\arcmin$} & \colhead{$15\arcmin$-$20\arcmin$} }

\startdata
$\leq18$ & \nodata &   2/2   &   8/8   &   9/9   \\
18-18.5  & \nodata & \nodata &   5/5   &   3/5   \\
18.5-19  &   1/1   &   4/4   &  13/13  &  10/12  \\
19-19.5  &   1/1   &   2/3   &  15/19  &  13/17  \\
19.5-20  &   3/3   &  16/17  &  23/27  &  39/44  \\
\enddata

\end{deluxetable}
\begin{deluxetable}{c|CCCC}
\tablecaption{Mrk~421 Completeness
\label{tab:mrk421}}

\tablehead{ \colhead{$g$} & \colhead{$\leq5\arcmin$} & \colhead{$5\arcmin$-$10\arcmin$} & \colhead{$10\arcmin$-$15\arcmin$} & \colhead{$15\arcmin$-$20\arcmin$} }

\startdata
$\leq18$ &   4/6   &   6/7   &  12/14  &  15/15  \\
18-18.5  & \nodata & \nodata & \nodata & \nodata \\
18.5-19  & \nodata &   1/2   & \nodata &   2/2   \\
19-19.5  &   1/1   & \nodata &   0/1   & \nodata \\
19.5-20  &   1/1   &   2/3   & \nodata & \nodata \\
\enddata

\end{deluxetable}

\clearpage
\begin{deluxetable}{c|CCCC}
\tablecaption{PG~0832+251 Completeness
\label{tab:pg0832}}

\tablehead{ \colhead{$g$} & \colhead{$\leq5\arcmin$} & \colhead{$5\arcmin$-$10\arcmin$} & \colhead{$10\arcmin$-$15\arcmin$} & \colhead{$15\arcmin$-$20\arcmin$} }

\startdata
$\leq18$ &   3/3   &   1/1   &   6/7   &  11/11  \\
18-18.5  &   1/1   &   1/1   &   1/1   &   6/6   \\
18.5-19  &   1/1   &   2/3   &   6/7   &  11/12  \\
19-19.5  &   4/4   &   2/2   &   8/12  &  12/13  \\
19.5-20  &   3/3   &   6/9   &   9/11  &  17/22  \\
\enddata

\end{deluxetable}
\begin{deluxetable}{c|CCCC}
\tablecaption{PG~0953+414 Completeness
\label{tab:pg0953}}

\tablehead{ \colhead{$g$} & \colhead{$\leq5\arcmin$} & \colhead{$5\arcmin$-$10\arcmin$} & \colhead{$10\arcmin$-$15\arcmin$} & \colhead{$15\arcmin$-$20\arcmin$} }

\startdata
$\leq18$ &   1/1   &   6/6   &   4/4   &  10/11  \\
18-18.5  &   1/1   &   4/5   &   4/4   &   2/4   \\
18.5-19  & \nodata &   3/4   &  13/13  &   7/9   \\
19-19.5  &   2/2   &   9/10  &   6/9   &  20/24  \\
19.5-20  &   5/5   &  10/13  &  16/20  &  15/25  \\
\enddata

\end{deluxetable}
\begin{deluxetable}{c|CCCC}
\tablecaption{PG~1001+291 Completeness
\label{tab:pg1001}}

\tablehead{ \colhead{$g$} & \colhead{$\leq5\arcmin$} & \colhead{$5\arcmin$-$10\arcmin$} & \colhead{$10\arcmin$-$15\arcmin$} & \colhead{$15\arcmin$-$20\arcmin$} }

\startdata
$\leq18$ & \nodata &   3/3   &   6/7   &  17/18  \\
18-18.5  &   1/1   &   6/6   &   7/8   &   8/8   \\
18.5-19  &   1/2   &   4/4   &  11/12  &  12/13  \\
19-19.5  &   1/1   &   8/8   &   9/9   &  14/14  \\
19.5-20  &   2/2   &  14/16  &  17/24  &  10/21  \\
\enddata

\end{deluxetable}
\begin{deluxetable}{c|CCCC}
\tablecaption{PG~1048+342 Completeness
\label{tab:pg1048}}

\tablehead{ \colhead{$g$} & \colhead{$\leq5\arcmin$} & \colhead{$5\arcmin$-$10\arcmin$} & \colhead{$10\arcmin$-$15\arcmin$} & \colhead{$15\arcmin$-$20\arcmin$} }

\startdata
$\leq18$ & \nodata &   4/4   &   1/2   &   4/5   \\
18-18.5  & \nodata &   4/4   &   1/1   &   2/2   \\
18.5-19  &   1/1   &   4/4   & \nodata &   6/6   \\
19-19.5  &   1/1   &   4/4   &  10/11  &  11/12  \\
19.5-20  &   2/3   &   5/6   &   9/11  &   5/8   \\
\enddata

\end{deluxetable}
\begin{deluxetable}{c|CCCC}
\tablecaption{PG~1115+407 Completeness
\label{tab:pg1115}}

\tablehead{ \colhead{$g$} & \colhead{$\leq5\arcmin$} & \colhead{$5\arcmin$-$10\arcmin$} & \colhead{$10\arcmin$-$15\arcmin$} & \colhead{$15\arcmin$-$20\arcmin$} }

\startdata
$\leq18$ & \nodata &   1/1   &   2/2   &   3/4   \\
18-18.5  & \nodata &   5/5   &   2/2   &   3/3   \\
18.5-19  &   2/2   &   4/4   &   7/7   &   3/3   \\
19-19.5  &   2/2   &   7/7   &  12/12  &   9/9   \\
19.5-20  &   3/3   &   6/7   &  12/14  &  10/10  \\
\enddata

\end{deluxetable}
\begin{deluxetable}{c|CCCC}
\tablecaption{PG~1116+215 Completeness
\label{tab:pg1116}}

\tablehead{ \colhead{$g$} & \colhead{$\leq5\arcmin$} & \colhead{$5\arcmin$-$10\arcmin$} & \colhead{$10\arcmin$-$15\arcmin$} & \colhead{$15\arcmin$-$20\arcmin$} }

\startdata
$\leq18$ &   1/1   &   4/4   &  11/11  &   5/5   \\
18-18.5  &   1/1   &   2/2   &   5/5   &   7/7   \\
18.5-19  &   3/3   &   4/4   &   9/9   &  11/11  \\
19-19.5  &   5/5   &  14/14  &  13/15  &  10/14  \\
19.5-20  &   5/6   &   8/8   &  22/29  &  11/26  \\
\enddata

\end{deluxetable}

\clearpage
\begin{deluxetable}{c|CCCC}
\tablecaption{PG~1121+422 Completeness
\label{tab:pg1121}}

\tablehead{ \colhead{$g$} & \colhead{$\leq5\arcmin$} & \colhead{$5\arcmin$-$10\arcmin$} & \colhead{$10\arcmin$-$15\arcmin$} & \colhead{$15\arcmin$-$20\arcmin$} }

\startdata
$\leq18$ &   1/1   &   1/2   &   2/2   &   4/4   \\
18-18.5  & \nodata &   3/3   & \nodata &   5/5   \\
18.5-19  &   2/2   &   6/6   &   8/10  &  10/10  \\
19-19.5  &   2/2   &   7/9   &  15/15  &  25/25  \\
19.5-20  &   6/6   &  12/12  &  20/22  &  25/30  \\
\enddata

\end{deluxetable}
\begin{deluxetable}{c|CCCC}
\tablecaption{PG~1216+069 Completeness
\label{tab:pg1216}}

\tablehead{ \colhead{$g$} & \colhead{$\leq5\arcmin$} & \colhead{$5\arcmin$-$10\arcmin$} & \colhead{$10\arcmin$-$15\arcmin$} & \colhead{$15\arcmin$-$20\arcmin$} }

\startdata
$\leq18$ &   1/1   &   4/5   &   4/4   &   4/4   \\
18-18.5  &   1/1   &   3/3   &   4/4   &   7/9   \\
18.5-19  &   1/1   &   6/6   &   4/4   &   5/6   \\
19-19.5  & \nodata &   9/9   &   9/10  &  12/12  \\
19.5-20  &   3/3   &   8/8   &   9/11  &  18/23  \\
\enddata

\end{deluxetable}
\begin{deluxetable}{c|CCCC}
\tablecaption{PG~1222+216 Completeness
\label{tab:pg1222}}

\tablehead{ \colhead{$g$} & \colhead{$\leq5\arcmin$} & \colhead{$5\arcmin$-$10\arcmin$} & \colhead{$10\arcmin$-$15\arcmin$} & \colhead{$15\arcmin$-$20\arcmin$} }

\startdata
$\leq18$ &   1/1   &   3/3   &   6/6   &   2/2   \\
18-18.5  &   1/1   & \nodata &   1/1   &   5/7   \\
18.5-19  &   1/1   &   3/3   &   7/7   &   7/8   \\
19-19.5  &   0/3   &   5/8   &   5/13  &   2/14  \\
19.5-20  &   2/4   &   6/9   &  17/29  &   6/33  \\
\enddata

\end{deluxetable}
\begin{deluxetable}{c|CCCC}
\tablecaption{PG~1259+593 Completeness
\label{tab:pg1259}}

\tablehead{ \colhead{$g$} & \colhead{$\leq5\arcmin$} & \colhead{$5\arcmin$-$10\arcmin$} & \colhead{$10\arcmin$-$15\arcmin$} & \colhead{$15\arcmin$-$20\arcmin$} }

\startdata
$\leq18$ &   2/2   &   4/4   &   1/3   &   7/7   \\
18-18.5  & \nodata &   6/6   &   3/3   &  10/11  \\
18.5-19  & \nodata &   4/4   &   8/8   &  18/18  \\
19-19.5  &   2/2   &   9/10  &  15/16  &  14/15  \\
19.5-20  &   6/6   &  18/19  &  21/26  &  31/35  \\
\enddata

\end{deluxetable}
\begin{deluxetable}{c|CCCC}
\tablecaption{PG~1424+240 Completeness
\label{tab:pg1424}}

\tablehead{ \colhead{$g$} & \colhead{$\leq5\arcmin$} & \colhead{$5\arcmin$-$10\arcmin$} & \colhead{$10\arcmin$-$15\arcmin$} & \colhead{$15\arcmin$-$20\arcmin$} }

\startdata
$\leq18$ &   1/1   &   4/5   &   5/5   &  10/10  \\
18-18.5  &   1/1   &   1/1   &   7/7   &  16/16  \\
18.5-19  &   1/1   & \nodata &   6/6   &   8/8   \\
19-19.5  &   0/1   &   2/3   &   7/10  &   6/17  \\
19.5-20  &   1/1   &   4/4   &   4/6   &   7/27  \\
\enddata

\end{deluxetable}
\begin{deluxetable}{c|CCCC}
\tablecaption{PG~1626+554 Completeness
\label{tab:pg1626}}

\tablehead{ \colhead{$g$} & \colhead{$\leq5\arcmin$} & \colhead{$5\arcmin$-$10\arcmin$} & \colhead{$10\arcmin$-$15\arcmin$} & \colhead{$15\arcmin$-$20\arcmin$} }

\startdata
$\leq18$ & \nodata &   5/6   &   9/9   &   6/7   \\
18-18.5  &   3/3   &   3/3   &   6/7   &  13/14  \\
18.5-19  &   1/1   &   6/7   &  12/13  &  12/12  \\
19-19.5  &   2/2   &   3/7   &   8/16  &  16/20  \\
19.5-20  &   4/4   &  14/16  &  14/19  &  15/25  \\
\enddata

\end{deluxetable}

\clearpage
\begin{deluxetable}{c|CCCC}
\tablecaption{PHL~1811 Completeness
\label{tab:phl1811}}

\tablehead{ \colhead{$g$} & \colhead{$\leq5\arcmin$} & \colhead{$5\arcmin$-$10\arcmin$} & \colhead{$10\arcmin$-$15\arcmin$} & \colhead{$15\arcmin$-$20\arcmin$} }

\startdata
$\leq18$ &   1/1   &   9/9   &  11/11  &  11/11  \\
18-18.5  &   3/3   &   6/6   &   4/4   &  13/13  \\
18.5-19  &   1/1   &   5/6   &  16/16  &  20/21  \\
19-19.5  &   6/6   &  12/14  &  26/30  &  19/27  \\
19.5-20  &   4/5   &  16/20  &  15/22  &  37/54  \\
\enddata

\end{deluxetable}
\begin{deluxetable}{c|CCCC}
\tablecaption{PKS~2005--489 Completeness
\label{tab:pks2005}}

\tablehead{ \colhead{$g$} & \colhead{$\leq5\arcmin$} & \colhead{$5\arcmin$-$10\arcmin$} & \colhead{$10\arcmin$-$15\arcmin$} & \colhead{$15\arcmin$-$20\arcmin$} }

\startdata
$\leq18$ &   9/18  &  16/16  &  23/23  &  32/42  \\
18-18.5  & \nodata &   2/2   &   3/3   &   9/11  \\
18.5-19  &   3/5   &   8/8   &   8/9   &  10/11  \\
19-19.5  &   4/6   &   8/9   &  10/11  &  18/19  \\
19.5-20  &   3/5   &  14/17  &  26/30  &  27/41  \\
\enddata

\end{deluxetable}
\begin{deluxetable}{c|CCCC}
\tablecaption{Q~1230+0115 Completeness
\label{tab:q1230}}

\tablehead{ \colhead{$g$} & \colhead{$\leq5\arcmin$} & \colhead{$5\arcmin$-$10\arcmin$} & \colhead{$10\arcmin$-$15\arcmin$} & \colhead{$15\arcmin$-$20\arcmin$} }

\startdata
$\leq18$ &   1/1   &   1/1   &   4/4   &   6/6   \\
18-18.5  & \nodata &   1/1   &   4/4   &   6/6   \\
18.5-19  &   1/1   &   2/2   &   1/1   &   4/5   \\
19-19.5  &   1/1   &   1/2   &   8/9   &   6/8   \\
19.5-20  &   1/1   &   4/8   &   6/7   &   8/11  \\
\enddata

\end{deluxetable}
\begin{deluxetable}{c|CCCC}
\tablecaption{RBS~711 Completeness
\label{tab:rbs711}}

\tablehead{ \colhead{$g$} & \colhead{$\leq5\arcmin$} & \colhead{$5\arcmin$-$10\arcmin$} & \colhead{$10\arcmin$-$15\arcmin$} & \colhead{$15\arcmin$-$20\arcmin$} }

\startdata
$\leq18$ &   2/2   &   5/5   &   7/7   &   8/8   \\
18-18.5  &   1/1   &   4/4   &   4/5   &  11/12  \\
18.5-19  &   1/1   &  12/14  &   7/8   &  13/17  \\
19-19.5  &   4/4   &  14/16  &  17/22  &  15/26  \\
19.5-20  &   4/5   &  11/14  &  24/37  &  20/30  \\
\enddata

\end{deluxetable}
\begin{deluxetable}{c|CCCC}
\tablecaption{RX~J0439.6--5311 Completeness
\label{tab:rxj0439}}

\tablehead{ \colhead{$g$} & \colhead{$\leq5\arcmin$} & \colhead{$5\arcmin$-$10\arcmin$} & \colhead{$10\arcmin$-$15\arcmin$} & \colhead{$15\arcmin$-$20\arcmin$} }

\startdata
$\leq18$ & \nodata &   7/9   &   9/11  &  18/18  \\
18-18.5  &   1/1   &   2/3   &   6/11  &   4/7   \\
18.5-19  &   2/3   &   3/6   &   9/15  &   9/15  \\
19-19.5  &   1/4   &   5/10  &  10/16  &  18/26  \\
19.5-20  &   4/5   &   7/16  &  17/27  &  28/47  \\
\enddata

\end{deluxetable}
\begin{deluxetable}{c|CCCC}
\tablecaption{RX~J2154.1--4414 Completeness
\label{tab:rxj2154}}

\tablehead{ \colhead{$g$} & \colhead{$\leq5\arcmin$} & \colhead{$5\arcmin$-$10\arcmin$} & \colhead{$10\arcmin$-$15\arcmin$} & \colhead{$15\arcmin$-$20\arcmin$} }

\startdata
$\leq18$ &   4/4   &   1/1   &  10/10  &   2/3   \\
18-18.5  & \nodata &   2/2   &   4/4   &   3/3   \\
18.5-19  &   2/3   &   4/4   &  12/13  &   7/9   \\
19-19.5  &   1/1   &   7/9   &  12/14  &  13/15  \\
19.5-20  &   1/1   &   8/14  &  13/24  &  15/27  \\
\enddata

\end{deluxetable}

\clearpage
\begin{deluxetable}{c|CCCC}
\tablecaption{S5~0716+714 Completeness
\label{tab:s0716}}

\tablehead{ \colhead{$g$} & \colhead{$\leq5\arcmin$} & \colhead{$5\arcmin$-$10\arcmin$} & \colhead{$10\arcmin$-$15\arcmin$} & \colhead{$15\arcmin$-$20\arcmin$} }

\startdata
$\leq18$ & \nodata &   4/4   &   6/6   &   9/9   \\
18-18.5  &   1/1   & \nodata &   3/3   &   4/7   \\
18.5-19  &   2/3   &   2/4   &   6/7   &   9/10  \\
19-19.5  &   0/3   &  11/18  &   6/17  &  11/17  \\
19.5-20  &   4/10  &   7/13  &  18/44  &  12/39  \\
\enddata

\end{deluxetable}
\begin{deluxetable}{c|CCCC}
\tablecaption{SBS~0956+509 Completeness
\label{tab:sbs0956}}

\tablehead{ \colhead{$g$} & \colhead{$\leq5\arcmin$} & \colhead{$5\arcmin$-$10\arcmin$} & \colhead{$10\arcmin$-$15\arcmin$} & \colhead{$15\arcmin$-$20\arcmin$} }

\startdata
$\leq18$ &   1/1   &   3/3   &  10/11  &   6/7   \\
18-18.5  &   3/4   &   1/1   &   4/5   &   2/3   \\
18.5-19  &   5/5   &   9/14  &   8/8   &   8/9   \\
19-19.5  &   7/8   &   4/5   &   7/9   &   4/7   \\
19.5-20  &   4/5   &   6/11  &  15/20  &   7/9   \\
\enddata

\end{deluxetable}
\begin{deluxetable}{c|CCCC}
\tablecaption{SBS~1108+560 Completeness
\label{tab:sbs1108}}

\tablehead{ \colhead{$g$} & \colhead{$\leq5\arcmin$} & \colhead{$5\arcmin$-$10\arcmin$} & \colhead{$10\arcmin$-$15\arcmin$} & \colhead{$15\arcmin$-$20\arcmin$} }

\startdata
$\leq18$ &   2/2   &   4/5   &   6/6   &   6/7   \\
18-18.5  &   4/4   &   1/1   &   5/6   &   8/8   \\
18.5-19  &   2/2   &   9/9   &  11/11  &  14/14  \\
19-19.5  &   2/2   &   7/7   &  11/13  &  17/19  \\
19.5-20  &   6/6   &  17/20  &  23/24  &  29/34  \\
\enddata

\end{deluxetable}
\begin{deluxetable}{c|CCCC}
\tablecaption{SBS~1122+594 Completeness
\label{tab:sbs1122}}

\tablehead{ \colhead{$g$} & \colhead{$\leq5\arcmin$} & \colhead{$5\arcmin$-$10\arcmin$} & \colhead{$10\arcmin$-$15\arcmin$} & \colhead{$15\arcmin$-$20\arcmin$} }

\startdata
$\leq18$ &   1/1   &   6/6   &  12/15  &   8/8   \\
18-18.5  & \nodata &   6/6   &   8/8   &   8/8   \\
18.5-19  &   2/2   &   4/4   &  13/13  &   8/8   \\
19-19.5  &   6/6   &  10/10  &  14/14  &  19/21  \\
19.5-20  &   1/1   &  12/14  &  25/30  &  26/30  \\
\enddata

\end{deluxetable}
\begin{deluxetable}{c|CCCC}
\tablecaption{SDSS~J1028+2119 Completeness
\label{tab:sdss1028}}

\tablehead{ \colhead{$g$} & \colhead{$\leq5\arcmin$} & \colhead{$5\arcmin$-$10\arcmin$} & \colhead{$10\arcmin$-$15\arcmin$} & \colhead{$15\arcmin$-$20\arcmin$} }

\startdata
$\leq18$ &   3/3   &   5/5   &   5/6   &   7/8   \\
18-18.5  &   1/2   &   7/9   &   1/2   &   4/6   \\
18.5-19  &   2/2   &   5/8   &   4/6   &   5/14  \\
19-19.5  &   3/3   &   5/10  &  11/17  &  11/21  \\
19.5-20  &   7/8   &   4/14  &  10/18  &  13/36  \\
\enddata

\end{deluxetable}
\begin{deluxetable}{c|CCCC}
\tablecaption{SDSS~J1333+4518 Completeness
\label{tab:sdss1333}}

\tablehead{ \colhead{$g$} & \colhead{$\leq5\arcmin$} & \colhead{$5\arcmin$-$10\arcmin$} & \colhead{$10\arcmin$-$15\arcmin$} & \colhead{$15\arcmin$-$20\arcmin$} }

\startdata
$\leq18$ &   1/1   &   3/3   &   2/2   &   1/2   \\
18-18.5  &   1/2   &   3/4   &   4/4   &   5/5   \\
18.5-19  &   2/3   &   6/7   &   7/8   &   4/6   \\
19-19.5  &   2/2   &   7/11  &  11/14  &  11/14  \\
19.5-20  &   4/4   &  12/15  &  21/24  &  28/42  \\
\enddata

\end{deluxetable}

\clearpage
\begin{deluxetable}{c|CCCC}
\tablecaption{SDSS~J1439+3932 Completeness
\label{tab:sdss1439}}

\tablehead{ \colhead{$g$} & \colhead{$\leq5\arcmin$} & \colhead{$5\arcmin$-$10\arcmin$} & \colhead{$10\arcmin$-$15\arcmin$} & \colhead{$15\arcmin$-$20\arcmin$} }

\startdata
$\leq18$ & \nodata &   4/4   &   5/5   &   3/3   \\
18-18.5  &   1/1   &   2/2   &   0/1   &   4/4   \\
18.5-19  &   1/1   &   2/2   &   3/4   &  17/17  \\
19-19.5  &   4/4   &   9/9   &  10/12  &  17/18  \\
19.5-20  &   3/3   &  19/21  &  29/34  &  28/34  \\
\enddata

\end{deluxetable}
\begin{deluxetable}{c|CCCC}
\tablecaption{SDSS~J1540--0205 Completeness
\label{tab:sdss1540}}

\tablehead{ \colhead{$g$} & \colhead{$\leq5\arcmin$} & \colhead{$5\arcmin$-$10\arcmin$} & \colhead{$10\arcmin$-$15\arcmin$} & \colhead{$15\arcmin$-$20\arcmin$} }

\startdata
$\leq18$ & \nodata &   4/4   &   3/6   &   1/1   \\
18-18.5  &   2/2   &   2/2   &   4/5   &   3/4   \\
18.5-19  &   1/1   &   6/7   &   8/9   &   6/7   \\
19-19.5  &   6/7   &  12/15  &   7/17  &   4/6   \\
19.5-20  &   6/8   &  16/26  &  14/21  &  11/16  \\
\enddata

\end{deluxetable}
\begin{deluxetable}{c|CCCC}
\tablecaption{Ton~236 Completeness
\label{tab:ton236}}

\tablehead{ \colhead{$g$} & \colhead{$\leq5\arcmin$} & \colhead{$5\arcmin$-$10\arcmin$} & \colhead{$10\arcmin$-$15\arcmin$} & \colhead{$15\arcmin$-$20\arcmin$} }

\startdata
$\leq18$ & \nodata &  14/15  &  12/12  &  15/16  \\
18-18.5  &   2/2   &   8/8   &   3/3   &  11/13  \\
18.5-19  &   1/1   &   8/8   &   6/6   &  13/14  \\
19-19.5  &   7/7   &  10/10  &  17/17  &  14/16  \\
19.5-20  &   6/7   &  16/16  &  23/29  &  37/49  \\
\enddata

\end{deluxetable}
\begin{deluxetable}{c|CCCC}
\tablecaption{Ton~580 Completeness
\label{tab:ton580}}

\tablehead{ \colhead{$g$} & \colhead{$\leq5\arcmin$} & \colhead{$5\arcmin$-$10\arcmin$} & \colhead{$10\arcmin$-$15\arcmin$} & \colhead{$15\arcmin$-$20\arcmin$} }

\startdata
$\leq18$ & \nodata &   1/1   &   2/2   &   3/3   \\
18-18.5  &   1/1   & \nodata &   6/6   &   6/6   \\
18.5-19  & \nodata &   3/3   &   1/2   &   5/5   \\
19-19.5  &   1/1   &   7/7   &   8/9   &  11/11  \\
19.5-20  &   3/3   &   6/7   &  13/15  &  14/15  \\
\enddata

\end{deluxetable}
\begin{deluxetable}{c|CCCC}
\tablecaption{Ton~1187 Completeness
\label{tab:ton1187}}

\tablehead{ \colhead{$g$} & \colhead{$\leq5\arcmin$} & \colhead{$5\arcmin$-$10\arcmin$} & \colhead{$10\arcmin$-$15\arcmin$} & \colhead{$15\arcmin$-$20\arcmin$} }

\startdata
$\leq18$ &   1/1   &   7/7   &   8/8   &  16/16  \\
18-18.5  &   1/1   &   1/1   &   3/3   &   5/5   \\
18.5-19  &   0/1   &   2/2   & \nodata & \nodata \\
19-19.5  & \nodata &   2/4   & \nodata &   6/7   \\
19.5-20  & \nodata &   2/3   &   3/5   &   0/1   \\
\enddata

\end{deluxetable}
\begin{deluxetable}{c|CCCC}
\tablecaption{VII~Zw~244 Completeness
\label{tab:viizw244}}

\tablehead{ \colhead{$g$} & \colhead{$\leq5\arcmin$} & \colhead{$5\arcmin$-$10\arcmin$} & \colhead{$10\arcmin$-$15\arcmin$} & \colhead{$15\arcmin$-$20\arcmin$} }

\startdata
$\leq18$ & \nodata & \nodata &   1/1   &   4/4   \\
18-18.5  & \nodata &   2/2   &   4/4   &   4/4   \\
18.5-19  & \nodata &   2/2   &   2/2   &  12/12  \\
19-19.5  &   2/2   &   3/3   &  10/10  &  14/23  \\
19.5-20  &   4/4   &  10/10  &  13/15  &  23/32  \\
\enddata

\end{deluxetable}

\end{document}